\documentclass{aa} 

\usepackage{graphicx}
\usepackage{xcolor}
\usepackage{txfonts}
\usepackage{hyperref}
\usepackage{natbib}
 \newcommand{\heii}{\ion{He}{ii}\xspace}
 \newcommand{\oi}{[\ion{O}{i}]\xspace}
 
 \newcommand{\oiii}{[\ion{O}{iii}]\xspace}
 \newcommand{\nii}{[\ion{N}{ii}]\xspace}
 \newcommand{\sii}{[\ion{S}{ii}]\xspace}
 \newcommand{\siii}{[\ion{S}{iii}]\xspace}
 \newcommand{\hiireg}{\textsc{H\,ii}~region\xspace}
 \newcommand{\hiiregs}{\textsc{H\,ii}~regions\xspace}

\begin{document}

   \title{MUSE crowded field 3D spectroscopy  in NGC\,300}

   \subtitle{I. First results from central fields\thanks{Based on observations obtained at the Very Large Telescope (VLT) 
          of the European Southern Observatory, Paranal, Chile (ESO Programme ID 094.D-0116(A), 094.D-0116(B), 095.D-0173(A))  }}

   \author{Martin M. Roth
          \inst{1}
          \and
          Christer Sandin\inst{1}
          \and
          Sebastian Kamann\inst{2,3}  
          \and
          Tim-Oliver Husser\inst{2}  
          \and
          Peter M. Weilbacher\inst{1}
          \and
          Ana Monreal-Ibero\inst{4,5}
          \and
          Roland Bacon\inst{7}  
          \and
          Mark den Brok\inst{1}
          \and
          Stefan Dreizler\inst{2}
          \and
          Andreas Kelz \inst{1}
          \and
          Raffaella Anna Marino\inst{6}
          \and
          Matthias Steinmetz\inst{1}      
          }

   \institute{Leibniz-Institut f\"ur Astrophysik Potsdam (AIP),
              An der Sternwarte 16, 14482, Potsdam, Germany
         \and
             Institute for Astrophysics, University of  G\"ottingen, Friedrich-Hund-Platz 1, 37077, G\"ottingen, Germany
         \and
             Astrophysics Research Institute, Liverpool John Moores University, 146 Brownlow Hill, Liverpool L3 5RF, United Kingdom
         \and
           Instituto de Astrof\'{\i}sica de Canarias (IAC), E-38205 La Laguna, Tenerife, Spain
        \and
           Universidad de La Laguna, Dpto.\ Astrof\'{\i}sica, E-38206 La Laguna, Tenerife, Spain 
         \and
            Department of Physics, ETH Z\"urich,Wolfgang$-$Pauli$-$Strasse\,27, 8093\,Z\"urich, Switzerland
         \and
            CRAL, Observatoire de Lyon, CNRS, Universit\'{e} Lyon 1, 9 avenue Ch. Andr\'{e}, 69561 Saint Genis-Laval Cedex, France\\
             }

   \date{Received 13 March, 2018; accepted 6 June, 2018}

 
  \abstract
   {}
   {As a new approach to the study of resolved stellar populations in nearby galaxies, 
   our goal is to demonstrate with a pilot study in NGC\,300 that integral field spectroscopy with high spatial resolution 
   and excellent seeing conditions reaches an unprecedented depth in severely crowded fields.}
   {MUSE observations with seven pointings in NGC\,300 have resulted in datacubes 
   that are analyzed in four ways:
   (1) PSF-fitting 3D spectroscopy with {\textsc PampelMUSE}, as already successfully pioneered in globular clusters, 
   yields deblended spectra of individually distinguishable stars, thus providing a complete inventory of 
   blue/red supergiants, and AGB stars of type M and C. The technique is also applicable to 
   emission line point sources and provides samples of planetary nebulae that are complete down to
   m$_{5007}$=28.
   (2) pseudo-monochromatic images, created at the wavelengths of the most important emission lines
   and corrected for continuum light by using the {\textsc P3D} visualization tool, provide maps of \hiiregs, 
   supernova remnants, and the diffuse interstellar medium at a high level of sensitivity, where also 
   faint point sources stand out and allow for the discovery of planetary nebulae, WR stars etc. 
   (3) The use of the {\textsc P3D} line-fitting tool yields emission line fluxes, surface brightness, and kinematic 
   information for gaseous objects, corrected for absorption line profiles of the underlying stellar 
   population in the case of H$\alpha$.
   (4) Visual inspection of the datacubes by browsing through the row-stacked-spectra image in {\textsc P3D} is 
   demonstrated to be efficient for data mining and the discovery of background galaxies and 
   unusual objects.}
   {We present a catalogue of luminous stars, rare stars such as WR and other emission line stars, 
   carbon stars, symbiotic star candidates, planetary nebulae, \hiiregs, supernova remnants, 
   giant shells, peculiar diffuse and filamentary emission line objects, and background galaxies, 
   along with their spectra.}
   {The technique of crowded-field 3D spectroscopy, using the {\textsc PampelMUSE} code, is capable 
   of deblending individual bright stars, the unresolved background of faint stars, gaseous nebulae, 
   and the diffuse component of the interstellar medium, resulting in unprecedented legacy value for 
   observations of nearby galaxies with MUSE.}


   \maketitle
%

\section{Introduction}
\label{section_introduction}
The quest for understanding the formation and evolution of galaxies has provided us with a wealth of data from imaging and spectroscopic surveys, most notably the Sloan Digital Sky Survey (SDSS), to quote just one prominent example \citep{york2000,alam2015}. As a more recent development, integral field spectroscopy (IFS) enables us to obtain spatially resolved information on stellar populations, gas, dust, and the star formation history, including spatially resolved kinematics across the entire face of a galaxy, rather than the limited apertures  of a single fibre, or a slit   \citep[see][]{sanchez2012}. The CALIFA survey is the first effort to fully exploiting this capability with a reasonably large sample of galaxies of all Hubble types \citep{walcher2014,husemann2013,garcia2015}.
Surveys with deployable integral field units (IFU) like SAMI \citep{cortese2014} and MaNGA \citep{bundy2015} are carrying this approach further to much larger sample sizes. However, given the integration of light from typically hundreds of thousands of individual stars over a spatial resolution  element (spaxel) of the respective integral field units, it is not entirely clear  whether the technique of stellar population synthesis for the resulting spectra is capable to uniquely determine stellar populations and their star formation history. The study of resolved stellar populations in nearby galaxies, where stars can be measured accurately on an individual basis, would be able to lift any degeneracies and allow to calibrate the integral light measurements of the more distant galaxies. Such work has been  attempted with Hubble Space Telescope (HST) photometry, e.g. in the ANGST survey \citep{dalcanton2009}, for the galaxy NGC\,300 studied in this paper see specifically \citet{gogarten2010}. However limitations of photometry, e.g. the age--metallicity degeneracy, sensitivity to extinction, or the lacking ability to distinguish objects such as Wolf-Rayet (WR) stars from normal O stars, carbon stars from M stars, etc. would rather call for spatially resolved spectroscopy. Except for conventional slit spectroscopy  of individual bright giants or supergiants in local volume galaxies, e.g. \citet{bresolin2001}, \citet{smartt2001} in the optical,  or \citet{urbaneja2005}, \citet{gazak2015} in the NIR, there exists to date no comprehensive spectroscopic dataset for stars in heavily  crowded fields of nearby galaxies beyond the Magellanic Clouds. Pioneering attempts to utilize the technique of IFS for crowded field spectroscopy, i.e. fitting the point spread function (PSF) to point sources through the layers of a datacube, ana\-lo\-gous to CCD photometry in a single image, e.g. {\textsc DAOPHOT} \citep{stetson1987}, have demonstrated that it is indeed possible to obtain deblended spectra for individual point sources in heavily  crowded fields in local group galaxies , e.g. \cite{becker2004}, \cite{roth2004}, \cite{fabrika2005}. These first attempts were limited by the small field-of-view (FoV) and the coarse spatial sampling provided by first generation IFUs, e.g. $8\times8$ arcsec$^2$ at 0.5 arcsec sampling in the case of PMAS \citep{roth2005}, the light collecting power of 4m class telescopes, and less than optimal image quality in terms of seeing. 

The advent of MUSE as a second generation instrument for the ESO Very Large Telescope (VLT)  has completely changed the situation: the IFU features a  $1\times1$~arcmin$^2$ FoV with spatial sampling of 0.2~arcsec, high throughput, excellent image quality and a one octave free spectral range with a spectral resolution of R=$1800\ldots3600$ \citep{bacon2014}. In preparation for new opportunities with MUSE, \cite{kamann2013a} developed the {\textsc PampelMUSE} software and validated the PSF fitting technique on the basis of PMAS data obtained at the Calar  Alto 3.5m telescope \citep{kamann2013b}, measuring for the first time accurate radial velocities of individual stars in the very central crowded regions of the globular clusters M3, M13, and M92, thus putting stringent limits on the masses of any putative intermediate-mass black holes in the cores of those clusters \citep{kamann2014}. These first experiments were taken to another level with commissioning data from MUSE, where the globular cluster NGC6397 was observed as a $5\times5$ mosaic of $\approx60$s  snapshot exposures, resulting in a spectacular dataset of spectra for more than 12000 stars. By fitting to a library of PHOENIX spectra \citep{husser2013} and with photometry, good results for the atmospheric parameters $T_\mathrm{eff}$ and log~$g$, as well as metallicities were obtained, such that a spectroscopic Hertzsprung-Russell diagram (HRD) could be presented for a globular cluster for the first time \citep{husser2016}. 
In a kinematic analysis of the same dataset, \cite{kamann2016} study again the radial velocities and velocity dispersion of individual stars, however with a sample size of more than 12000 objects. Despite the low spectral resolution of MUSE, it is shown that for spectra with sufficient signal-to-noise ratio (S/N), the radial velocities are measured to an accuracy of 1~km~s$^{-1}$.

\begin{figure}[t]
    \centering
    \includegraphics[width=\hsize,bb=35 50 750 600,clip]{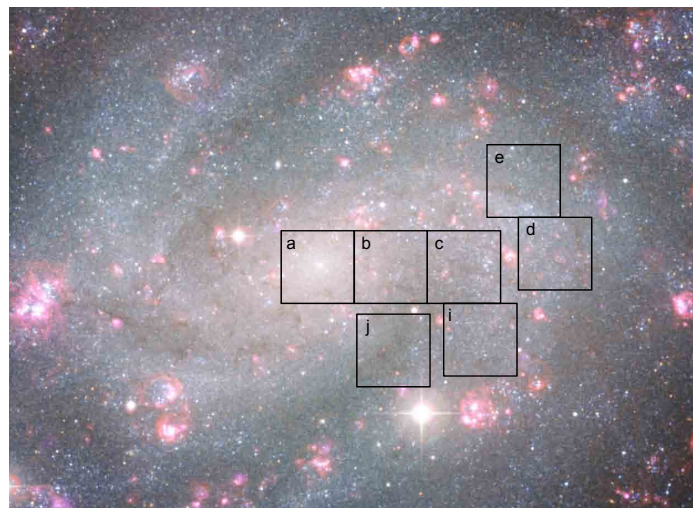}
    \caption{Footprint of MUSE pointings  (a), (b), $\ldots$, (j) in NGC\,300, covering the nucleus (a), different fractions of the      
     spiral arm that extends to the NW of the center (b), (c), (d), (e), (i), and the inter-arm region (j). 
     Orientation: N up, E left. (image credit: ESO).}
     \label{footprint}
\end{figure}

\begin{table}[!bh]
 \caption{Basic parameters of NGC\,300. References: [1] \citet{devaucouleurs1991}, [2] this work, [3] \citet{gieren2005}, [4] \citet{bresolin2005}, 
               [5] \citet{carignan1985}, [6] \citet{westmeier2011}, [7] \citet{burstein1984}  }           
\label{NGC300_data}
  
\centering
\begin{normalsize}         
\begin{tabular}{  l  c  c  }     
\hline\hline
Parameter                                                          &          value                    &     reference \\
\hline
Morphological type                                            &                   SA(s)d          &         [1]        \\
R.A. nucleus (J2000.0)                                      &            00:54:53.389       &         [2]        \\    
Decl. nucleus (J2000.0)                                     &            $-37$:41:02.23   &         [2]        \\
Distance                                                            &                 1.88\,Mpc       &         [3],[4]        \\
Angular scale                                                    &     9.2 pc/arcsec            &        [3],[4]        \\
R$_{25}$                                                           &            9.8\,arcmin          &         [5]         \\
Inclination                                                          &    39.8$^\circ$                 &        [6]          \\
Position angle of major axis                              &     114.3$^\circ$             &         [6]         \\
Barycentric radial velocity                                 &        $144\pm2$ km~s$^{-1}$      &        [6]          \\
 Max. rotation velocity                                       &        $98.8\pm3.1$ km~s$^{-1}$  &        [6]          \\
Corr. apparent magnitude B$^0$$_T$             &          8.38 mag                &        [5]           \\
Corr. absolute magnitude M$^0$$_T$             &       $-18.01$  mag           &        [5]           \\
Foreground extinction   E(B-V)                         &        0.025 mag                &        [7]          \\
 \hline
\end{tabular}
\end{normalsize}   
\end{table}

\begin{table*}[!th]
\caption{Journal of observations.}
\centering
\label{journal}
\begin{tabular}{lrlllllll}\hline\hline\\[-1.8ex]
 & \# &\multicolumn{1}{c}{Date} & \multicolumn{1}{c}{Coordinates} & \multicolumn{1}{c}{T$_{\text{exp}}$} & Airmass & \multicolumn{1}{c}{FWHM$_{\text{DIMM}}$} & \multicolumn{1}{c}{FWHM$_{\text{Moffat}}$} & STD\\\hline\\[-1.8ex]
$a$ & 1 & 2014-09-26 06:48:26 & 00:54:53.6 -37:41:05.1 & 1800 & 1.10--1.16 & $0\farcs66$--$0\farcs80$ & $0\farcs56$--$0\farcs66$ & EG 21\tablefootmark{a}\\
    & 2 & 2014-09-26 07:29:49 &                        & 1800 & 1.18--1.28 & $0\farcs61$--$0\farcs70$ & & EG 21\tablefootmark{a}\\
    & 3 & 2014-09-26 08:12:15 &                        & 1800 & 1.30--1.44 & $0\farcs64$--$0\farcs69$ & & EG 21\tablefootmark{a}\\
$b$ & 1 & 2014-10-28 01:03:47 & 00:54:48.5 -37:41:05.2 & 1800 & 1.08--1.14 & $1\farcs16$--$1\farcs38$ & $0\farcs76$--$1\farcs05$ & Feige 110\tablefootmark{b}\\
    & 2 & 2014-10-28 01:49:23 &                        & 1800 & 1.04--1.08 & $1\farcs06$--$1\farcs39$ & & Feige 110\tablefootmark{b}\\
$c$ & 1 & 2014-10-29 02:36:39 & 00:54:43.4 -37:41:05.1 & 1800 & 1.03       & $0\farcs99$--$1\farcs32$ & $0\farcs85$--$0\farcs96$ & Feige 110\tablefootmark{c}\\
    & 2 & 2014-10-29 03:27:27 &                        & 1800 & 1.03--1.05 & $1\farcs03$--$1\farcs05$ & & Feige 110\tablefootmark{c}\\
    & 3 & 2014-10-30 00:23:44 &                        & 1800 & 1.13--1.21 & $1\farcs13$--$1\farcs21$ & & Feige 110\tablefootmark{d}\\
$i$ & 1 & 2014-10-30 02:23:39 & 00:54:42.3 -37:42:05.0 & 1800 & 1.03--1.04 & $0\farcs61$--$0\farcs98$ & $0\farcs47$--$0\farcs59$ & Feige 110\tablefootmark{d}\\
    & 2 & 2014-11-25 01:10:12 &                        & 1800 & 1.03       & $0\farcs77$--$1\farcs03$ & & Feige 110\tablefootmark{e}\\
    & 3 & 2014-11-26 00:16:15 &                        & 1800 & 1.03--1.05 & $0\farcs57$--$0\farcs70$ & & Feige 110\tablefootmark{e}\\
$j$ & 1 & 2014-11-26 01:03:38 & 00:54:48.1 -37:42:13.7 & 1800 & 1.03       & $0\farcs65$--$0\farcs82$ & $0\farcs64$--$0\farcs77$ & Feige 110\tablefootmark{e}\\
    & 2 & 2014-11-28 03:28:50 &                        & 1800 & 1.20--1.30 & $1\farcs04$--$1\farcs15$ & & Feige 110\tablefootmark{f}\\
$d$ & 1 & 2015-08-23 04:49:11 & 00:54:37.0 -37:40:52.6 & 1800 & 1.18--1.22 &                          & $0\farcs63$--$0\farcs73$ & EG 274\tablefootmark{g}\\
    & 2 & 2015-08-23 05:36:44 &                        & 1800 & 1.08--1.13 &                          & & EG 274\tablefootmark{g}\\
    & 3 & 2015-09-08 04:34:36 &                        & 1800 & 1.07--1.11 & $0\farcs89$--$1\farcs41$ & & GJ 754.1A\tablefootmark{h}\\
$e$ & 1 & 2015-09-13 05:35:16 & 00:54:39.4 -37:39:50.3 & 1800 & 1.03       & $0\farcs82$--$1\farcs23$ &$0\farcs48$--$0\farcs61$ & GJ 754.1A\tablefootmark{d}\\[0.3ex]
\hline
\end{tabular}
\tablefoot{Observations of NGC 300 using MUSE in the extended mode. Col.~1, field identifier (see Fig.~\ref{footprint}); Col.~2, observation block; Col.~3, date and time of the observations (UTC); Col.~4, right ascension and declination coordinates (J2000) at the field center; Col.~5, exposure time (s); Col.~6, the airmass; Col.~7, the recorded DIMM seeing (arcsec); Col.~8, the FWHM of the Moffat function measured in the data using \textsc{PampelMUSE} (arcsec); and Col.~9, the used standard star (STD). The value ranges refer to the full wavelength range using the extended mode, 4650--9300{\AA}, where lower values are measured for redder wavelengths.\\
\tablefoottext{a}{Observed 2015-09-15 09:14:48, at an airmass of 1.46.}\tablefoottext{b}{2015-10-27 23:16:04, 1.22.}\tablefoottext{c}{2015-10-28 23:31:10, 1.22.}\tablefoottext{d}{2015-10-29 23:30:53, 1.21.}\tablefoottext{e}{2015-11-25 23:49:44, 1.06.}\tablefoottext{f}{2015-11-27 23:48:57, 1.06.}\tablefoottext{g}{2015-08-23 23:11:08, 1.03.}\tablefoottext{h}{The standard star was observed 2015-09-15 01:11:10 as no standard star was observed for the MUSE extended mode during the same night of the observations.}}
\end{table*}

Given these encouraging first results, we are now attempting to extend the technique of crowded field 3D spectroscopy to nearby galaxies with the goal of providing a new element to the current state of the art of stellar population synthesis, namely individual spectroscopy of the brightest stars: AGB and RGB stars, the most massive O stars, blue and red supergiants, luminous blue variable (LBV) and WR stars. As opposed to the Milky Way, where severe selection effects are at work, e.g. dust extinction, small numbers of known objects, lack of accurate parallaxes \citep[see][]{crowther2007}, one can hope to obtain complete samples down to a given limiting magnitude (although some of these stars are rare), and to then study their distribution across the face of the galaxy. Unlike the classical technique of broad-/narrowband filter imaging, and then follow-up spectroscopy \citep[see][]{massey2015}, IFS allows one to obtain images and spectra homogeneously in a single exposure. Filters with arbitrary transmission curves can be devised in the process of data analysis {\it after} the observation  to reconstruct images from the data cube, which not only enables one to define extremely narrow filter bandwidths and then obtain a very high sensitivity for continuum background limited emission line objects, but also to create efficient notch filters to mask out disturbing emission lines from reconstructed broadband images. Next to the stars, we are then also targeting planetary nebulae (PN) as potentially useful indicators of the underlying stellar population \citep{ciardullo2010,arnaboldi2015a,arnaboldi2015b}. Modelling internal dust extinction, the discovery of supernova remnants (SNR) on the basis of H$\alpha$/{\sii} and {\oi}/H$\beta$ line ratios \citep{fesen1985}, the distribution and physical properties of \hiiregs as well as the discovery of compact \hiiregs, in particular, and the diffuse and filamentary interstellar medium (ISM) are further pieces of information that can be retrieved from the datacube. 

As part of the coordinated guaranteed observing time (GTO) program of the MUSE consortium, we have launched a pilot study on the galaxy NGC\,300, that is located  in the foreground of the Sculptor group, in order to demonstrate the feasibility of the proposed approach. Basic parameters of this galaxy are listed in Table~\ref{NGC300_data}. Here we report the first results of our GTO observations in ESO periods P93, P94, P95, also demonstrating new legacy value opportunities provided by MUSE. The reader shall be convinced that any data\-cube hosts copious amounts of information in addition to the objects of the science case proper that has lead to the observation in the first place. For the purpose of a proof-of-principle, we present first selected results in order to underpin this claim. In future follow-up papers we shall explore the completed data set with a focus on different science cases (see section 6).\\

The paper is organized as follows: section 2 presents details of the observations and data reduction, section 3 describes the data analysis, section 4 summarizes the results with catalogues of stars, planetary nebulae, \hiiregs, SNRs, and serendipitous discoveries, along with a discussion within each class of objects. The paper closes with conclusions and an outlook to the next steps of this project.

\section{Observations and data reduction}
\label{section_observations}

Observations were made with the multi unit spectroscopic explorer instrument (MUSE; Bacon et al. 2014), which is placed at the Nasmyth focus of the UT4 8.2m telescope at the Very Large Telescope observatory (VLT) of the European Southern Observatory (ESO) in Chile. NGC 300 was observed as part of guaranteed time observations of the MUSE instrument-building consortium during the three periods P93, P94, and P95. All observations were performed in the extended mode, which includes the wavelength range 4650--9300\,{\AA} at a resolution of about 1800--3600 and a dispersion of 1.25 $\text{\AA}\,\text{pixel}^{-1}$. Effects of second-order stray light are clearly seen with this mode in continua with blue emission for wavelengths $\lambda\ga8000\,${\AA}.

The fields were chosen to cover as much as possible of the center regions of NGC 300, while avoiding the brightest \hiiregs seen in the H$\alpha$ image of \cite{bresolin2009}. Our aim was to observe each field for a total of 5400\,s, which was achieved with three observation blocks (OBs) of 1800\,s each, and each OB was in turn split in two exposures of 900\,s. To allow for an accurate normalization of the data, the IFU was shifted off the center and rotated between the exposures. 
Only the first four (two) values were used when only two (one) OB was observed. The sky was observed for two minutes in each OB using the same field on the sky, centered 7{\arcmin} West and 8{\arcmin}41{\arcsec} South of field $a$. Observations of the standard stars EG 21, Feige 110, EG 274, and GJ 754.1A allowed a spectrophotometric calibration of the data. Daytime calibrations of the morning after the observing night were used. The footprint of the selected fields is outlined in Fig.~\ref{footprint}, and the observations journal is shown in Table~\ref{journal}.


Data were reduced with the MUSE pipeline version 1.0 \citep{weilbacher2012} through the MUSE-WISE framework \citep{vriend2015}. The standard procedure was followed for the basic calibration. The first three steps are: ten bias images are combined to make a master bias image, five continuum exposures are combined to make a master flat-field image, and one exposure of each of the three arc lamps (HgCd, Xe, and Ne) are used to derive a wavelength solution. Eleven flat-field exposures of the sky, taken during the evening twilight preceding the science exposures, were combined and used to create a three-dimensional correction of the illumination where $4650\la\lambda\la8000$\,{\AA}. Detector defects were removed by means of a bad-pixel table.

We applied the calibration products to the science exposures as well as the standard star exposures. %
The average atmospheric extinction curve of Patat et al. (2011; included in the pipeline) was used when creating the sensitivity functions. The sensitivity functions and the astrometric calibration were applied to each exposure individually. The pipeline automatically corrects the data for atmospheric refraction (using the equation of \citealt{filippenko1982}) and the barycentric velocity offset. The sky exposures in each OB were used to create a sky spectrum as an average of the total of 90000 spectra of the sky cube, that was subtracted from the extracted data set of the respective OB. The two to six extracted data sets of each field were combined into a single cube. Owing to the IFU rotation of the observing strategy, the data were affected by the derotator wobble, which is why each exposure needed to be repositioned. Each field contains plenty of stars that could be used to this purpose. The resulting cubes were created using the standard sampling of $0\farcs2\times0\farcs2\times1.25${\AA}.

\section{Analysis} 
\label{section_analysis}
The data analysis was done in two complementary approaches: automatic source detection of point sources (stars, PNe) on the one hand, and visual inspection for the discovery and characterization of spatially extended objects and emission line sources on the other hand. The former was done with the {\textsc PampelMUSE} PSF-fitting code for IFS \citep{kamann2013a}, whereas the latter was based on inspection and processing using the {\textsc P3D} tool, and maps individually extracted from datacubes using DS9, respectively. {\textsc P3D} is an open source IFS software package developed at the AIP and publicly available under GPLv3 from SourceForge\footnote{\url{http://p3d.sourceforge.net/}.} \citep{sandin2010}. Previous applications of {\textsc P3D} for the VIMOS and FLAMES IFUs at the VLT were presented in \cite{sandin2011}. In the following paragraphs we describe in more detail how these approaches were accomplished.

\begin{table}[h]
\caption{Synthetic Filter Definitions}             
\label{filters}      
\centering 
\begin{small}
\begin{tabular}{l c r c c} 
\hline\hline 
Filter    & $\lambda_{c}^{'} $  &   $\Delta\lambda$ &  $\lambda_{0}-\lambda_{1}$  &  $\lambda$-bins\\    
\hline 
{\heii}                       &  4687.16   &        5.00    &  4687.35$-$4691.10      &  70$-$73              \\  
{\heii}$_{c_{1}}$      &                  &       70.00   &   4612.35$-$4681.10      &  10$-$65              \\  
{\heii}$_{c_{2}}$         &                 &       62.50   &   4763.60$-$4824.85      &  131$-$180          \\  
H$\beta$                  &  4862.69  &       6.25     &   4862.35$-$4866.10      &  210$-$213          \\  
H$\beta_{c_{1}}$     &                 &      62.50    &   4788.60$-$4849.85      &  151$-$200          \\  
H$\beta_{c_{2}}$     &                 &      62.50    &   4888.60$-$4949.85      &  231$-$280          \\  
{\oiii}                        &  5008.24  &       6.25     &   5007.35$-$5012.35      &   326$-$330          \\   
{\oiii}$_{c}$              &                 &   126.25     &   5028.60$-$5153.60      &   343$-$443          \\   
H$\alpha$                &  6564.61  &       6.25     &   6564.85$-$6569.85      &  1572$-$1576       \\  
H$\alpha_{c_{1}}$   &                 &      92.50    &   6371.10$-$6462.35      &  1417$-$1490       \\  
H$\alpha_{c_{2}}$   &                 &    103.75    &   6609.85$-$6712.35      &  1608$-$1690       \\  
{\nii}$_{1}$               & 6549.86  &       5.00     &   6549.85$-$6553.60      &  1560$-$1563         \\ 
{\nii}$_{2}$               & 6585.27  &       5.00     &   6586.10$-$6589.85      &  1589$-$1592         \\ 
{\nii}$_{c_{1}}$         &                &     92.25     &    6371.10$-$6462.35      &  1417$-$1490         \\ 
{\nii}$_{c_{2}}$         &                &   103.75     &    6609.85$-$6712.35      &  1608$-$1690         \\ 
{\sii}$_{1}$               & 6720.29  &       5.00     &    6718.41$-$6722.16      &  1695$-$1698          \\ 
{\sii}$_{2}$               & 6734.66  &       3.75     &    6733.41$-$6735.91      &  1707$-$1709          \\ 
{\sii}$_{c_{1}}$         &                &     75.0       &   6626.10$-$6699.85       &  1621$-$1680         \\ 
{\sii}$_{c_{2}}$         &                &      62.5      &   6763.60$-$6824.85       &  1731$-$1780         \\ 
{\siii}                         & 9071.1    &       5.00     &   9073.25$-$9076.10       &  3578$-$3581         \\ 
{\siii}$_{c}$              &                 &      46.25    &   9016.10$-$9061.10      &  3533$-$3569          \\ 
V                              & 5149.66  &    1101.25   &  4599.66$-$5699.66      &   0$-$880                 \\ 
R                              & 6475.29  &    1550.00   &  5700.91$-$7249.66      &   881$-$2120            \\ 
I                               &  8299.66  &    2098.75   &  7250.91$-$9348.41      &   2121$-$3799         \\ 
\hline                                   
\end{tabular}
\tablefoot{{\nii}$_{c_{1}}$  identical to H$\alpha_{c_{1}}$, {\nii}$_{c_{2}}$ identical to H$\alpha_{c_{2}}$}
\end{small}
\end{table}

\subsection{Mapping stars and gaseous nebulae}
\begin{figure*}[h]
   \centering
    \includegraphics[width=\hsize,bb=35 20 750 570,clip]{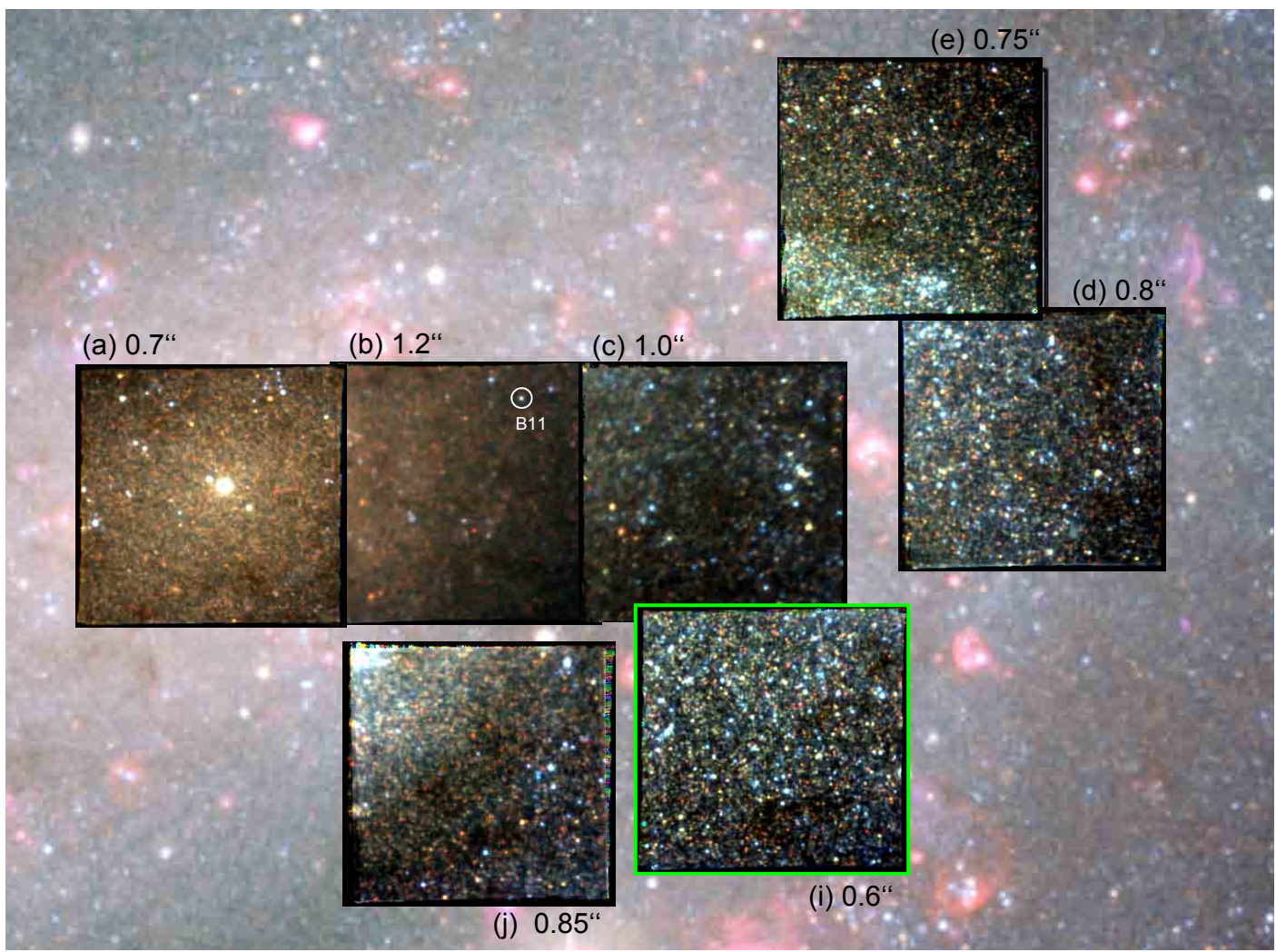}
    \caption{VRI images reconstructed from MUSE datacubes for pointings (a), (b), $\ldots$, (j). The seeing FWHM for each    
     pointing is indicated, the best image quality with 0.6" (green frame) in (i). Fields (b) and (c) are affected by poor seeing 
     (1.2'' and 1.0'', respectively). Owing to a reddish colour, an old stellar population is apparent in the nuclear region 
     sampled in (a) and (b), whereas the spiral arm fields (c), (d), (e), and (i) are dominated by blue stars. The interarm 
     region in (j) features extinction by dust lanes. The background image is from the ESO/MPI 2.2m WFI (credit: ESO).
    }
     \label{NGC300_stars}
\end{figure*}

The inventory of stars, gas, and dust is most intuitively mapped by coadding a number of suitable wavelength bins of the data cube to then form broadband  or narrowband images. As the MUSE free spectral range spans (in the extended  mode) one octave from 465 to 930\,nm, there is an almost unlimited choice of synthetic filter curves that can be realized. We have chosen to define three broadband transmission curves V, R, I, that are similar to the Bessell filter curves \citep{bessell1990} in that they cover approximately the same wavelength intervals. We have chosen top-hat functions for simplicity and did not attempt to mimic the exact slopes of the transmission curves. Fig.~\ref{NGC300_stars} shows colour composite images of our fields, overlaid onto a wide-field colour composite image obtained with the ESO 2.2m telescope.

Furthermore, we defined narrowband filters for important emission lines such as H$\alpha$, H$\beta$, {\oiii}, {\nii}, {\sii}, {\oi}, etc. with central wavelengths adjusted to the redshift of NGC\,300, and wide enough to cover the instrumental profile of MUSE, respectively, in order to not lose any flux. The filter width, however, was chosen small enough to suppress continuum background light as much as possible. With a typical narrowband filter width of 5 spectral bins (6.25 $\AA$) MUSE is therefore superior in sensitivity to any conventional narrowband imaging camera based on interference filters with a full width at half-maximum (FWHM) of typically $\approx 30\,\AA$. In order to account for the residual continuum flux collected by these filters, a continuum correction was applied to the emission line fluxes on the basis of continuum estimates shortward and longward of the line center. The synthetic filter parameters for the emission lines and corresponding continuum estimates used for our analysis, as well as for the broad band filters, are listed in Table~\ref{filters}. 

\subsection{Extracting point source spectra}
\label{extracting}
To extract spectra of bright individual stars and emission-line objects in our MUSE data, we used the tool \textsc{PampelMUSE} \citep{kamann2013b}. The tool relies on a reference catalog with relative positions and (initial-value) brightnesses of the objects.  In a first approach, we used the available stellar-source catalog of the ACS nearby galaxy survey (ANGST\footnote{We used the catalog {\it wide3 f475w-606w st} from the web page\\ \url{https://archive.stsci.edu/prepds/angst/}.}). While some of our fields are completely (fields $b$ and $c$) -- or nearly completely (fields $a$ and $j$) -- covered by the ANGST catalog, three fields are only partially covered (fields $d$, $e$, and $i$), which is why an alternative approach was needed to extract spectra in all parts of all fields. We used the \textsc{FIND} tool of \textsc{DAOPHOT} \citep{stetson1987} to determine the centroids of stellar images and create a source catalog. We again used \textsc{FIND} in both a blue and a red part of the continuum to locate as many point sources as possible in each data cube. We set the intensity of each object to a faint value and manually added a set of additional objects with a brighter value that were used as PSF stars. All object coordinates were tied to those of the ANGST catalog, which covers at least a few sources in each field. Notably, the alternative approach only allows a spectrum extraction of the brightest blue and red giant stars, while all fainter appearing stars are part of the background. 

Concerning emission line point sources, we again used  \textsc{find} to specify the positions of potential planetary nebulae or compact \hiiregs after summing up five pixels on the dispersion axis about $[\ion{O}{iii}]\,\lambda5007$ in the respective data cube; additional faint blobs that \textsc{find} missed -- which could also be sources -- were added to the resulting catalog manually after a visual inspection of the respective data cube.

The extraction of the stellar spectra is performed in a procedure that includes several steps. The analysis begins with an initial guess of the MUSE data PSF, which is modeled as an analytic Moffat profile with up to four free parameters: the FWHM, the kurtosis $\beta$, the ellipticity $e$, and the position angle $\theta$, all of which may depend on the wavelength. Initially, the FWHM is set to the seeing of the observations, $\beta=2.5$, and $e=0$. 
By combining the PSF model with the stellar positions and brightnesses in the reference catalog, a mock MUSE image is created for a pre-defined filter. Another image is created by integrating the MUSE cube in wavelength direction over the same filter curve. By cross-correlating the two images, initial guesses for all catalogued sources are obtained. The subset of those sources for which meaningful spectra can be extracted is then identified. This is done by estimating the S/N of each source based on its magnitude in the reference catalog, the PSF, and the variances of the MUSE data. In addition, the density of brighter sources around the source in consideration is determined. Only those sources are used in the further analysis where S/N> 5, and where the density determination yields less than 0.4 sources of similar or greater brightness per resolution element. The brightest and most isolated of the selected sources are flagged as PSF sources. They are used in the actual extraction process to model the PSF parameters and the coordinate transformation from the reference catalogue to the cube as a function of wavelength.

The spectrum extraction is carried out in a layer-after-layer approach, starting at the central wavelength and then progressing alternatingly to the red and blue ends of the cube. In each layer, a sparse matrix is created, containing the model fluxes of one star (according to the current estimate of the PSF and its position) per column. Next to the stars, local background estimates are included in the matrix in order to account for the non-negligible surface brightness of unresolved stars or diffuse nebular emission.  {\textsc PampelMUSE} allows for the definition of such background elements as square tiles with a user-selectable size. Via matrix inversion, the fluxes of all (stellar and background) sources are fitted simultaneously. Afterwards, all sources except those identified as isolated enough to model the PSF are subtracted and the parameters of the PSF and the coordinate transformation are refined. The new estimates are then used in another simultaneous flux fit. The procedure is iterated until convergence is reached on the source fluxes and the analysis of the next wavelength bin is started. Each new wavelength bin uses the resulting values of the previous bin as an initial guess.

After all wavelength bins are processed this way, a final PSF model is derived for all of the data cubes. To this aim, the values of the PSF parameters obtained in the individual wavelength bins are fitted with low-order polynomials. The object coordinates are also fitted with polynomials along the dispersion axis to reduce the effect of small random jumps between wavelength bins and thereby increase the S/N. The use of polynomials for this task is justified because ambient characteristics such as atmospheric refraction or the seeing should result in a smooth change of the PSF and the source coordinates with wavelength. Notably, while the FWHM should always show a monotonic decrease with wavelength, which is the theoretically expected behaviour, we found that it instead increases where $\lambda\ga8000$\,{\AA}. However, such a behavior is expected owing to contamination from second-order scattered light of bluer wavelengths, when using the extended mode of MUSE. Meanwhile, $\beta$ did not vary strongly with wavelength. In the last step, the final spectra are extracted by traversing all bins of the cube once more, using the fitted estimates of the PSF and the object coordinates. This was done again by a simultaneous flux fit to all stars and background components. After convergence of the fitting process is reached, the stellar fluxes and the local background estimates are available as individual spectra for further analysis. The background spectra turned out to be very useful for the measurement of extremely faint diffuse gas emission, in particular at the wavelengths of H$\alpha$ and H$\beta$ where the nebular emission coincides with stellar absorption lines. Stars with successfully extracted spectra are referenced in what follows through the {\textsc PampelMUSE} input catalogue number per field as listed in column (1) of Table~\ref{stars-i}.

\subsection{Fitting stellar spectra}
Two major goals of the project were to demonstrate, as a proof of principle for crowded fields in NGC\,300, that PSF-fitting IFS is capable of extracting spectra of individual stars with sufficient quality to derive trustworthy spectral type classifications, and to measure radial velocities, even with moderate to low signal-to-noise ratios (S/N). To this end, we have fitted the extracted spectra both to an empirical library of stellar spectra, in what follows MIUSCAT, and to a grid of models computed with the Phoenix code \citep{husser2013}, henceforth GLIB (G\"ottingen Library). MIUSCAT is an (unpublished) extension of the MILES library \citep{sanchez-blazquez2006,cenarro2007,falcon2011} that was kindly provided to us by Alexandre Vazdekis. It was created to reach up to the calcium triplet region and covers the gap between the blue and the red spectral ranges of MILES and CaT with data from the indo-U.S. library \citep{valdes2004} as a basis for the stellar population synthesis models from \citet{vazdekis2012}. Unlike other empirical libraries, MIUSCAT covers the entire free spectral range of MUSE in the extended mode. 

The stellar spectra that could be extracted successfully by {\textsc PampelMUSE} were fitted to MIUSCAT by means of the ULySS code\footnote{\url{http://ulyss.univ-lyon1.fr}}, that was originally developed to study stellar populations of galaxies and star clusters, as well as atmospheric parameters of stars \citep{koleva2009}. An advantage for our application is the fact that ULySS, as adapted from pPXF \citep{cappellari2004}, fits a spectrum as a {\it linear combination} of non-linear components, multiplied by a polynomial continuum, thus helping to identify unresolved blended stars, as opposed to a single best guess for the spectral type in question. Especially for our application, where we are confronted with spectra of low S/N, the output of several (up to 10) library spectra with their respective weights supports a proper judgement of the quality of the fit and the identification of potential problems. As a practical disadvantage, the ULySS output requires a significant level of human interaction, i.e. a decision for each individual input spectrum in how far the linear combination of library spectra with different weights allows to make a good guess for the spectral type of the observed star. However, any less than plausible spectra can be ruled out immediately, e.g. foreground stars, or main sequence stars, the latter of which, at a distance of  1.9 Mpc, would be far below the detection limit of MUSE. In order to assist with this criterion, we searched the Simbad database for photometry of the MIUSCAT library stars and shifted their magnitudes to the most recent, cepheid-based NGC\,300 distance modulus of (m-M)$_{0}$=26.37 \citep{gieren2004,gieren2005} for comparison with the apparent magnitudes from the ANGST catalogue, neglecting extinction. 

\begin{figure}[t]
    \centering
    \includegraphics[width=\hsize,bb=45 50 400 575,clip]{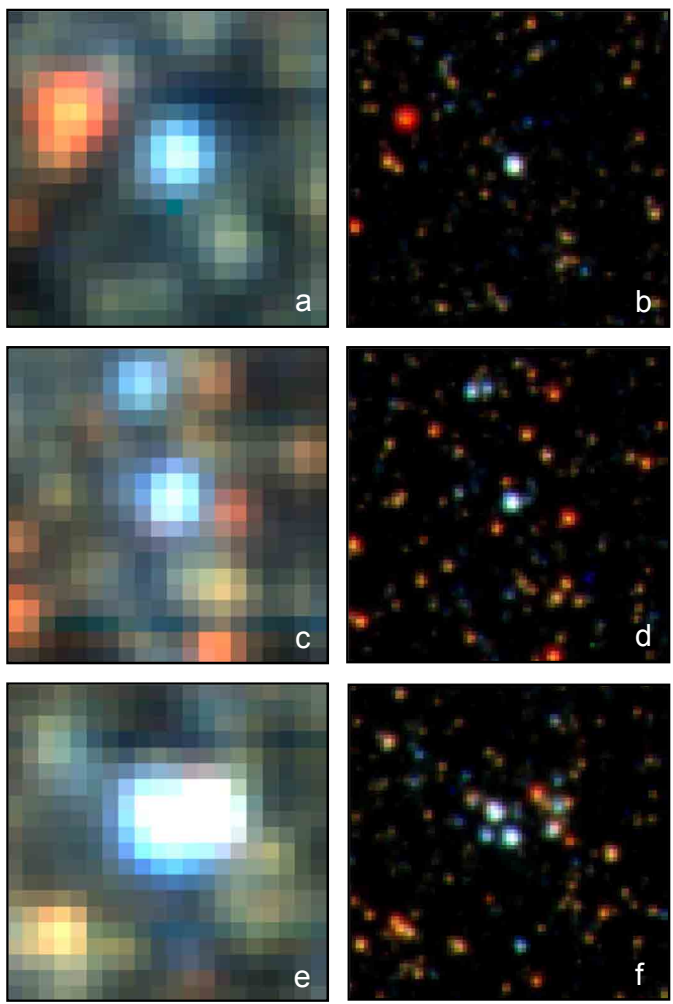}
    \caption{MUSE versus HST images of $4\times4$ arcsec$^2$ size from field (i), as examples of the star in the
    center being affected by different severity of crowding. a/b: ID1154, A1$\,$Ib star, unblended. c/d: ID1179, B3$\,$III star,
    negligible blending. e/f: ID995, central star cluster in bright \hiireg, spectrum heavily affected by blending and nebular emission,
    no reliable classification possible.}
     \label{poststamps}
\end{figure}

An important aid in assessing the validity of the classification is the inspection of images to find any evidence of blending. To this end, we plotted for each object a reconstructed VRI-map from the MUSE datacube as a post-stamp-like image of a size of 4$\times$4~arcsec$^2$, accompanied with an HST ACS image of the same region, colour coded from the
 filter combination F475W+F606W+F814W \citep{dalcanton2009}, see Fig.~\ref{poststamps}. Together with a blend flag issued by {\textsc PampelMUSE} the probability and severity of  blending effects was assessed and recorded as a quality flag.

As an alternative to ULySS fits to the MIUSCAT library, we used the technique of \citet{husser2013}, that was initially developed  for globular cluster stars, to fit GLIB spectra to our objects. In this case, the result is a set of stellar parameters for a single locus in the HRD, with $T_\mathrm{eff}$, log {\it g}, and metallicity. Both the MIUSCAT and GLIB approaches also yielded measurements of the radial velocity. 

For the final assessment of spectral type and radial velocity, we performed a visual comparison of the fits relative to the measured spectra, inspecting whether or not important absorption lines would be in accord with the fit and stand out from the noise, checked the resulting stellar parameters for plausibility, and listed the outcome with a set of quality flags to indicate (a) the quality and plausibility of the fits, based on the visual inspection of critical absorption lines and the photometry, (b) the agreement between the MIUSCAT and GLIB fits, any apparent effects of blending from nearby stars, and the plausibility of the measured radial velocities. In a final step, the most probable spectral type (or a range of spectral types) was determined, as well as the most probable radial velocity, and a global quality flag for the final result. The adopted outcome of the fits and visual inspection were recorded in individual  log files and summarized in a catalogue file. In order to reduce the elements of subjective judgement on a steep learning curve, the whole procedure was actually exercised twice, and only the results of the final assessment were retained.  The quality of spectral type classification is summarized in Table~\ref{distrib-stars-i} and discussed in \S~\ref{subsection_stars}. An excerpt of the catalogue is presented in Table~\ref{stars-i}.

\subsection{Extracting non-stellar sources}
\label{extract-gas}
As expected from the outset, visual inspection has shown indeed  that our MUSE datacubes contain valuable information about objects that are not discovered as continuum point sources with the methods described above. This is particularly true for emission line objects like \hiiregs, supernova remnants (SNR), supershells, diffuse ionised interstellar gas (DIG), planetary nebulae (PN), etc. Also faint background galaxies reveal their presence though redshifted emission lines that stand out from the foreground galaxy continuum. The search and classification of such objects was done by visually inspecting the emission line maps described in {\S}3.1 with DS9. \hiiregs were readily identified on the basis of spatial extension and brightness. PNe were detected by blinking {\oiii} vs. H$\alpha$, as well as {\heii} in order to find high excitation objects. SNR, superbubbles and supershells were distinguished from \hiiregs  by the technique of ionization parameter mapping of \citet{pellegrini2012}. Emission line point source objects were identified by measuring the FWHM of their images through a Gaussian fit and comparison with the PSF obtained for stars in the same datacube with {\textsc PampelMUSE}. By blinking H$\alpha$ against continuum V, R, and I images the emission line point sources that were found to coincide with stars, yielded candidates for emission line stars as a valuable complement and cross-check for the detection of such objects made with {\textsc PampelMUSE}. All visually detected emission line objects are referenced with an ID given by a leading character for fields (a)...(i), followed by a serial detection number (regardless of type of object) as listed in Column (1) of Tables~\ref{PN-list}, \ref{cHII-list}, \ref{emStar-list}, \ref{HII-list}, and \ref{SNR-list}.

Subsequent analysis using the line fitting capability of the {\textsc P3D} visualization tool allowed to measure emission line fluxes and radial velocities for all kinds of emission line objects. As most of the objects exhibit several sufficiently bright emission lines suitable for the line fitting tool, we typically measured all sufficiently bright lines and determined a resulting Doppler shift for the radial velocity estimate from a flux-squared weighted average of all lines. The uncertainty estimate was obtained from the scatter of the contributing lines. Broad line profiles that are indicative of strong stellar winds were measured for point sources, which were thus identified as emission line stars. In extended sources, the {\sii}{/H${\alpha}$ line ratio was used to discriminate SNR against \hiiregs, which is particularly important for SNR candidates that are too faint for the ionization parameter mapping technique to be applied. A prerequisite for good flux measurements is an accurate subtraction of the background surface brightness that is composed of contributions from continuum light of faint unresolved stars and of diffuse or filamentary emission line flux arising from the DIG, ancient SNR shells, etc. Background estimates were obtained with {\textsc P3D} by  defining an aperture for the object in question and a surrounding annulus for the background, where either strictly circular, or otherwise arbitrary user-defined geometries can be defined in order to accomodate complex surface brightness distributions. For the special case of recording DIG intensities, we used the unresolved background estimates as output from {\textsc PampelMUSE} to correct {\textsc P3D} flux measurements.  Background galaxies were discovered by browsing the row stacked spectra available in the visualization tool of {\textsc P3D} and searching for emission features at unusual wavelengths. As a map is automatically displayed when the cursor moves through the suspicious wavelengths, a summed spectrum for the affected spaxels is readily created. This technique proved to be extremely efficient for data mining redshifted background galaxies.

In what follows, we describe how we have exploited the high level of sensitivity obtained with MUSE for the discovery of emission line objects like PNe, emission line stars, compact, normal, and giant \hiiregs, SNR, superbubbles, giant shells, and DIG. 

To this end, we first of all searched for point sources using {\textsc DAOPHOT FIND} as described in \S~\ref{extracting} and extracted their spectra with {\textsc PampelMUSE}, similar to the procedure with stars. We also visually examined all of the fields in H$\alpha$ and recorded extended and point sources down to very low contrast levels to create a provisional initial catalogue, without consulting the {\textsc PampelMUSE} catalogue to avoid any subjective bias in the detection process. By blinking against the {\oiii} images we identified high excitation objects such as PN candidates, finding also objects that would not be bright enough to appear in the H$\alpha$ image. The {\heii} maps allowed us to readily identify high excitation PNe and to discover a WR star (see below). 

We then inspected each object from the initial catalogue one by one using the {\textsc P3D} visualization tool, and measured emission line fluxes with aperture spectrophotometry, assisted by an interactive background subtraction feature of {\textsc P3D} which allows one to define object and background apertures  as standard regions of interest (ROI) of different geometries (circular, elliptical, rectangular), and a mouse-controlled editor to quickly add or remove spaxels. The {\textsc P3D} line fitting tool was also used to measure the central wavelength of emission lines and the corresponding line-of-sight radial velocity from the Doppler shift with respect to laboratory wavelengths. 

The uncertainty of emission line fluxes was estimated using the equation by \citet{gonzalez1994}, however adding an extra term to account for flat fielding and flux calibration uncertainties:
\begin{equation}
\hspace{1cm} \sigma_l = [ ~ \sigma_{cont}^2 ~ N ~ ( 1 + EW / N \Delta) + (0.05 \times F_l)^2 ~ ] ^{1/2}
\end{equation}
where  $\sigma_{cont}$ is the standard deviation of the continuum near the emission line, N is number of spectral bins used to measure the line, $\Delta$ is the reciprocal dispersion in $\AA$/bin, EW is the equi\-valent width of the line, and $F_l$ is the flux measured over the N spectral bins.

We also employed the PhAst tool \citep{mighell2012} to perform aperture photometry in the narrowband images to double-check the {\textsc P3D} flux measurements, and to estimate the FWHM for any point-like object that was found from the visual inspection. We finally merged the results from the two different approaches and classified the detected objects on the basis of emission line fluxes, line ratios, and the FWHM of point-like sources or otherwise the size of extended objects as further discussed in the following section.

\section{Results and discussion}
\subsection{Stars}
\label{subsection_stars}
{\bf Results:} For the demonstration purpose of this paper, we selected field~(i) as the best pointing in terms of seeing (FWHM=0.6"), as highlighted in Fig.~\ref{NGC300_stars},  covering a fraction of the north-western spiral arm at a galactocentric distance of $\approx$1.5 kpc. We adopted both of the two procedures introduced in ${\S}$3.2, i.e. the input of stellar centroid priors from high spatial resolution HST images, and alternatively from a search in the datacube using {\textsc DAOPHOT FIND}. {\textsc PampelMUSE} was run with both catalogues, resulting in a total of 3540 and 552 extracted spectra, respectively. It should be noted that the HST coverage is only 2/3 of the MUSE field~(i). In order to reduce the number of poor quality spectra that would not allow conversion of the fitting procedure, we set a threshold of S/N=3 for the estimate computed by {\textsc PampelMUSE} and forwarded only spectra above the threshold to the ULySS code. From successful ULySS fits to the MIUSCAT library we obtained a total of 345 and 392 results for the HST and FIND input catalogues, respectively. For the same number of spectra, the fitting procedure was repeated using GLIB and the code from \citet{husser2013}. All of the spectra and corresponding fits were inspected visually, along with VRI images of the stars extracted from the datacube, as well as from HST, where available, for comparison and assessment of blending. Following the determination of spectral type, radial velocity v$_\mathrm{rad}$, and quality parameters as described in ${\S}$3.2, we created the final catalogues for the two sets of spectra. An excerpt of the catalogue for field (i) with HST input is presented in Table~\ref{stars-i}. The complete catalogue for field (i) will be made publicly available through CDS. 

 \begin{figure}[th]
   \centering
    \includegraphics[width=\hsize,bb=45 150 755 570,clip]{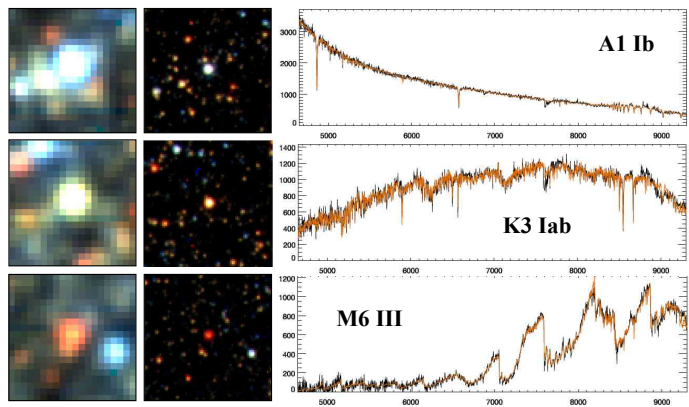}
    \caption{Example spectra from field (i). Images in the two columns to the left are analogous to Fig.~\ref{poststamps}. The spectra cover the full extended MUSE wavelength range of $4600 \ldots 9300~\AA$. Flux is plotted in units of $10^{-20}$erg cm$^{-2}$ s$^{-1}$ $\AA^{-1}$. Black curves: observed spectra, orange: ULySS fits.}
    \label{examplespectra}
 \end{figure}

To illustrate the quality of spectra that can be obtained with MUSE, Fig.~\ref{examplespectra} presents three examples  extracted from the stars in field (i) that  are centered on the poststamp images (left panel: MUSE, right: HST). The spectrum extracted from the upper panel  star numbered ID119 has a S/N=19 and is classified A1Ib with v$_\mathrm{rad}$=149$\pm$18~km~s$^{-1}$, the one in the middle from ID260 has S/N=14 and is classified K3Iab with v$_\mathrm{rad}$=171$\pm$17~km~s$^{-1}$, the one in the bottom panel from ID1379 has S/N=6 and is classified M6III with v$_\mathrm{rad}$=174$\pm$7~km~s$^{-1}$.  The plot range covers the full MUSE free spectral range from 460\,nm to 930\,nm. Black lines correspond to observed spectra, and the orange ones to the best MIUSCAT fits, that are for the most part almost indistinguishable from the measured data.

 \begin{figure}[th]
   \centering
    \includegraphics[width=\hsize,bb=30 210 755 580,clip]{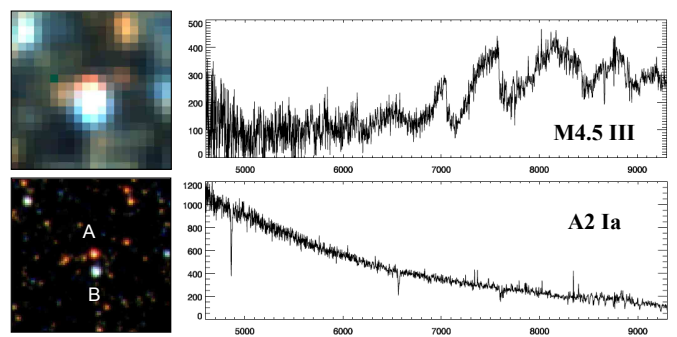}
    \caption{Deblending of two overlapping stellar images in field (i). The MUSE and HST maps to the left are analogous to Fig.~\ref{poststamps}, for units of spectra see 
                  Fig.~\ref{examplespectra}.}
    \label{deblending}
 \end{figure}

Fig.~\ref{deblending} demonstrates how the PSF fitting technique works for an example of two stars in field (i) that are separated by 0.6" as determined from the corresponding HST image. The separation amounts to the FWHM of the MUSE PSF. The blend consists of a red star A to the north (ID1539, F606W=23.26), and a blue star B to the south (ID633, F606W=21.84). Deblending with {\textsc PampelMUSE} delivers the spectra as shown in the right panels:  A is classified as an M4.5 subgiant, clearly identifiable through its prominent TiO absorption bands, wheras B is classified as A2Ia supergiant with relatively strong Balmer lines and a noticeable Paschen series. No obvious cross-talk is seen in any of the two deblended spectra, remarkably similar to the showcase for deblending of globular cluster stars shown in Fig.~2 of \citet{husser2016}. This capability is in stark contrast to the limitations of fiber-based spectroscopy, where an isolation criteria must be imposed, e.g. less then 20\% contamination from neighboring stars \citep{massey2016}. The HST input catalogue stars that have delivered useful spectra cover a magnitude range in the F606W filter of 20\ldots25 mag. 

Fig.~\ref{deblending_Ha} presents a case that is even more extreme: a faint blue star (A) (F606W=24.43) separated by 1.4" from K5III star (B) (ID682,  F606W=22.49). Due to its faint magnitude and a blend with group (C) of five faint K stars, the case of star A is too difficult to allow for the extraction of a meaningful spectrum with {\textsc PampelMUSE}. The object, however, was discovered as an emission line point source (i101) close to the detection limit of the field (i) H$\alpha$ map. Subsequent inspection using the {\textsc P3D} tool revealed that i101 is slightly offset to the north from group (C), thus most probably associated with the faint blue star B visible in the HST image. This is supported by the fact that the emission line is quite broad with FWHM(H$\alpha$)=6.5\AA, i.e. not of nebular origin. The  H$\alpha$ flux was measured to $4.0\times10^{-18}$ erg/cm$^2$/s, and the radial velocity as 144~km~s$^{-1}$. The collapsed broad-band image from the MUSE cube shows merely a vague blue hue and illustrates the fact that for ground based observations only integral field spectroscopy opens a chance to discover such an object.

 \begin{figure}[th]
   \centering
    \includegraphics[width=\hsize,bb=30 210 755 580,clip]{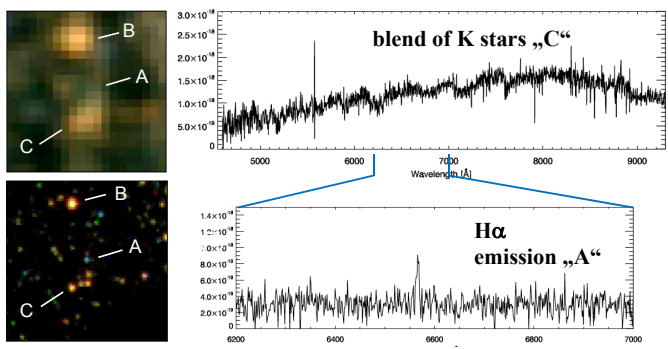}
    \caption{Deblending of the extremely faint emission line star (A) from an unresolved group of K stars (C). For explanation see Fig.~\ref{deblending} and text below. The insert   
                  spectrum covers the wavelength range  $6200 \ldots 7000~\AA$, flux is plotted in units of erg cm$^{-2}$ s$^{-1}$ $\AA^{-1}$ (continuum background not 
                  subtracted).}
    \label{deblending_Ha}
 \end{figure}

\begin{figure}[!h]
   \centering
    \includegraphics[width=\hsize,bb=45 65 755 500,clip]{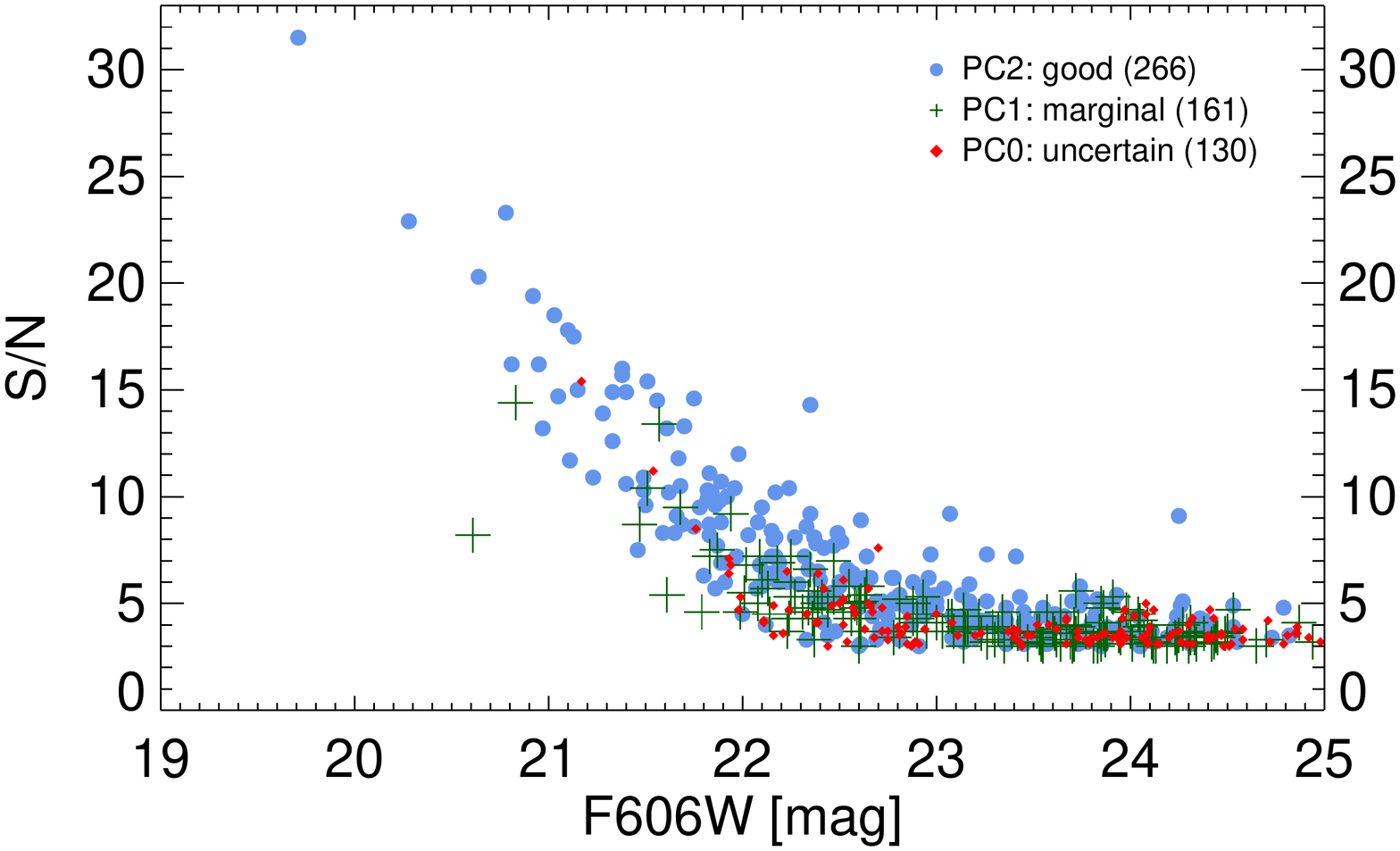}
    \includegraphics[width=\hsize,bb=45 65 755 500,clip]{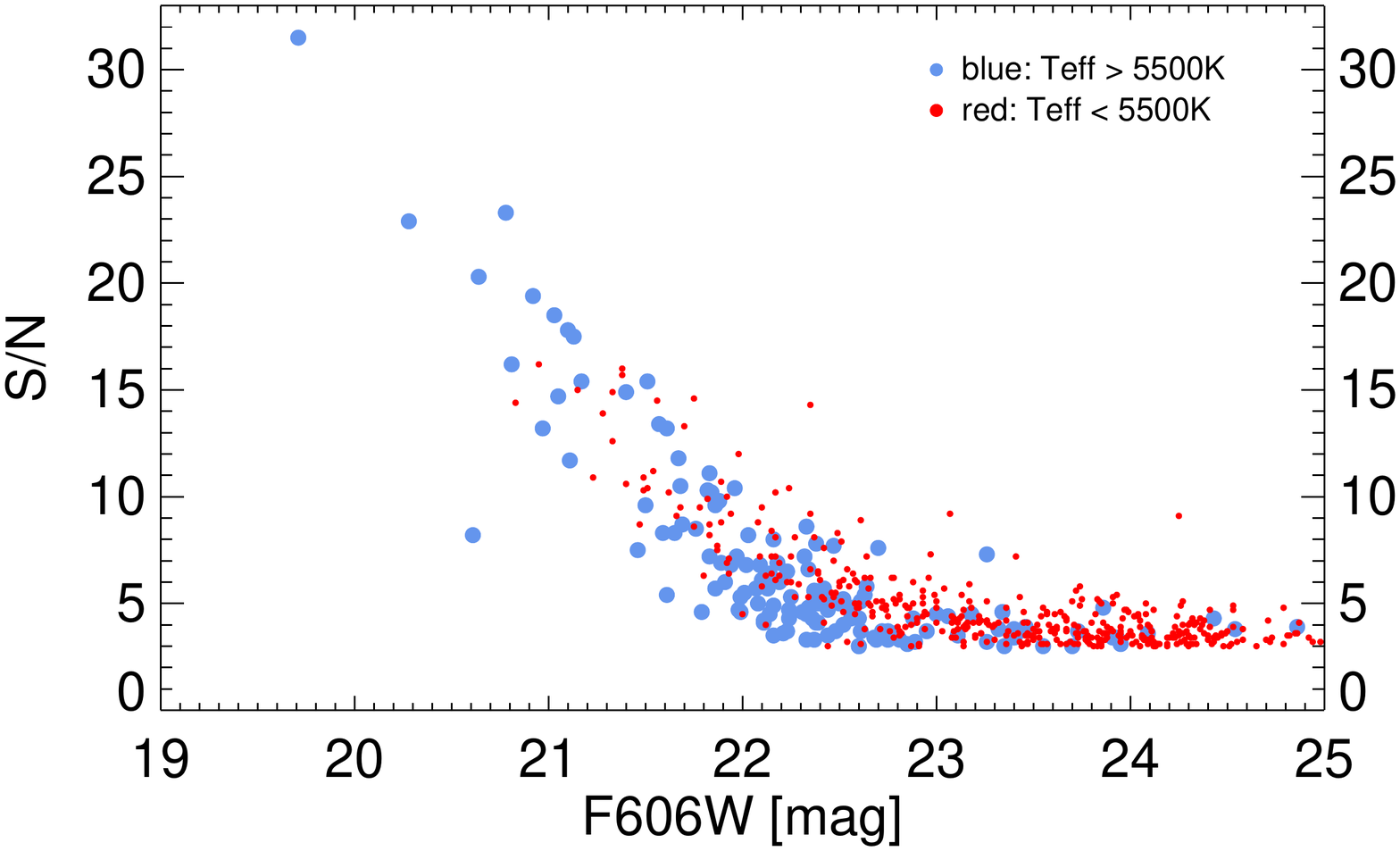}
    \caption{Distribution of S/N as a function of F606W magnitude in field (i), selectively by quality flag (top), 
         and $T_\mathrm{eff}$ (bottom).}
    \label{SN-ratio}
 \end{figure}

In Fig.~\ref{SN-ratio} we plot the S/N of extracted spectra of field (i) as a function of magnitude, broken down into three groups of different classification quality. The top panel illustrates the effect of the final estimate of spectral type classification with plausiblity flag PC2 \ldots PC0, where apparently good fits are plotted with blue circles (PC2), marginal fits as green crosses (PC1), and uncertain cases as red circles (PC0). First of all, the latter ones are found mostly at S/N levels below 4. Secondly, marginally plausible fits show a similar behaviour, with only a handful of cases at brighter magnitudes and higher S/N. Thirdly, the distribution of the majority of good fits suggests completeness down to a F606W magnitude of 22.5. The lower panel shows the distribution broken down by effective temperature, with blue symbols corresponding to stars hotter than 5500~K, and red symbols to stars cooler than this temperature. This plot shows that in the range of F606W=22\ldots23 the cool stars tend to exhibit a higher S/N than the hot stars, reflecting the fact that the F606W magnitude is not a good measure for the red flux of cool giants, which show spectra with high equivalent width absorption lines, e.g. the calcium triplet, or in the case of M stars, very pronounced molecular bands, in a spectral region where MUSE is most sensitive, whereas the cut-off of MUSE in the extended mode at 460\,nm limits the sensitivity for hot stars with regard to diagnostic lines in the blue.  Moreover, the wavelength-dependence of seeing leads to a stronger susceptibility to blending in the blue than it does in the red, e.g. measured as FWHM=0.60'' at 460\,nm vs. 0.48'' at 850\,nm for field (i). One could also argue that dust extinction leads to a selection effect that is more important for blue stars, however it is beyond the scope of this paper to address this issue quantitatively.

\begin{figure}[!h]
   \centering
    \includegraphics[width=\hsize,bb=90 65 750 550,clip]{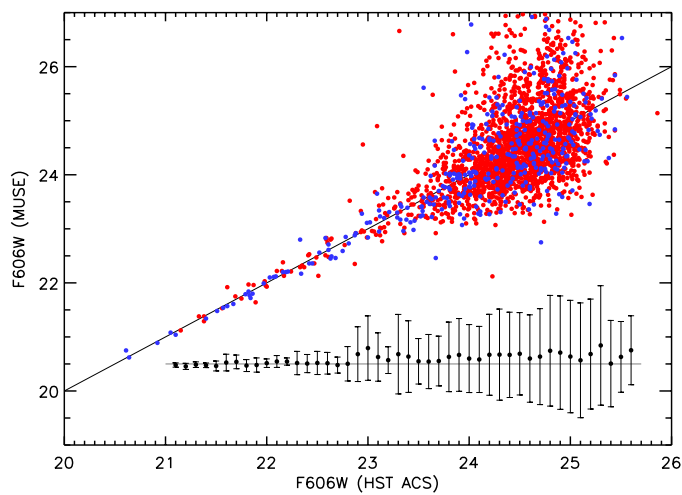}
    \caption{MUSE F606W photometry versus HST ACS magnitudes, broken down into cool stars (red dots), and hot stars (blue dots).
    Mean residuals against the 1:1 relation in 0.2 dex mag bins are shown as black dots. The error bars indicate the standard deviation while the gray   
    horizontal line indicates zero.}
    \label{ACS_MUSE_photometry}
 \end{figure}

We have tested the validity of our approach to spectral type classification in two ways. First of all, we checked whether the MUSE spectrophotometry correctly reproduces the ACS photometry. Artificial star tests as commonly used for CCD photometry were not found to be useful because {\sc PampelMuse} works already with a catalogue of stars obtained at high angular resolution (HST) with the benefit of accurately pin-pointing stellar centroids, regardless of magnitude. We therefore used the information from that catalogue as a reference. To this end, we convolved our flux-calibrated MUSE spectra with the ACS F606W filter curve as provided at the Spanish Virtual Observatory website \footnote{http://svo.cab.inta-csic.es}. In Fig.~\ref{ACS_MUSE_photometry} the resulting MUSE magnitudes (with a zeropoint of 36.5 mag) are plotted versus ACS magnitudes, whose errors amount to 0.01 mag for F606W= 22\ldots23, and a scattered distribution between 0.01 and 0.1 mag for fainter stars with F606W= 23\ldots25.5. The red plot symbols represent cool stars with colors (F606W-F814W)$\geq$0.4, blue symbols hot stars with (F606W-F814W)<0.4.

Both datasets show a very similar distribution, with a tight correlation at bright magnitudes, and a rapid degradation towards large errors at faint magnitudes. Residuals against the 1:1 relation in magnitude bins of 0.2~mag are shown in the lower right part of the graph (with an offset of 20.5 mag). The error bars indicate the standard deviation in each bin, with values of order 0.1~mag for bright stars up to F606W=22.7, and an abrupt increase to 0.3~mag and larger for stars fainter than F606W=22.8. At this magnitude, the distribution begins to become asymmetrical, and there is the onset of a bias to larger magnitudes. With reference to   Fig.~\ref{SN-ratio}, we interpret the branch towards bright magnitudes as the object photon shot noise dominated regime with a slope of -2, as expected. On the other hand, the branch towards fainter magnitudes must be source confusion limited, where the subtraction of blends and of the background of unresolved stars introduces errors at a level comparable with the poissonian noise of the object spectrum.

The second test was performed on the basis of seven template spectra from the MIUSCAT library, chosen such as to cover the relevant range of effective temperature and luminosity of the stars we would expect to discover in NGC300 (M1a-ab, K4III, K0III, G3III, F6Iab, A2Ia, B3Ib). These spectra were modulated with random noise resulting in S/N~= 30, 20, 10, 7.5, 5, 4, 3, 2, 1, with 10 random realizations per S/N value, and a sample of in total 630 simulated spectra. The ULySS code was applied to each spectrum to recover the input spectral type from the noisy simulation. Fig.~\ref{simTeff} shows the outcome of the exercise as the recovered $T_\mathrm{eff}$ for each simulation versus the corresponding S/N value as provided as an output parameter by the ULySS code, color-coded for the different spectral types.

\begin{figure}[!h]
   \centering
    \includegraphics[width=\hsize,bb=67 65 740 550,clip]{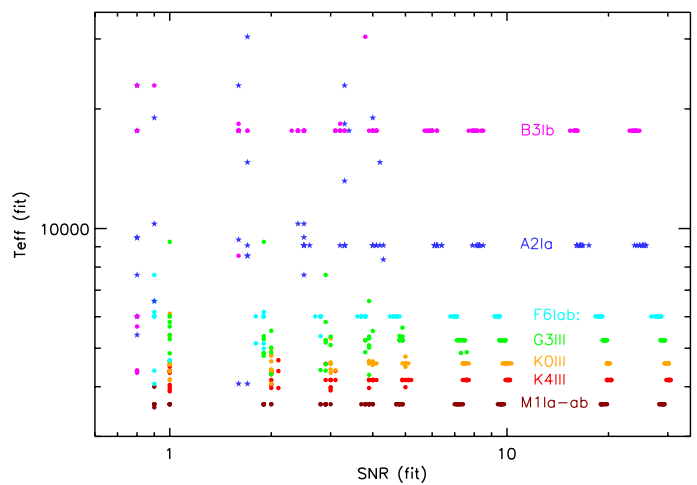}
    \caption{Recovered  $T_\mathrm{eff}$ as a function of S/N for a total of 630 simulated spectra, as determined by ULySS. For details, see text.}
    \label{simTeff}
 \end{figure}

One can immediately see that at S/N $\geq$7.5 all of the spectra are perfectly well recovered (with the exception of two G3III outliers at S/N=7.5).  Below that noise level, the reliability of the fit is strongly depending on the spectral type. For example, K0 and K4 subgiants show still agreement down to S/N=3, however with some outliers involving  offsets of a few 100~K in $T_\mathrm{eff}$. The M1 supergiant spectra, owing to the pronounced molecular band features, are even recognized unequivocally down to S/N=2. This is in stark contrast to G3III spectra, that scatter across a range of almost 2000~K at S/N=4 and below. The F supergiant simulations are uniquely recovered again down to S/N=4, whereas the hot A and B supergiant spectra begin to scatter at S/N=4. The different robustness of spectral type identification against noise depends on the dominant strength of absorption features characteristic for different temperatures, e.g. the calcium triplet for cool stars, molecular bands for  M stars, the Balmer and Paschen series for hotter stars, etc., and the spectral energy distribution with regard to those features.  It is noteworthy that for the global input noise levels imposed by the simulations, ULySS returns  S/N estimates that tend to be slightly lower than the input value, depending on spectral type. This points to the fact that a global S/N value is not a uniquely determining parameter for spectra over a broad free spectral range, such as with MUSE. For the sake of simplicity, we have not attempted to correct for these deviations.

\begin{table}[h]
 \caption{Number of stars with SNR>5 (4) for different magnitudes}             
\label{limitmag-i}      
\centering          
\begin{tabular}{ c   r   r  r   r   r    r   r   r  }     
\hline\hline
         F606W       &  N$_{5}^{b}$   & f$_{5}^{b}$ & N$_{4}^{b}$ & f$_{4}^{b}$  & N$_{tot}^{b}$&    N$_{4}^{r}$ & f$_{4}^{r}$  & N$_{tot}^{r}$ \\
\hline
      21.0--21.5  &       8      &     100      &     8     &   100    &    8    &    11   &   100   &    11   \\
      21.5--22.0  &     24      &       89      &     27   &   100    &   27   &    24   &   100   &    24   \\
      22.0--22.5  &     25      &       56      &     38   &     84    &   45   &    41   &    98    &    42   \\
      22.5--23.0  &       6      &       25      &     11   &     46    &   24   &    55   &    79    &    70   \\
      23.0--23.5  &       1      &         7      &     5     &     33    &   15   &    38   &    54    &    70   \\
      23.5--24.0  &       1      &       14      &     1     &     14    &    7    &    30   &    33    &    90   \\
\hline     
\end{tabular}
\end{table}

By combining our findings from photometry and the simulations, we conclude that the ULySS fits provide robust spectral type estimates for spectra with S/N $\geq$5, regardless of $T_\mathrm{eff}$. For a S/N below that level, hot and cool star show a different behaviour, chiefly in the sense that M and K stars still provide meaningful $T_\mathrm{eff}$ estimates around S/N=3\ldots4, where A and B stars are already suffering significant scatter. From Table~\ref{limitmag-i}, we estimate a limiting magnitude for completeness at F606W$\approx$22.5 for blue stars, and at F606W$\approx$23.5 for red stars, in the sense that more than 50\% of stars within a magnitude bin yield a reliable fit (nomenclature: N$_{4}^{b}$ is the number of blue stars with SNR>4, N$_{tot}$ is the total number of stars within a magnitude bin, f indicates the useful fraction of the total per magnitude bin in \%). These limits are dictated by the onset of crowding and the associated additional noise contributions for any given star -- hence they are no sharp thresholds, but rather depend on the environment (i.e. amount of blending) in each individual case.

\begin{table}[b]
 \caption{Distribution and quality of spectral type classification: field (i)}             
\label{distrib-stars-i}      
\centering          
\begin{tabular}{ c  c  r r r r  }     
\hline\hline
spectral type &  $T_\mathrm{eff}$ [K]   &\bf{$\Sigma$}  &    PC2     &    PC1    &   PC0    \\
\hline
O   &       > 30000                   &     38       &      9        &    2         &    27      \\
B    &       10282\ldots30000   &     32      &      9        &   15        &      8       \\
A    &       7715\ldots9703       &     39       &    24        &   14        &      1       \\
F    &       5688\ldots7715       &     30       &    14        &   11        &      5       \\
G   &       4709\ldots5680        &    12       &      6        &     4        &      2       \\
K   &       3895\ldots4696        &   164     &     65       &    64       &     35       \\
M  &        < 3810                     &   179     &    115      &    50      &     14        \\
C  &        <  3000                    &     23      &      23      &     0      &       0        \\
\hline
     &                                             &      517      &   265         &  160     &      92        \\
\hline     
\end{tabular}
\end{table}

Table~\ref{distrib-stars-i} summarizes the distribution of spectral types as classified by the procedure explained above, comprising roughly 2/3 of the spectra extracted from the HST input catalogue, and complemented for the remaining 1/3 with {\textsc DAOPHOT FIND} detections. The total number of classifications is thus 517, of which 265 were assigned highly plausible  (PC2, 51\%), 160 marginal (PC1, 31\%), and 92 uncertain (PC0, 18\%). The distribution of stars (subgiants, giants, and supergiants) in terms of temperature  reflects the stellar population of a spiral arm region, that was selected to not be dominated heavily by ongoing star formation, i.e. a minimal number of bright \hiiregs, in order to maximize the detection rate of faint PNe. We find an appreciable number of O star candidates, including a significant fraction of hot emission line stars (11 out of 38, i.e. 29\%), albeit the problem of our current approach using the MIUSCAT and GLIB libraries that do not allow for an accurate classification of hot stars due to the lack of O template stars in MIUSCAT, and the lack of non-LTE models for GLIB --- a shortcoming that we are planning to resolve with an improved approach in the future. For the remaining spectral types there is good coverage such that we expect to having obtained a complete sample of spectral types B...M brighter than or equal to luminosity class III.


\begin{figure}[h]
   \centering
    \includegraphics[width=\hsize,bb=45 75 750 500, clip]{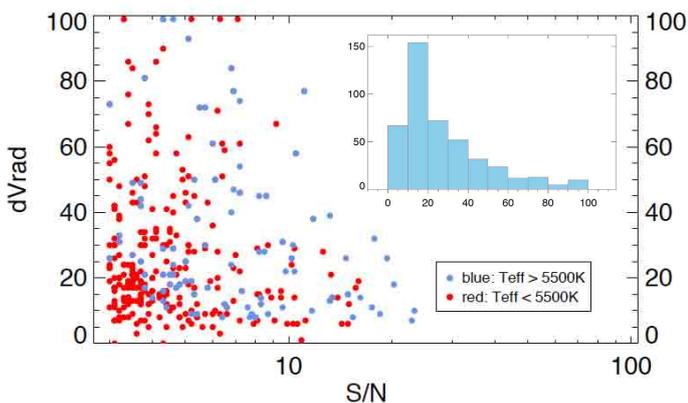}
    \caption{Radial velocity uncertaintites $\Delta$v$_\mathrm{rad}$ plotted against signal-to-noise ratio. The sample is segregated into hot and cool stars as in Fig.~\ref{SN-ratio}. The insert shows a histogram of the $\Delta$v$_\mathrm{rad}$ distribution, the 50th percentile at
    20~km~s$^{-1}$.}
    \label{hist_vrad}
 \end{figure}

 \begin{table*}[t]
 \caption{Catalogue of stars in field (i) with spectral type classification}             
\label{stars-i}      
\centering          
\begin{tabular}{ r c    c    c      r   c  r  c   c   c   r   l  c }     
\hline\hline     
  ID &  m$_{F606W}$ &  RA  &  Dec           &  S/N & B & $T_\mathrm{eff}$ & log {\it g} & [Fe/H] &  v$_r$ & $\Delta$v$_r$ & spectral type & P \\
\hline
     80 &  20.28 &   0:54:40.2 & -37:41:54.9 &  22.9 & 1 &   7643 &   1.50 &   0.10 &   191 &    7 &            F0II & 2 \\
   119 &  20.64 &   0:54:44.0 & -37:41:50.9 &  20.3 & 1 &   9377 &   1.50 &  -0.10 &   149 &   18 &      A1Ib..A5II & 2 \\
   123 &  20.61 &   0:54:41.2 & -37:41:34.2 &   8.2 & 2 &  17629 &   2.70 &  -0.20 &   186 &   28 &            B3Ib & 1 \\
   151 &  20.78 &   0:54:43.4 & -37:41:51.1 &  23.3 & 1 &   9377 &   1.90 &  -0.10 &   180 &   10 &            A1Ib & 2 \\
   192 &  20.92 &   0:54:42.7 & -37:42:07.3 &  19.4 & 1 &   9377 &   1.90 &  -0.10 &   157 &   26 &            A1Ib & 2 \\
   201 &  21.15 &   0:54:44.7 & -37:41:48.3 &  15.0 & 1 &   4175 &   0.80 &  -0.30 &   159 &   12 &           K3Iab & 2 \\
   209 &  21.03 &   0:54:41.0 & -37:41:54.3 &  18.5 & 1 &   8357 &   1.80 &   0.10 &   147 &    9 &      A5II..A1Ib & 2 \\
   216 &  21.28 &   0:54:43.0 & -37:41:50.5 &  13.9 & 0 &   3843 &   0.47 &  -0.12 &   173 &   14 &  M1Ia-ab..K3Iab & 2 \\
   223 &  21.05 &   0:54:42.1 & -37:41:36.1 &  14.7 & 1 &   9377 &   1.90 &  -0.10 &   175 &   26 &            A1Ib & 2 \\
   243 &  21.38 &   0:54:44.2 & -37:42:23.8 &  16.0 & 2 &   3845 &   0.34 &   0.10 &   168 &   19 &  K3Iab..M1Ia-ab & 2 \\
   246 &  21.33 &   0:54:44.8 & -37:41:46.9 &  14.9 & 2 &   3838 &   0.55 &  -0.11 &   157 &    6 &  K3Iab..M1Ia-ab & 2 \\
   247 &  21.10 &   0:54:43.2 & -37:42:00.7 &  17.8 & 1 &   9377 &   1.90 &  -0.10 &   157 &   32 &            A1Ib & 2 \\
   252 &  21.13 &   0:54:41.5 & -37:41:41.1 &  17.5 & 1 &  17629 &   2.70 &  -0.20 &   187 &   11 &            B3Ib & 2 \\
   260 &  21.38 &   0:54:41.7 & -37:41:54.0 &  15.7 & 1 &   3878 &   0.00 &  -0.30 &   171 &   17 &      K3Iab..K5I & 2 \\
   283 &  21.17 &   0:54:40.2 & -37:41:53.3 &  15.4 & 0 &  30000 &   0.00 &   0.00 &   999 &  999 &              OB & 0 \\
   301 &  21.56 &   0:54:42.1 & -37:41:51.7 &  14.5 & 2 &   3836 &   0.70 &  -0.40 &   174 &   13 &  K5III..M1Ia-ab & 2 \\
   352 &  21.40 &   0:54:41.2 & -37:41:36.8 &  14.9 & 1 &   8357 &   1.80 &   0.10 &   146 &   15 &     A5II...A1Ib & 2 \\
   395 &  21.75 &   0:54:40.8 & -37:41:40.6 &  14.6 & 2 &   3894 &   0.40 &  -0.30 &   183 &    2 &  K3Iab..M1Ia-ab & 2 \\
   402 &  21.46 &   0:54:44.9 & -37:42:30.7 &   7.5 & 1 &  30000 &   0.00 &   0.00 &   999 &  999 &            OBem & 2 \\
   407 &  21.70 &   0:54:43.5 & -37:42:14.1 &  13.3 & 1 &   3775 &   0.80 &  -0.30 &   166 &   21 &    K2III..M1Iab & 2 \\
   435 &  21.51 &   0:54:41.6 & -37:41:58.6 &  15.4 & 1 &   9377 &   1.90 &  -0.10 &   162 &    8 &            A1Ib & 2 \\
   462 &  21.57 &   0:54:44.1 & -37:41:45.2 &  13.4 & 2 &   8357 &   1.80 &   0.10 &   152 &   14 &     A5II..F6Iab & 1 \\
   468 &  21.61 &   0:54:41.7 & -37:41:50.4 &  13.2 & 1 &   8357 &   1.80 &   0.10 &   166 &   39 &            A5II & 2 \\
   479 &  21.87 &   0:54:41.1 & -37:41:59.3 &   7.7 & 1 &   4175 &   0.80 &  -0.30 &   161 &   25 &   K2IIIb..K3Iab & 2 \\
   487 &  21.61 &   0:54:42.8 & -37:41:38.8 &   5.4 & 1 &   6011 &   1.50 &   0.10 &   184 &    9 &           F6Iab & 1 \\
   488 &  21.82 &   0:54:39.9 & -37:41:54.5 &   9.9 & 0 &   4576 &   1.00 &   0.00 &   181 &    6 &      K2I..M1Iab & 2 \\
 $ \cdots$ &       &                   &                    &          &    &            &           &           &          &       &                     &   \\
 1003 &  22.82 &   0:54:43.3 & -37:41:53.7 &     0  & 0 &   0         &   9.99 &   9.99 &   999 &  999 &           PER & 0 \\
 $ \cdots$ &       &                   &                    &          &    &            &           &           &          &       &                     &   \\
 26377 &  24.33 &   0:54:43.6 & -37:41:41.2 &   3.4 & 2 &   3969 &   1.30 &  -0.30 &   168 &   14 &     K4III-K5III & 1 \\
 26433 &  24.27 &   0:54:42.5 & -37:41:40.1 &   4.0 & 1 &   3810 &   1.10 &   0.00 &   157 &   17 &           M0III & 1 \\
 26596 &  24.26 &   0:54:43.9 & -37:41:55.7 &   3.2 & 2 &   3915 &   1.46 &   0.25 &   193 &   10 &           K5III & 0 \\
 26762 &  24.34 &   0:54:41.5 & -37:42:01.8 &   3.1 & 1 &   3810 &   1.10 &   0.00 &   207 &   32 &           M0III & 1 \\
 26771 &  24.05 &   0:54:42.8 & -37:41:53.2 &   4.2 & 1 &   4379 &   2.60 &  -0.10 &   197 &    8 &          K2IIIb & 1 \\
 27053 &  24.19 &   0:54:44.6 & -37:42:14.0 &   3.0 & 2 &   4159 &   1.90 &   0.10 &   151 &   19 &           K4III & 1 \\
 27597 &  24.20 &   0:54:44.5 & -37:41:45.9 &   3.4 & 2 &   3900 &   1.60 &  -0.40 &   999 &  999 &           K5III & 0 \\
 27794 &  24.32 &   0:54:42.8 & -37:42:02.0 &   4.3 & 0 &   3939 &   1.80 &  -0.30 &   183 &   90 &         K3.5III & 0 \\
 28087 &  24.09 &   0:54:43.2 & -37:42:09.3 &   3.1 & 0 &      0 &   9.99 &   9.99 &   999 &  999 &            none & 0 \\
 29133 &  24.36 &   0:54:44.2 & -37:41:45.5 &   3.8 & 0 &   3810 &   1.10 &   0.00 &   176 &   51 &           M0III & 2 \\
 29326 &  24.44 &   0:54:43.8 & -37:42:17.7 &   3.5 & 2 &   3915 &   1.46 &   0.25 &   175 &   19 &           K5III & 1 \\
 29932 &  24.30 &   0:54:42.6 & -37:41:57.5 &   3.0 & 2 &   3900 &   1.60 &  -0.40 &   171 &   58 &           K5III & 1 \\
 30791 &  24.52 &   0:54:43.4 & -37:41:59.9 &   3.1 & 0 &   3915 &   1.46 &   0.25 &   166 &   17 &           K5III & 0 \\
 30801 &  24.94 &   0:54:41.4 & -37:41:39.8 &   3.2 & 2 &   3810 &   1.10 &   0.00 &   101 &   48 &           M0III & 1 \\
 30878 &  23.88 &   0:54:42.8 & -37:42:06.6 &   3.4 & 1 &   3244 &   0.20 &   0.00 &   169 &   14 &           M6III & 2 \\
 32224 &  24.25 &   0:54:43.3 & -37:41:54.3 &   9.1 & 1 &   3481 &   4.40 &  -0.70 &    66 &    7 &              K5V & 2 \\
 32522 &  24.87 &   0:54:41.1 & -37:41:57.7 &   4.1 & 1 &   3665 &   1.20 &  -0.20 &   163 &   35 &           M3III & 1 \\
   $\cdots$ &      &                    &                    &          &    &            &           &           &          &       &                     &   \\
     \hline                  
\end{tabular}
\tablefoot{Excerpt of catalogue for field (i) with spectral type classification for spectra extracted with \textsc{PampelMUSE}, covering
ranges of brighter and fainter F606W magnitudes (full table available online). Column~1, ANGST catalogue ID; Col.~2, F606W magnitude;
Col.~3, right ascension (J2000); Col.~4, declination (J2000); Col.~5, signal-to-noise ratio estimate from \textsc{PampelMUSE};
Col.~6, flag ''blending'': 2=not obvious, 1=minor, 0=significant; Col.~7, effective temperature of best fit [K]; Col.~8, gravity of best fit; Col.~9, metallicity of best fit; Col~10, radial velocity [km~s$^{-1}$]; Col.~11, radial velocity uncertainty [km~s$^{-1}$]; Col~12, spectral type of best fit, or probable range of  spectral type for ambiguous cases; Col.~13: flag "plausibility of fit": 2=very plausible, 1=marginal, 0=uncertain. ''999'' in Col.~10 and 11 indicate uncertain results. Hot stars assigned OB, $T_\mathrm{eff}$ is set to 30000. PER: peculiar red star visible in HST image, not detected in MUSE cube.
} 
\end{table*}

The radial velocities determined by the MIUSCAT and GLIB fitting procedures were found in the majority of cases to be both plausible and in good agreement with each other, even for spectra with S/N as low as 3, where sometimes important diagnostic absorption lines such as e.g. the calcium triplet were only marginally discernible from the noise. The measurements merely failed in cases where excessive residuals from very strong nebular emission line background, that was too bright to be removed by {\textsc PampelMUSE}, prevented a reliable determination. We attribute this result to the robustness of our fitting procedures that are essentially based on a multitude of features over a free spectral range as large as an entire octave, rather than looking only into a few selected absorption lines. The values of v$_\mathrm{rad}$ all seem to be realistic in that they are close to the systemic velocity of 144~km~s$^{-1}$, with a mean of 169~km~s$^{-1}$ and a dispersion of 23~km~s$^{-1}$ in the case of field (i). Preliminary results of radial velocities from the remaining pointings indicate that the increase of median radial velocity per field with growing galactocentric distance in the range of 140...200 km~s$^{-1}$ is indeed sampling the rotation curve of NGC\,300. This finding is further discussed in $\S$\ref{discuss_kinematics} and Fig.~\ref{kinematics} below. 

Fig.~\ref{hist_vrad} illustrates the scatter of formal radial velocity errors
$\Delta$v$_\mathrm{rad}$ determined by ULySS versus S/N,  plotted with red and blue symbols for hot and cool stars, respectively, as in Fig.~\ref{SN-ratio}. The cumulative histogram for the blue and red subsample shows that half of the radial velocity errors are below 20~km~s$^{-1}$, and the 78th percentile at 40~km~s$^{-1}$. The plot also illustrates that even faint red giant and supergiant spectra down to 24$\ldots$25~mag with S/N$\approx$3 or less have yielded useful radial velocity information. 

An unexpected finding was the presence of spectra of cool stars that were not fitted at all, neither with MIUSCAT, nor with GLIB, despite a reasonable S/N. By comparison with spectra from the literature, e.g. \citet{vanLoon2005}, these stars turned out to be carbon stars that are as yet not covered by the libraries we use. We find 23 such stars in field (i) with a high level of confidence, rendering the C to M star ratio C/M=0.13, which is reasonably in line with the study of \citet{hamren2015} in M31, who find C/M$\approx0.15\ldots0.32$ at [O/H]=8.6. In NGC\,300, this compares to a metallicity  for field~(i) at a galactocentric distance of 1.5 kpc (R/R25=0.3) of [O/H]=8.5$\pm$0.1 dex, following \citet{bresolin2009}.

\begin{figure*}[h]
     \centering
    \includegraphics[width=\hsize,bb=10 325 740 570,clip]{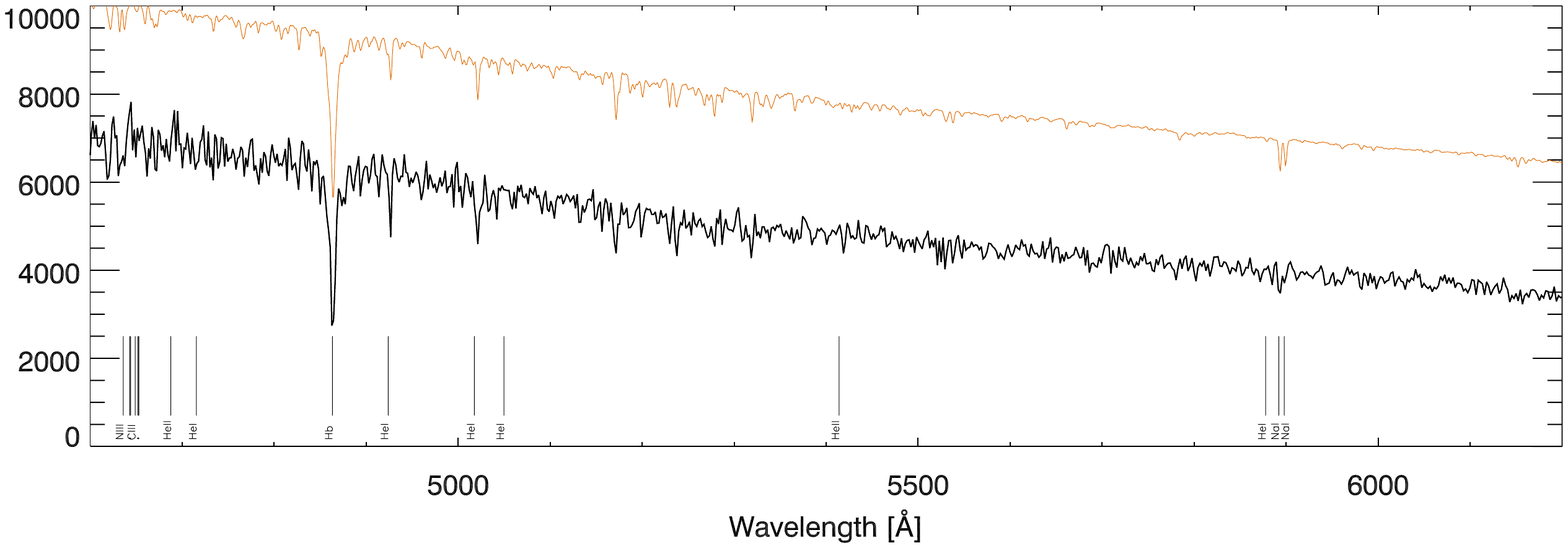}
    \includegraphics[width=\hsize,bb=10 325 740 570,clip]{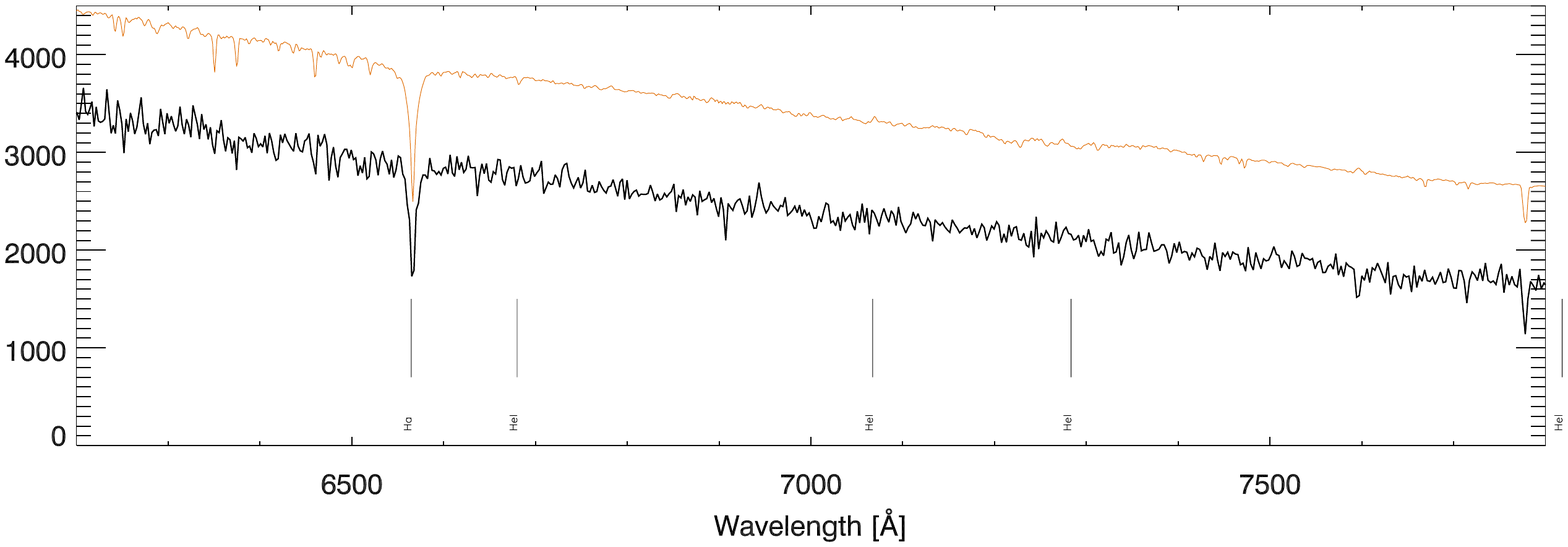}
    \includegraphics[width=\hsize,bb=10 325 740 570,clip]{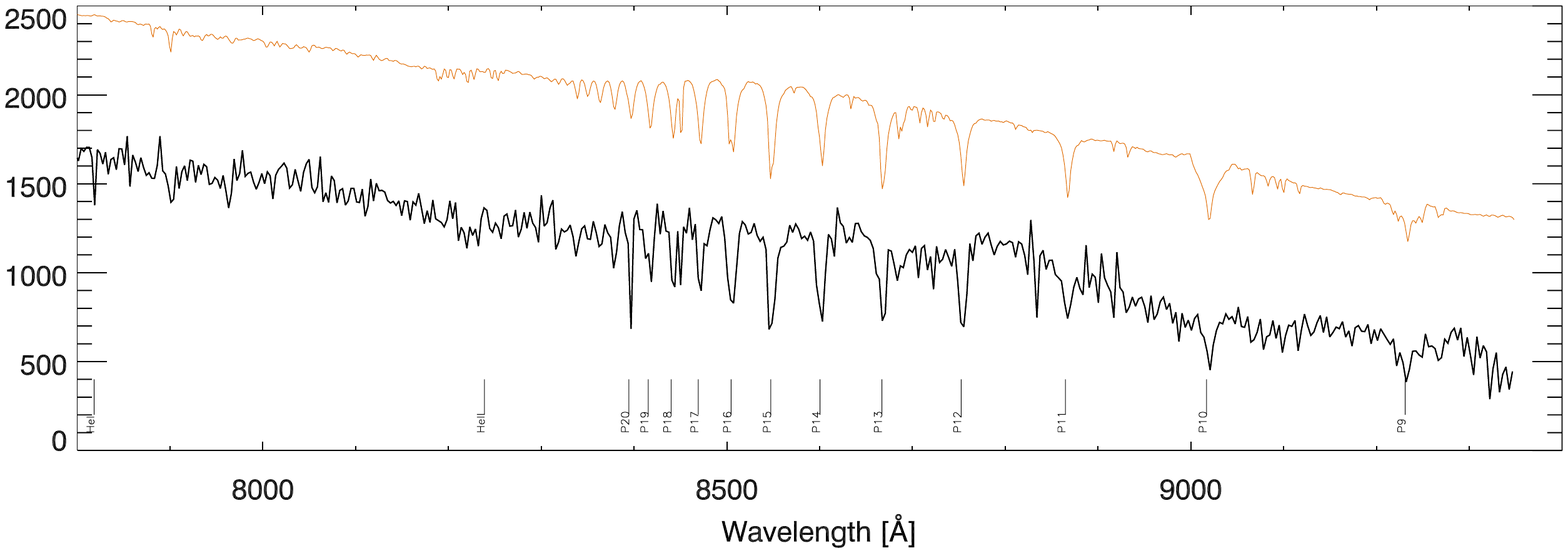}
    \caption{MUSE spectrum of A5 supergiant B11 from the sample of \citet{bresolin2002}, seen in field (b), ID25. 
    This star has a V magnitude of 19.95. The MIUSCAT fit is plotted in orange with an offset for clarity. 
    Wavelengths in units of \AA, flux in units of $10^{-20}$erg cm$^{-2}$ s$^{-1}$ $\AA^{-1}$.
    The reference wavelengths of a number of absorption lines (hydrogen Balmer and Paschen series, helium lines) are indicated. }
   \label{Supergiant_B11}  
\end{figure*}

\begin{figure*}[h]  
   \centering
    \includegraphics[width=\hsize,bb=10 325 740 570,clip]{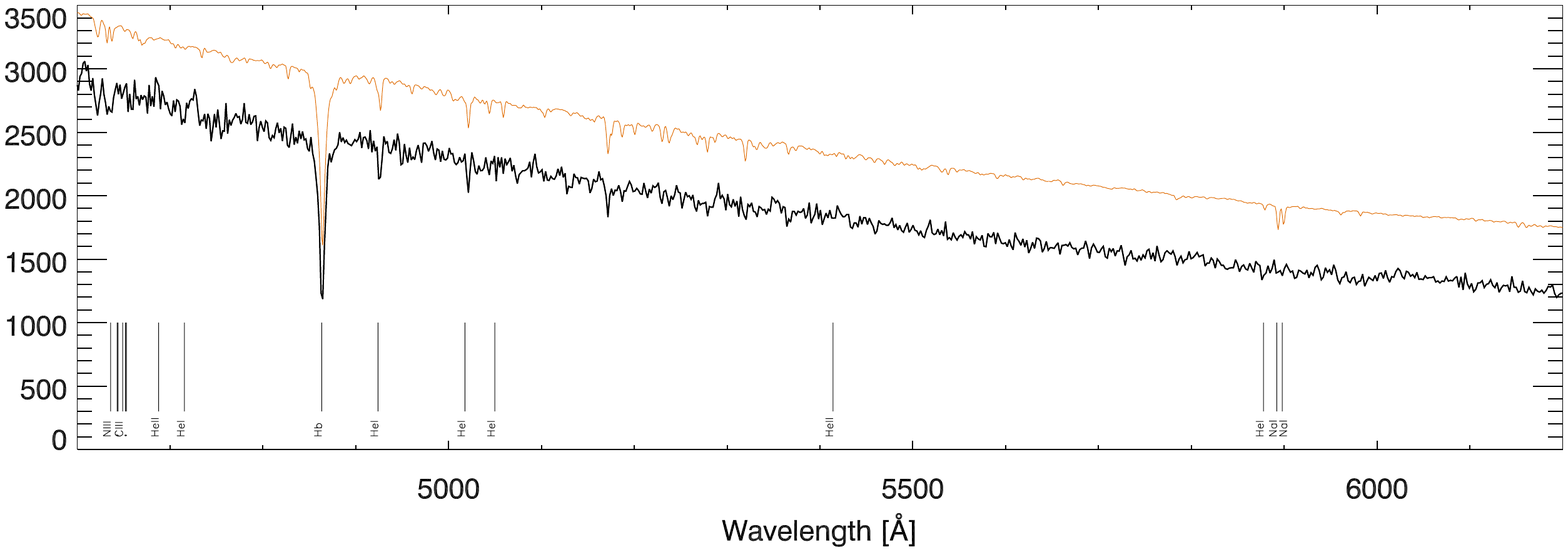}
    \includegraphics[width=\hsize,bb=10 325 740 570,clip]{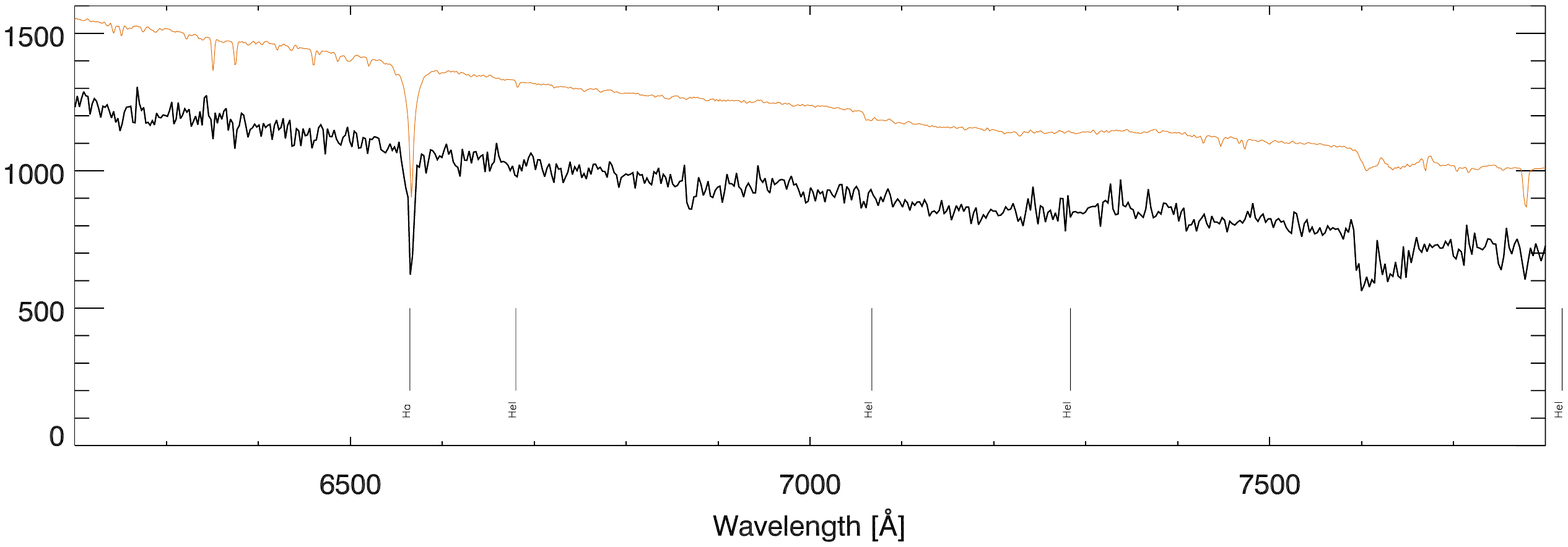}
    \includegraphics[width=\hsize,bb=10 325 740 570,clip]{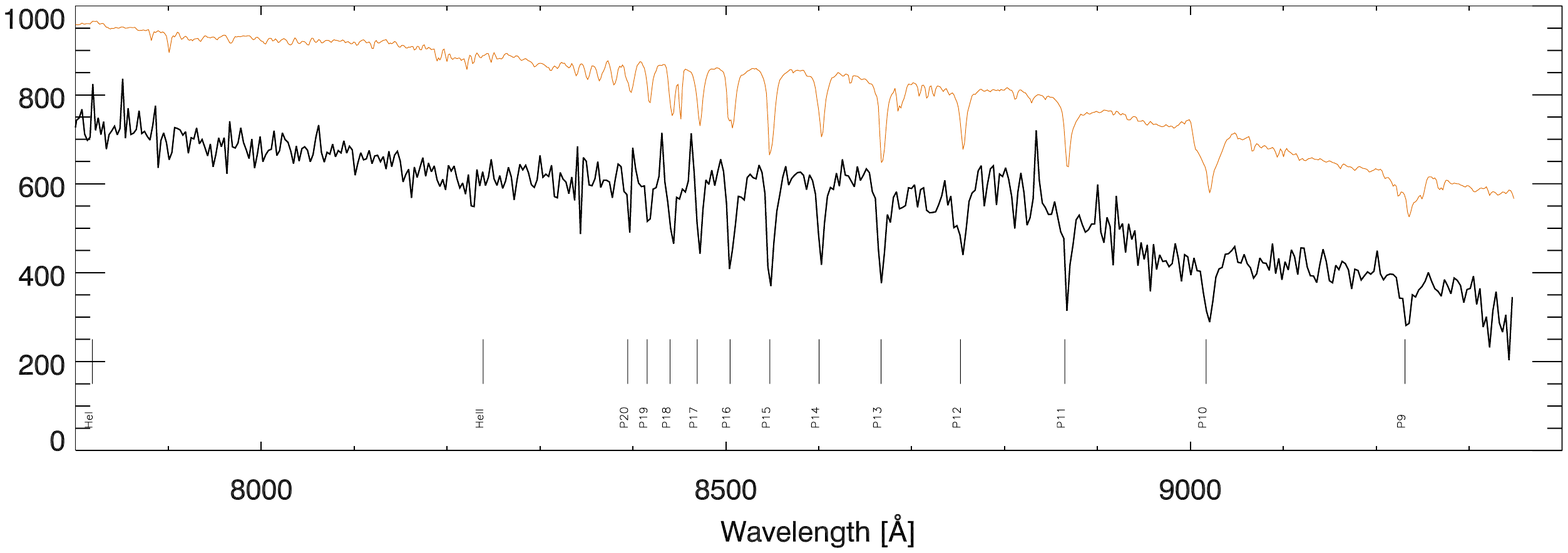}
    \caption{MUSE spectrum of AIa star of magnitude m$_{F606W}$=20.78 in field (i), ID151. 
    The MIUSCAT fit is plotted in orange with an offset for clarity.
    Wavelengths in units of \AA, flux in units of $10^{-20}$erg cm$^{-2}$ s$^{-1}$ $\AA^{-1}$.}
    \label{AIa_i_ID151}
\end{figure*}

\begin{figure*}[h]
   \centering
    \includegraphics[width=\hsize,bb=10 325 740 570,clip]{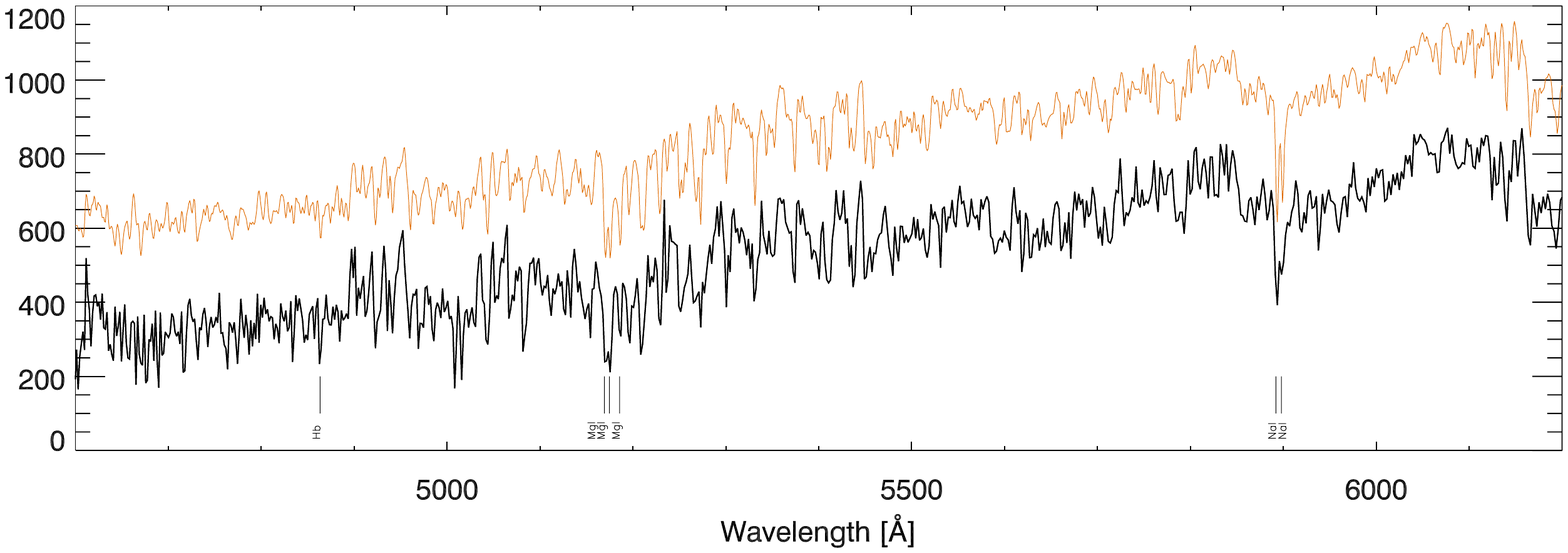}
    \includegraphics[width=\hsize,bb=10 325 740 570,clip]{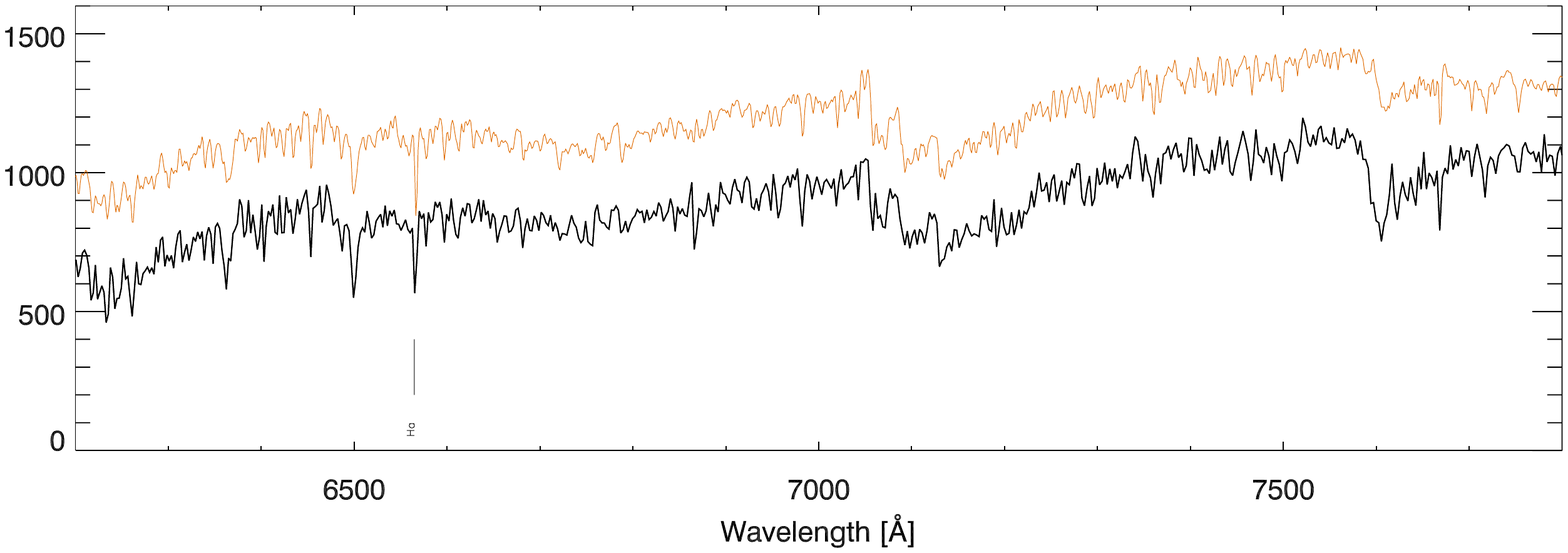}
    \includegraphics[width=\hsize,bb=10 325 740 570,clip]{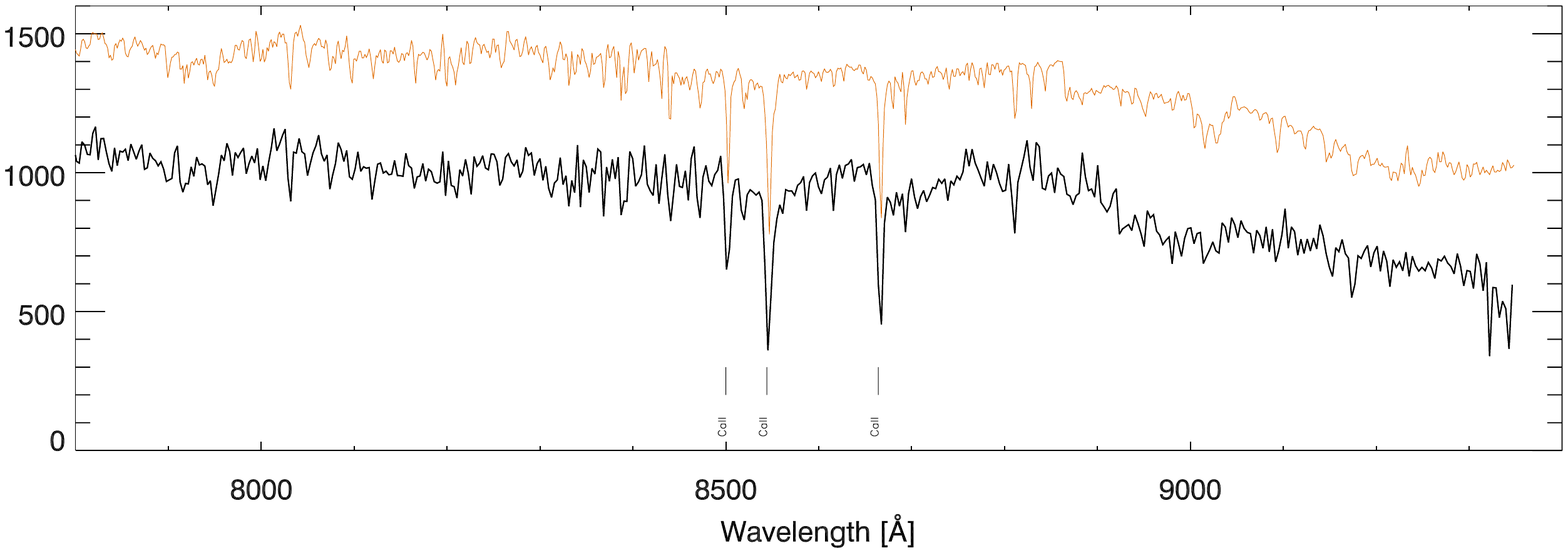}
    \caption{MUSE spectrum of M1Ia star of magnitude m$_{F606W}$=21.56 in field (i), ID301. 
    The MIUSCAT fit is plotted in orange with an offset for clarity.
    Wavelengths in units of \AA, flux in units of $10^{-20}$erg cm$^{-2}$ s$^{-1}$ $\AA^{-1}$.
    The reference wavelengths of absorption lines of calcium, magnesium, sodium, H$\alpha$ 
    and H$\beta$ are indicated.}
   \label{MIa_i_ID301}
\end{figure*}

We have also discovered a number of emission line stars from the visual inspection after automatic processing for MIUSCAT and GLIB fits. Their spectra are typically showing broad H$\alpha$ and H$\beta$ emission lines that are characteristic of hot stars with strong stellar winds. As our current fitting procedure is not sensitive enough to distinguish well enough between different classes of such stars, we have assigned, for the time being, merely a classification OBem. An alternative, more complete approach to discover and measure emission line stars and the discovery of a WR star is discussed in \S~\ref{emStars}.

For resolved stellar population studies one has to correct for contamination of the sample under study by foreground stars which are difficult to identify merely on the basis of  photometry.  The spectrum of star A in Fig.~\ref{foreground}, ID 32224 in Table~\ref{stars-i} with F606W=24.25,  is classified a K5 dwarf, translating to an apparent R magnitude of 33.3 for the distance modulus of NGC\,300, which is obviously by far too faint to be observable. The radial velocities determined from the MIUSCAT and GLIB fits are $66\pm6$~km~s$^{-1}$ and $69\pm3$~km~s$^{-1}$, respectively, i.e. almost identical, each with a very small uncertainty. This velocity is far away from the systemic velocity of NGC\,300 and therefore, together with the photometric evidence, indicative of a foreground star. In this particular case, the situation is somewhat more complicated as it turns out that in the HST image star A is resolved into two distinct stars. From the ULySS fit, we conclude that there is a weak blend from an M6III star, which explains the less than perfect fit of the spectrum. Nevertheless, the facts remain that for the main sequence K spectrum component, such a star is not visible at the distance of NGC\,300, and that the well-constrained radial velocity is incompatible with stars within this galaxy. Using the new Besancon Galaxy model \citep{czekaj2014} we have estimated the number of stars that one would expect in the direction towards NGC\,300 down to V = 20 to 0.5 stars  per 1$\times$1~arcmin$^2$. Extrapolation to V=24.25 predicts approximately 4500 star per square degree, i.e. 1.25 stars per MUSE pointing, which seems to be in accord with our observation.

{\noindent \bf Discussion:} Our first results from a complete analysis of field (i) have shown that the application of PSF-fitting crowded field 3D spectroscopy as previously exercised in globular clusters could successfully be established in the more challenging case of nearby galaxies, where, owing to a much larger distance, we can no longer expect to reach down to the main sequence (except for the most massive O stars). However, we have demonstrated that we are able to sample the population of giants, subgiants, and supergiants, from which we obtain spectra with sufficient S/N to make a spectral type classification, as well as to measure radial velocities with accuracies around 20~km~s$^{-1}$. We have also demonstrated that the deblending technique works well, even for stars as faint as m$_{F606W}\approx23$, provided we can rely on an input catalogue of HST stars.

We note that, thanks to the high efficiency of MUSE, the spectra of blue, yellow, and red supergiants in our sample have good enough S/N, despite the relatively short total exposure time of 1.5~h of the current data set, to allow for quantitative spectroscopy along the lines of, e.g. \citet{bresolin2002}, \citet{Kudritzki2008}, considering numerous absorption lines of the Balmer and Paschen series of hydrogen, of Mg, Fe, Na, Ca, etc., that are immediately apparent in the spectra. We have one star in common with the \citet{bresolin2002} sample that is suitable to make a direct comparison: their object B11, classified as A5  supergiant (marked in Fig.~\ref{NGC300_stars}). This star was observed with FORS1 at the VLT with approximately $5~\AA$ resolution, a total exposure time of 3.75~h, and seeing conditions ranging from 0.4'' to 0.7''. The MUSE parameters for comparison are: $2.5~\AA$ resolution, 1.5~h exposure time, 1.2" seeing. The star is listed as ID25 from a preliminary analysis of field (b). The MIUSCAT-based classification of this star is A1Iab...A5II, in agreement with \citet{bresolin2002}, despite a lower S/N due to poorer observing conditions and a shorter exposure time. Fig.~\ref{Supergiant_B11} shows our spectrum of this star, along with the best ULySS fit (with an offset for clarity). Unfortunately, the overlap between the FORS and MUSE spectra is restricted to a narrow interval in the H$\beta$ region of 4600\dots4900~\AA. However, the comparison is good enough to qualitatively demonstrate that MUSE performs at least as well as FORS, considering the less favourable observing conditions for the MUSE spectrum (exposure time, seeing),
and despite of the fact that in this wavelength region the efficiency curve of MUSE experiences a steep drop towards the blue. 

The importance of the statistics of red and blue supergiants was stressed by \citet{massey2017}. From a review on massive stars by \citet{massey2003}, and a more recent paper by \citet{massey2016} describing a survey in M31 and M33, the amount of work necessary
to safely detect and classify massive stars is quite apparent. However, the exercise to create a census of massive stars is a prerequisite to understand stellar evolution at the high mass end. The method is based, firstly, on broadband and narrowband photometry, to identify candidate stars. As photometry in the optical merely samples the Rayleigh-Jeans tail of the spectral energy distribution of hot stars, it is, secondly, necessary to perform follow-up spectroscopy, which is a tedious task with conventional multi-object spectrographs. As we have shown with a proof-of-principle in field (i), such a two-step procedure is no longer necessary with IFS. MUSE allows, for the first time over a reasonable FoV, to combine the different survey strategies into one: imaging and spectroscopy combined in a single datacube under identical observing conditions. Concerning the imaging capabilities, it is worthwhile noting that the discovery of evolved massive stars with heavy mass loss and fast stellar winds such as LBVs or WR stars is much facilitated by narrowband images that are readily extracted from a datacube at the wavelengths of interest, e.g. H$\alpha$, or {\heii} $\lambda$4686, see for comparison \citet{massey2015}. Although this capability was already demonstrated with PMAS by \citet{relano2010} and \citet{monreal2011}, who found WR stars in M33, it is only now with the advent of MUSE that we are able to analyse such data over a wide FoV.

The argument also holds for stars with spectral types other than OB. \citet{hamren2016}, for example, have studied carbon stars in the satellites and halo of M31 using 10 years worth of data from the SPLASH survey \citep{guhathakurta2006AJ,tollerud2012}. The ratio (C/M) of carbon-rich to oxygen-rich AGB stars has been used to study the evolution of AGB stars, concerning observational constraints on the thermally pulsating AGB phase, dredge-up processes, opacities, and mass loss during this phase. C/M ratios have also been used to study the galactic environment in which the stars have formed. The data of \citet{hamren2016} include 14143 stellar spectra taken with DEIMOS at the Keck-II 10m telescope. The spectra were obtained from 151 individual DEIMOS masks for 60 separate fields, with a typical exposure time of 3600~s per mask. The observations were targeting the environment of M31, i.e. dwarf spheroidals, dwarf ellipticals, the smooth virialized halo, halo substructure, and M32. From this entire dataset, 41 unambiguous carbon stars were identified. It is interesting to note that spectroscopy is apparently essential for obtaining a reliable identification. We take as an example the dwarf elliptical galaxy NGC\,147: based on NIR photometry, \citet{davidge2005} found 65 carbon stars, \citet{sohn2006} reported 91 carbon stars using the same technique, while \citet{hamren2016} confirmed merely 12 stars on the basis of photometry and spectroscopy in the visual. For comparison, our 1.5~h exposure in field (i) of NGC\,300 has yielded immediately 23 detections of carbon stars. While this is not really a fair comparison to the sample of \citet{hamren2016} owing to effects of metallicity and the size of the underlying stellar population, we note that a similar study in the disk of M31 by \citet{hamren2015}, which would be more comparable to NGC\,300, has yielded a total of 103 carbon star identifications from a sample of 10,619 spectra, again taken with DEIMOS and the circumstances described above.

Surveys of stars in external galaxies like the ones cited above must be corrected for foreground contaminant stars. This is usually accomplished spectroscopically, e.g. by constraining spectral type and luminosity class, and thereafter comparison with photometry, but also by comparing the measured radial velocity with the systemic velocity of the galaxy under study. We emphasize that through our automated procedure described in ${\S}$3.3, spectral type and radial velocity are directly reported as part of the process, so such contaminant stars are immediately identified without further effort. The example in ${\S}$4.1, Fig.~\ref{foreground}, illustrates how the foreground star detection is an integral part of the data analysis pipeline.
As shown in Fig.\ref{SN-ratio} and Table~\ref{distrib-stars-i}, the extracted spectra for field (i) span a range of S/N with a total of 265 that are flagged "good". Of those latter spectra several tens have a S/N$>$10. We have already referred to the blue supergiant B11 whose spectrum in Fig.~\ref{Supergiant_B11} is a less than ideal example due to the mediocre seeing of the observation in field (b). The S/N estimate for this spectrum is 17, decreasing somewhat towards the blue. Despite the modest S/N, a number of absorption lines can be seen. A higher quality spectrum is presented for the A1 supergiant ID151 of field (i) in Fig.~\ref{AIa_i_ID151} that exhibits a S/N of 22 ---  despite the fact that this star is about a magnitude fainter than B11, illustrating the importance of high image quality for this kind of observation. In contrast to the blue supergiants with relatively few identifiable lines, our example of the late K to early M supergiant ID301 in field (i) with m$_{F606W}$=21.56 and S/N=14 shows a multitude of absorption lines, including TiO bands, that are reproduced by the MIUSCAT fit (Fig.~\ref{MIa_i_ID301}). 

It is clear that with longer exposure times, the quality of such spectra will be improved. We stress again, however, that excellent seeing is a prerequisite. Insofar the installation of the adaptive optics (AO) facility at VLT-UT4 and the GALACSI module in front of MUSE \citep{stuik2006} will even further improve the deblending performance of our technique in crowded fields. For the time being, we have focused our effort on the automatic processing of hundreds of spectra with a pipeline for the purpose of spectral classification and the determination of radial velocities. The results from a provisional analysis for all fields (a)\ldots(i) are discussed later in $\S$\ref{discuss_kinematics}. 

Obviously, the comparison with model atmospheres and the determination of abundances is the next logical step. However, we have as yet only begun to develop the technique of crowded field 3D spectroscopy at 2 Mpc distance. In the first instance, we required to visually inspect all spectra one by one in order to gain confidence in the procedure. This turned out to be a very time-consuming procedure. For the future, the goal is to develop a fully automated, robust pipeline with very little, if any, human interaction. Also, as a weakness of the current scheme, the MIUSCAT and GLIB libraries do not support well the analysis of hot stars. This is an issue that we are planning to address in the near future. 

In summary, we believe that the first analysis of spectra in field (i) presented in this paper has validated crowded field 3D spectroscopy with MUSE as a powerful tool for quantitative spectroscopy of individual stars in nearby galaxies. 

\begin{figure*}[!th]
   \centering
    \includegraphics[width=\hsize,bb=35 20 750 570,clip]{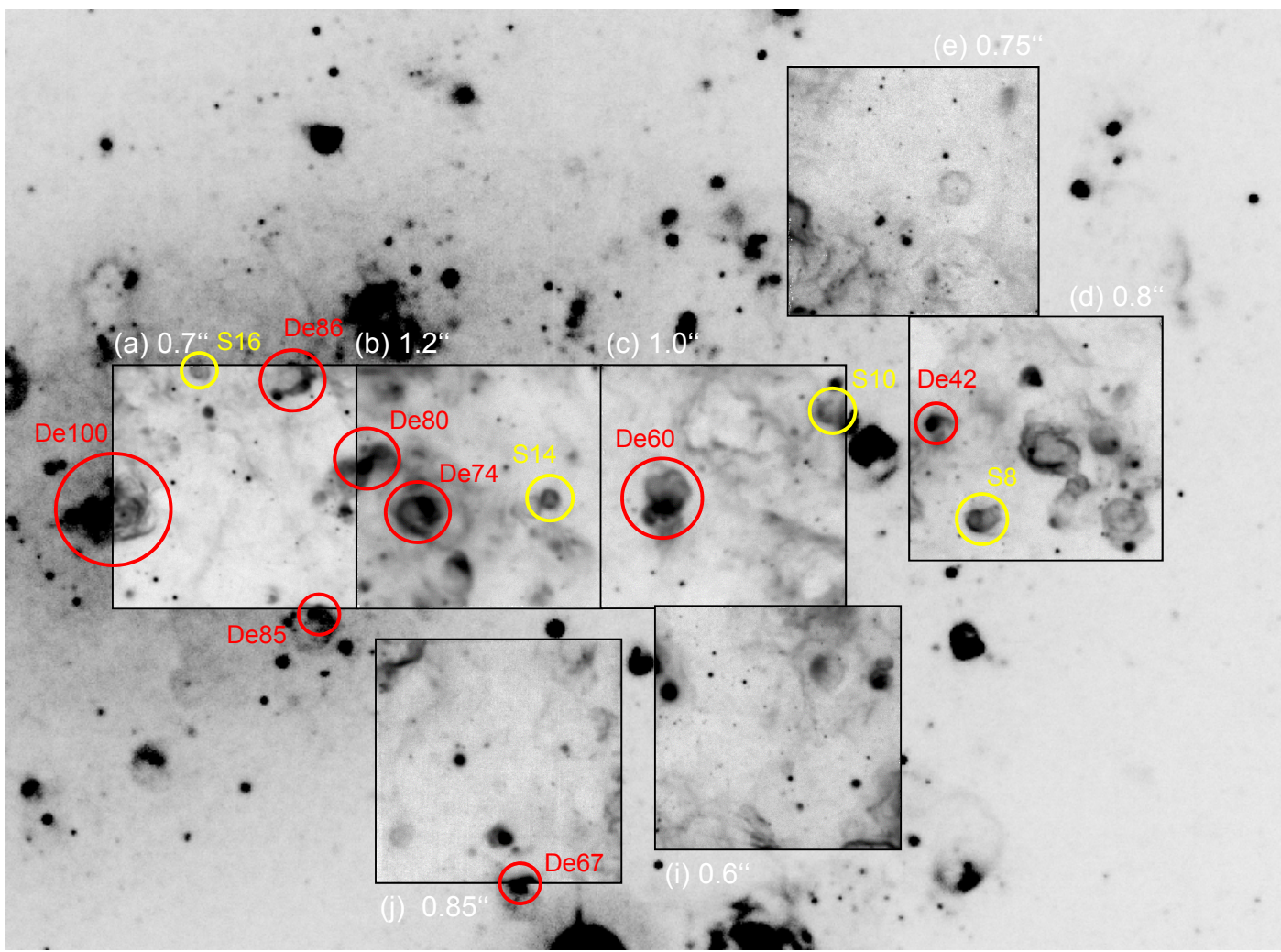}
    \caption{H$\alpha$ images reconstructed from MUSE datacubes for pointings (a), (b), $\ldots$, (j). The background image is from the ESO/MPI 2.2m WFI (credit: F. Bresolin). \hiiregs   in common with the catalogue of \citet{deharveng1988} are marked with red circles. Yellow circles denote the supernova remnants recorded by \citet{blair1997}.
    }
     \label{NGC300_Ha}
\end{figure*}

\subsection{Planetary nebulae and compact \hiiregs}
\label{Results-PNe}

 \begin{table*}[t]
 \caption{Catalogue of confirmed and candidate planetary nebulae (45 objects)}             
\label{PN-list}      
\centering          
\begin{small}
\begin{tabular}{ c  r  r     r r    c     c  c      c  c       l         l  }     
\hline\hline     
 ID  & S96 & P12 & x &  y & RA       &   Dec           &  m$_{5007}$  &  F(5007) & F(H$\alpha$) &  v$_r$   &  comment \\
\hline
a01 &       &      & 240.19 & 306.84 & 00:54:52.330 & -37:40:34.78 & 27.18 &    4.3$\pm$0.3   &   0.9$\pm$0.1    & 181$\pm$20    & ?  \\
a03 &       &      & 301.51 & 290.84 & 00:54:51.299 & -37:40:37.97 & 28.88 &    0.9$\pm$0.2   &   ...   & 197$\pm$9      & ? \\
a16 &  14 & 42 & 186.88 & 215.31 & 00:54:53.230 & -37:40:53.08 & 24.75 &  40.0$\pm$2.2   &  11.3$\pm$0.7   & 154$\pm$9      &  PN \\
a36:&       &      & 110.01 & 132.60 & 00:54:54.525 & -37:41:09.70 & 27.30 &   3.8$\pm$0.6    &   1.3$\pm$0.3    & 146$\pm$3      &  PN\tablefootmark{~(4)} \\
a53 &   5  & 51 &  64.77 &  46.09  & 00:54:55.311  & -37:41:27.29 & 23.18 & 171.0$\pm$8.7  &  64.0$\pm$3.3   & 133$\pm$7      &  PN \\
a60 &   7  & 48 &  88.68 &  23.81  & 00:54:54.903 & -37:41:31.75 & 23.31 & 152.0$\pm$7.7   &  48.0$\pm$2.4   & 139$\pm$5      &  PN \\
a61:&       &      & 130.04 &  14.35 & 00:54:54.209 & -37:41:33.70 & 27.35 &   3.7$\pm$0.3     &   0.7$\pm$0.1    & 150$\pm$12    & PN\tablefootmark{~(4)} \\     
a62:&       &      & 252.96 &  26.80 & 00:54:52.116  & -37:41:31.21 & 27.52 &   3.1$\pm$0.2    &   1.1$\pm$0.1    & 140$\pm$4      &  ? \tablefootmark{~(1),}\tablefootmark{(4)}  \\
a72:&       &      & 313.70 & 112.04 & 00:54:51.097 & -37:41:13.73 & 27.70 &   2.6$\pm$0.3    &   ...    & 150$\pm$20    &  ? \tablefootmark{~(4)} \\
b06:&       &      & 108.95 & 270.53 & 00:54:49.492 & -37:40:42.12 & 25.25 &  25.4$\pm$1.5   &   4.8$\pm$0.5    & 244$\pm$3      &  PN\tablefootmark{~(4)} \\ 
b12:&       &      & 187.37 & 239.96 & 00:54:48.171 & -37:40:48.23 & 26.06 &  12.0$\pm$0.8   &  12.0$\pm$0.7   & 145$\pm$3      &  PN\tablefootmark{~(4)} \\ 
b19:&       &      & 163.89 & 193.00 & 00:54:48.567 & -37:40:57.63 & 27.55 &   3.0$\pm$2.8    &   1.2$\pm$1.3    & 178$\pm$6      &  PN\tablefootmark{~(4)} \\ 
b41:&       &      & 304.33 & 202.30 & 00:54:46.201 & -37:40:55.76 & 28.00 &   2.0$\pm$0.3    &   0.9$\pm$0.2    & 168$\pm$9      &  PN\tablefootmark{~(4)} \\ 
b42:&       &      & 241.86 &  84.01  & 00:54:47.253 & -37:41:19.42 & 28.33 &   1.5$\pm$0.2    &   1.3$\pm$0.1    & 176$\pm$5      &  PN\tablefootmark{~(4)} \\ 
b47:&       &      &   42.05 &  39.27  & 00:54:50.620 & -37:41:28.37 & 26.86 &   5.8$\pm$0.6    &   ...    & 159$\pm$8      &  PN\tablefootmark{~(4)} \\ 
b50:&       &      & 233.44 &  34.84  & 00:54:47.395 & -37:41:29.26 & 26.31 &   9.6$\pm$0.8    &   3.2$\pm$0.4    & 162$\pm$14    &  PN\tablefootmark{~(4)} \\ 
c29 &   2  & 25 & 107.13 &   37.60 & 00:54:44.406 & -37:41:29.02 & 23.24 & 161.0$\pm$8.2  &  36.1$\pm$2.3   & 175$\pm$8      &  PN \\
c30:&       &      & 236.42 &   78.47 & 00:54:42.227 & -37:41:20.85 & 28.17 &   1.7$\pm$0.4    &   ...    & 170$\pm$6      &  PN\tablefootmark{~(4)} \\
c32 &       &      &   23.16 & 252.19 & 00:54:45.820 & -37:40:46.10 & 28.09 &   1.9$\pm$0.3    &   0.7$\pm$0.2    & 141$\pm$11     & PN \\
c33:&       &      & 186.66 &  57.59 & 00:54:43.066  & -37:41:25.02 & 27.04 &   4.9$\pm$0.6    &   1.5$\pm$0.6    & 169$\pm$14     & PN\tablefootmark{~(4)} \\ 
c35:&       &      &   75.58 & 245.97 & 00:54:44.937 & -37:40:47.35 & 27.81 &   2.4$\pm$0.4    &   ...    & 201$\pm$13     &  PN\tablefootmark{~(4)} \\
c40 &       &      & 108.02 & 107.35 & 00:54:44.391 & -37:41:15.07 & 25.65 &  17.5$\pm$1.4   &   ...    & 163$\pm$6       &  PN\tablefootmark{~(2)} \\ 
d41:&       &      & 236.90 & 105.79 & 00:54:35.799 & -37:41:02.56 & 29.02 &   0.8$\pm$0.1    &   ...    & 182$\pm$16     &  PN\tablefootmark{~(4),}\tablefootmark{(7)} \\ 
d46:&       &      &  13.35  & 100.88 & 00:54:39.566 & -37:41:03.54 & 26.12 &  11.4$\pm$0.7   &   4.0$\pm$0.3     & 186$\pm$11     &  PN\tablefootmark{~(4),}\tablefootmark{(8)} \\ 
d63 &       & 11 & 121.10 &   27.10 & 00:54:37.752 & -37:41:18.32 & 26.82 &   6.0$\pm$0.4    &   2.1$\pm$0.2     & 139$\pm$11     &  PN\tablefootmark{~(2)} \\  
d71:&       &      &  57.61  & 230.54 & 00:54:38.820 & -37:40:37.61 & 28.43 &   1.4$\pm$0.2    &   2.1$\pm$0.1     & 140$\pm$8       &  ?\tablefootmark{~(4)} \\ 
d79 &       &      &   45.54 &  66.90  & 00:54:39.023 & -37:41:10.34 & 28.33 &   1.5$\pm$0.1    &   0.3$\pm$0.1     & 144$\pm$7       &  PN\\
e01 &       & 14 & 196.00 & 201.92 & 00:54:38.845 & -37:39:41.36 & 22.76 & 251.0$\pm$12.7 &  71.5$\pm$3.6   & 210$\pm$3       &  PN \\
e02 &       & 13 & 242.50 & 210.42 & 00:54:38.062 & -37:39:39.66 & 25.81 &  15.2$\pm$0.9   &   4.3$\pm$0.3     & 145$\pm$4       &  PN \\
e07 &       &      &  54.41  & 215.78 & 00:54:41.210 & -37:39:38.58 & 27.95 &   2.1$\pm$0.2    &   1.2$\pm$0.1     & 174$\pm$6       &  PN \\
e08:&       &      &  47.12  & 205.50 & 00:54:41.353 & -37:39:40.64 & 29.67 &   0.4$\pm$0.1    &   ...     & 197$\pm$20     &  ?\tablefootmark{~(4)} \\ 
e11:&       &      & 144.13 & 137.54 & 00:54:39.718 & -37:39:54.23 & 29.41 &   0.6$\pm$0.1    &   0.2$\pm$0.1     & 169$\pm$20     &  ?\tablefootmark{~(4)} \\ 
e14:&       &      &  31.91  &   33.36 & 00:54:41.609 & -37:40:15.07 & 29.03 &   0.8$\pm$0.1    &   0.4$\pm$0.1     & 175$\pm$20     &  ?\tablefootmark{~(4)} \\
e15 & 12  & 20 &   35.94 &  11.45  & 00:54:41.541 & -37:40:19.45 & 24.21 &  66.0$\pm$3.4   &  97.8$\pm$4.9    & 184$\pm$4       &  PN\tablefootmark{~(3)} \\
e16 &       &      &   52.04 &   35.60 & 00:54:41.270 & -37:40:14.62 & 26.23 &  10.3$\pm$0.6   &   4.9$\pm$0.3     & 185$\pm$11     &  PN \\
e17:&       &      &   96.30 & 110.71 & 00:54:40.524 & -37:39:59.60 & 29.11 &   0.7$\pm$0.1    &   0.4$\pm$0.1     & 175$\pm$5      &  ?\tablefootmark{~(4)} \\
e18 &       & 19 & 129.38 & 108.03 & 00:54:39.967 & -37:40:00.13 & 26.75 &   6.3$\pm$0.4   &   1.8$\pm$0.2      & 207$\pm$7      &  PN \\
e20 &       &      & 198.24 & 121.89 & 00:54:38.807 & -37:39:57.36 & 27.92 &   2.2$\pm$0.2   &   1.2$\pm$0.1      & 143$\pm$3      &  PN \\
e22 &       & 12 & 256.81 &   47.22 & 00:54:37.821 & -37:40:12.29 & 24.17 &  68.6$\pm$3.5  &  13.3$\pm$0.7     & 165$\pm$9      &  PN \\
e23 &       &      & 185.72 &   28.89 & 00:54:39.018 & -37:40:15.96 & 27.40 &   3.5$\pm$0.3   &   1.4$\pm$0.2      & 181$\pm$8      &  PN\tablefootmark{~(2)} \\  
i02  &       & 18 & 314.17 & 312.27 & 00:54:39.773 & -37:41:33.66 & 26.85 &   5.8$\pm$0.4   &   1.6$\pm$0.2      & 163$\pm$13     &  PN \\
i19 &   8   & 24 &   79.67 & 227.78 & 00:54:43.724 & -37:41:50.56 & 23.19 & 169.0$\pm$8.5 &  32.2$\pm$1.6     & 165$\pm$6      &  PN \\
i84 &        & 26 &   15.51 &   45.23 & 00:54:55.806 & -37:42:27.07 & 25.04 &  30.7$\pm$1.6  &   7.5$\pm$0.4      & 108$\pm$8      &   PN\tablefootmark{~(5)} \\
j01 &        &      &   24.26 & 292.81 & 00:54:50.572 & -37:41:46.85 & 26.34 &   9.3$\pm$0.6   &   2.2$\pm$0.2      & 137$\pm$20     &  PN\tablefootmark{~(2)} \\ 
j12:&        &      & 157.16 & 210.64 & 00:54:48.333 & -37:42:03.29 & 27.89 &   2.2$\pm$1.1 &   0.6$\pm$0.8        & 172$\pm$5      &  PN\tablefootmark{(4)}\\  
\hline
\end{tabular}
\end{small}
\tablefoot{Column~1, name; Col.~2, ID in \citet{soffner1996}; Col.~3, ID in  \citet{pena2012};
Col.~4, datacube spaxel x-coordinate; Col.~5, datacube spaxel y-coordinate; Col.~6, right ascension (2000); Col.~7, declination (2000); Col.~8, {\oiii} magnitude (5007~\AA); Col.~9, observed {\oiii} flux (5007~\AA) in units of $10^{-17}$erg cm$^2$ s$^{-1}$; Col.~10, observed H$\alpha$ flux, same units ; Col~11, radial velocity [km~s$^{-1}$], Col~12, classification: bona fide planetary nebulae are marked "PN", uncertain PN candidates are flagged "?". \\
\tablefoottext{1}{aligned with M star;}
\tablefoottext{2}{bright, variable nebular background;}
\tablefoottext{3}{high extinction;}
\tablefoottext{4}{visual detection, not detected by FIND, also flagged as ":" in Col.~1;}
\tablefoottext{5}{retrograde velocity;}
\tablefoottext{6}{edge of field;}
\tablefoottext{7}{blend with compact \hiireg d97, located 0.5~arcsec to the SW;}
\tablefoottext{8}{blend with bright star at same position.}
} 
\end{table*}

{\bf Results:} The analysis of our datacubes for the discovery and measurement of point source and extended emission line objects began with the set of narrowband images for the most important wavelengths as listed in Table~\ref{filters}. An example for the wavelength of H$\alpha$ is shown in Fig.~\ref{NGC300_Ha} as a montage of fields (a)\ldots(j) over the H$\alpha$ image  from \citet{bresolin2009}, obtained with the  ESO/MPI 2.2m WFI. Not surprisingly, the MUSE data obtained at the VLT show much finer detail and go significantly deeper than the WFI image, which can be appreciated well e.g.\  from the giant \hiireg, marked De100, that is cut between WFI and MUSE at the eastern edge of field (a), or the variety of giant shells and filaments in fields (a)\ldots(j) that can be found nowhere at this level of detail in the WFI image. 

One of the immediate scientific goals motivating our NGC\,300 observations was to measure the {\oiii} planetary nebula luminosity function (PNLF) down to unprecedented faint magnitudes in order to infer its diagnostic value for  stellar population studies. \citet{Jacoby1989} introduced the m$_{5007}$ magnitude, based on the line flux of {\oiii} at $5007~\AA$ that is conventionally used to represent the PNLF:
\begin{equation}
\hspace{2cm} \mbox{m}_{5007} = -2.5 \log{F_{5007}} -13.74
\end{equation}
The bright end of the PNLF was initially proposed as a standard candle for extragalactic distance determinations by \citet{Jacoby1989} and collaborators, and subsequently utilized to measure distances to more than 60 galaxies; see review by 
\citet{mendez2016}. However, it was also suggested that the exact shape of the PNLF at fainter magnitudes than the bright cutoff could potentially unravel useful information on intermediate age stars of the underlying stellar population \citep{ciardullo2010}, and eventually help to understand the apparent invariance of the cutoff in the first place \citep{kwitter2014}. 

Beyond the classical narrow\-band filter techniques that have been employed at 4m class telescopes to typically sample the brightest $\sim$1.5 mag of the PNLF, which is sufficient for distance determinations, MUSE offers the advantage of providing filter maps with extremely narrow width at many wavelengths in parallel, combined with the VLT 8.2m light collecting power and excellent seeing, thus extremely high sensitivity. Moreover, we were able to apply the {\textsc PampelMUSE} PSF-fitting technique to extract PN spectra, deblended them from their crowded stellar and nebular environments, and then measure accurate  emission line fluxes and radial velocities.

\begin{figure}[h]
   \centering
    \includegraphics[width=\hsize,bb=60 70 760 560,clip]{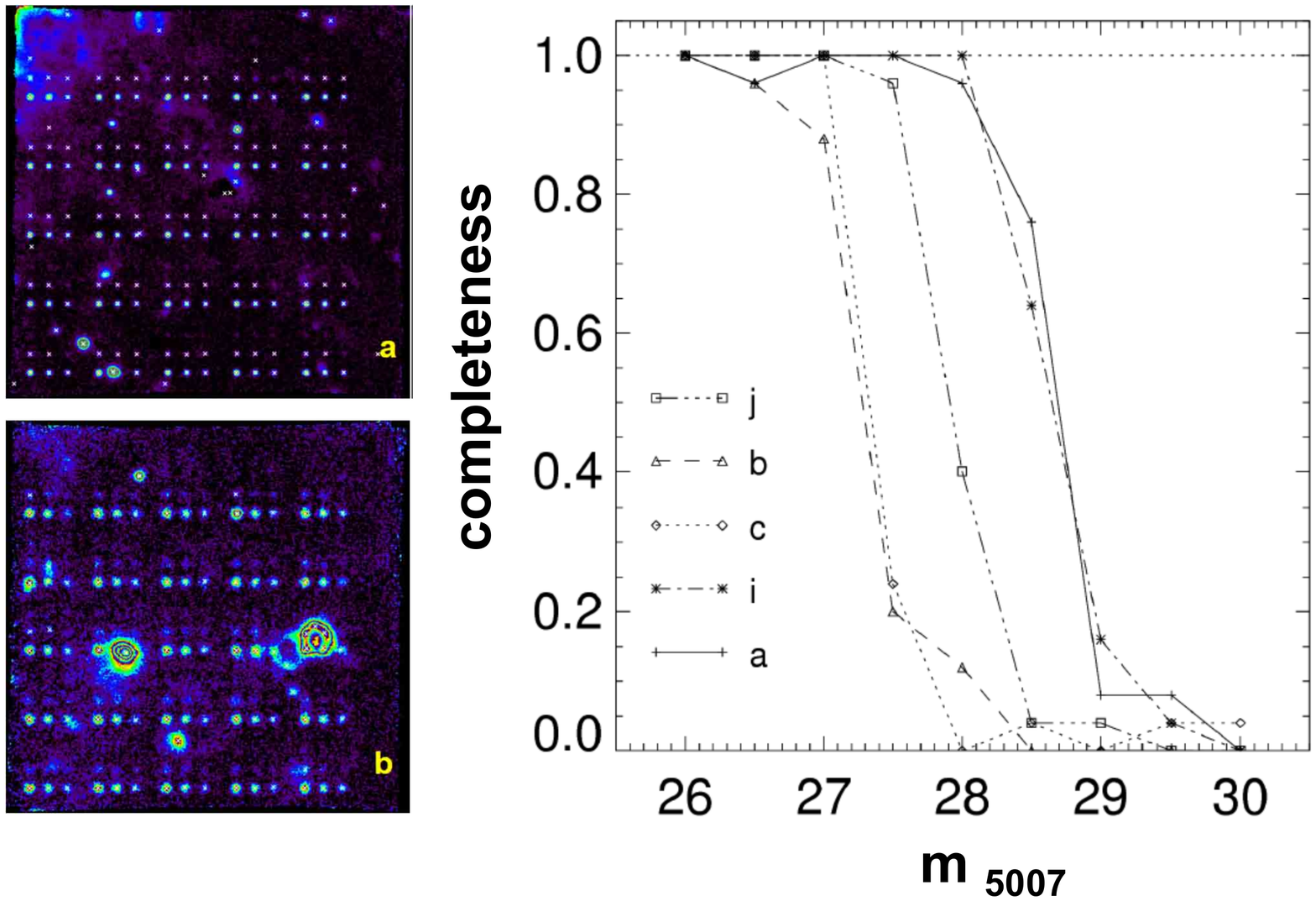}
    \caption{Left: examples for artificial PN point sources inserted in data\-cubes of fields (a) and (b), with seeing of 0.7" and 1.2", respectively. Right: fraction of artificial PN detections as a function of m$_{5007}$, obtained with {\textsc DAOPHOT FIND}, for fields (a), (b), (c), (i), (j), and seeing of 0.7", 1.2", 1.0", 0.6", 0.85", respectively.}
    \label{artificialPN}
 \end{figure}
 
In order to assess the sensitivity that we can reach with MUSE, we created sequences of artificial PN point sources in {\oiii} $5007~\AA$ with decreasing brightness that were inserted in our observed data\-cubes (Fig.~\ref{artificialPN}). We were able to demonstrate for all fields (a),\ldots,(j), representing different seeing conditions of 0.6 ''\ldots1.2'', that we indeed detect these objects using FIND, and in the best case reach completeness down to m$_{5007}$=28.0, which is about 6 magnitudes fainter than the PNLF cutoff. The dia\-gram in Fig.~\ref{artificialPN} shows that the completeness limit, as expected, depends strongly on the image quality, ranging from the best case at 0.6" and m$_{5007}$=28.0 for field (i) to 1.2" and m$_{5007}$ =26.0 for field (b). The faintest PNe that we detected in the experiment have m$_{5007}$ =30.0.

After having convinced ourselves that the data quality of our datacubes would be sufficient to search for faint PNe, we performed the search and classification procedure as described above. 
Typically, extragalactic PN candidates appear as point sources that are brighter in {\oiii} than in H$\alpha$. This fact used to be the canonical criterion for the selection of PN candidates for
 PNLF distance determinations of galaxies, practically being applied by blinking narrowband images in {\oiii} versus H$\alpha$. However, a problem with reliably classifying PNe consists in the possible confusion with compact \hiireg (cHII), that may appear as {\oiii} bright point sources at the distance of NGC\,300. In adopting the argument of \citet{stasinska2013}, we have therefore excluded all PN candidates whose spectra show a {\sii} line ratio  F(6716)/F(6731) > 1 as indicative of the low density limit, and in such cases classified them cHII. Also, H$\beta$ brighter than {\oiii} $\lambda$ 4959 is a good indicator for cHII, being not affected by extinction in contrast to the {\oiii}/H$\alpha$ line ratio.
 The results from  the analysis is presented in Table~\ref{PN-list} where we list a total of 45 PNe, out of which 34 are new discoveries, and 11 were already known from the work of \citet{soffner1996}, and \citet{pena2012}. Nine extremely faint objects (m$_{5007}\approx27\ldots29$)  are considered PN candidates on the basis that we could measure the {\oiii} line at the expected wavelength, however other lines fell below the detection limit. All of the known objects from the literature were easily rediscovered in our data, whereas almost all of the new detections are as expected significantly fainter than the objects found with the NTT  by \citet{soffner1996}), and VLT-FORS by \citet{pena2012}. Our faintest PN candidate e08 was measured to  have m$_{5007}$ =29.67. The analysis of results also  lead to the unexpected observation that the visual detection method is surprisingly sensitive: all of the FIND detections were recovered, whereas a total of 19 PN where found only by visual examination, without detections from FIND -- which is in line with practical experience from the early PNLF distance determination work (R. Ciardullo, priv. communication). 
 
 Table~\ref{PN-list} also lists the radial velocity estimates that were derived from the emission line fits with {\textsc P3D}. The overall distribution matches well the one of stars, thus giving confidence to indeed observing the rotation of the galaxy -- however with one exception: PN i84 exhibits a velocity of $108\pm13$ km~s$^{-1}$ which is far off the median in field (i), and also blueshifted from the systemic velocity of 144~km~s$^{-1}$, thus either indicating an extreme retrograde motion and the fact that the central star of this PN may be a runaway or a halo star, or that the object does not belong to NGC\,300.

Apart from the fact that cHII present a challenge to identify a clean sample of PNe, they are interesting objects on their own right. We have therefore compiled a listing in Table~\ref{cHII-list} with essentially the same entries as for the PNe in Table~\ref{PN-list}. We note that we have, again, recovered all of the objects found by \citet{soffner1996}, and \citet{pena2012}. However, upon inspection of the spectra we concluded that as many as 7 of their PN candidates must be classified cHII, as they show a {\sii} line ratio > 1 as discussed above. Similar to the PNe, the cHII kinematics in the seven fields (a),...,(j) match well the stellar kinematics in those fields. 

 \begin{table*}[!th]
 \caption{Catalogue of compact \hiireg candidates (52 objects)}             
\label{cHII-list}      
\centering          
\begin{small}
\begin{tabular}{ c    c c    r r     c  c      r r r r       l     }     
\hline\hline     
ID  & S96 & P12 &  x~~~     &  y~~     & RA           &  Dec         &   F(H$\beta$)   &     F(5007)     &   F(H$\alpha$)  &     F(6717)    &  v$_\mathrm{rad}$   \\
\hline
a06 &     &     &  255.3 &  272.2 & 00:54:52.077 & -37:40:41.71 &    2.6$\pm$0.2 &    ... &   10.2$\pm$0.5 &    ... & 179$\pm$3 \\
a14 &     &     &   61.2 &  230.9 & 00:54:55.348 & -37:40:49.97 &    0.7$\pm$0.2 &    ... &    2.1$\pm$0.1 &    0.5$\pm$0.1 & 143$\pm$6 \\
a21 &     &     &  295.9 &  186.6 & 00:54:51.394 & -37:40:58.83 &    1.0$\pm$0.1 &    0.9$\pm$0.1 &    3.1$\pm$0.2 &    0.5$\pm$0.1 & 160$\pm$10 \\
a29 &     &     &  279.3 &  166.9 & 00:54:51.673 & -37:41:02.77 &    2.8$\pm$0.3 &    5.1$\pm$0.4 &    8.2$\pm$0.4 &    ... & 146$\pm$3 \\
a30 &     & 49 &   84.7 &  165.1 & 00:54:54.951 & -37:41:03.12 &   14.4$\pm$0.9 &   16.7$\pm$1.1 &   43.6$\pm$2.2 &    ... & 159$\pm$5 \\
a32 &     &     &  302.4 &  154.0 & 00:54:51.284 & -37:41:05.34 &    0.5$\pm$0.2 &    1.6$\pm$0.2 &    1.8$\pm$0.2 &    0.2$\pm$0.1 & 178$\pm$5 \\
a39 &     &     &  286.9 &  106.9 & 00:54:51.546 & -37:41:14.76 &    1.4$\pm$0.2 &    ... &    5.7$\pm$0.3 &    1.5$\pm$0.1 & 146$\pm$6 \\
a51 &     &     &   41.6 &   54.5 & 00:54:55.678 & -37:41:25.25 &    1.3$\pm$0.3 &    8.6$\pm$0.6 &    3.8$\pm$0.2 &    0.7$\pm$0.1 & 135$\pm$5 \\
b07 &     &     &  157.2 &  276.7 & 00:54:48.679 & -37:40:40.88 &    9.6$\pm$0.8 &    ... &   33.1$\pm$1.7 &    8.5$\pm$0.5 & 164$\pm$8 \\
b25 &     &     &   25.7 &   61.0 & 00:54:50.896 & -37:41:24.03 &    3.7$\pm$0.9 &    ... &   11.4$\pm$0.8 &    3.6$\pm$0.3 & 138$\pm$6 \\
b28 &     &     &  231.7 &   98.2 & 00:54:47.425 & -37:41:16.59 &    7.2$\pm$0.8 &   11.0$\pm$0.9 &   30.0$\pm$1.6 &    6.4$\pm$0.4 & 151$\pm$8 \\
b33 &     &     &   23.0 &    9.6 & 00:54:50.941 & -37:41:34.31 &    2.8$\pm$0.3 &    ... &    9.8$\pm$0.5 &    2.2$\pm$0.2 & 141$\pm$2 \\
c05 & 15 & 21& 255.1 &  269.5 & 00:54:41.913 & -37:40:42.64 &    4.1$\pm$0.7 &   29.9$\pm$1.7 &   14.8$\pm$1.2 &    2.8$\pm$0.4 & 192$\pm$8 \\
c09 &     &     &  131.6 &  180.7 & 00:54:43.994 & -37:41:00.41 &    2.8$\pm$0.8 &    ... &   11.8$\pm$1.1 &    2.2$\pm$0.4 & 179$\pm$9 \\
c16 &     &     &   47.6 &  129.6 & 00:54:45.410 & -37:41:10.64 &    2.8$\pm$0.5 &    ... &    7.6$\pm$0.7 &    2.4$\pm$0.3 & 163$\pm$8 \\
c27 & 22 & 28 & 26.7 &   30.5 & 00:54:45.764 & -37:41:30.44 &    1.9$\pm$0.5 &   14.1$\pm$0.9 &    6.2$\pm$0.6 &    1.7$\pm$0.2 & 161$\pm$10 \\
c41 &     &     &   84.0 &   96.7 & 00:54:44.795 & -37:41:17.20 &    7.2$\pm$0.8 &    9.4$\pm$0.9 &   22.6$\pm$1.5 &    4.2$\pm$0.4 & 167$\pm$7 \\
d14 &     &     &  220.8 &  268.1 & 00:54:36.076 & -37:40:30.10 &   10.5$\pm$0.7 &    1.8$\pm$0.3 &   33.9$\pm$1.7 &    0.8$\pm$0.1 & 169$\pm$10 \\
d35 &     &     &   75.5 &  149.6 & 00:54:38.519 & -37:40:53.79 &    1.5$\pm$0.2 &    ... &    3.6$\pm$0.2 &    0.7$\pm$0.1 & 203$\pm$15 \\
d43 &     &     &  249.4 &  100.0 & 00:54:35.588 & -37:41:03.72 &    2.7$\pm$0.4 &    2.9$\pm$0.3 &   10.5$\pm$0.6 &    3.5$\pm$0.2 & 192$\pm$6 \\
d44 &     &     &   63.4 &  123.2 & 00:54:38.721 & -37:40:59.12 &    0.9$\pm$0.2 &    1.4$\pm$0.2 &    3.5$\pm$0.2 &    1.0$\pm$0.1 & 186$\pm$8 \\
d45 &     &     &   31.2 &  112.5 & 00:54:39.264 & -37:41:01.21 &    5.3$\pm$0.5 &    ... &   18.6$\pm$1.0 &    4.1$\pm$0.2 & 183$\pm$6 \\
d47 &     &     &   66.6 &  108.0 & 00:54:38.669 & -37:41:02.11 &    1.2$\pm$0.2 &    ... &    5.0$\pm$0.3 &    1.4$\pm$0.1 & 185$\pm$7 \\
d50 &     &     &  245.4 &   81.7 & 00:54:35.656 & -37:41:07.39 &    1.3$\pm$0.2 &    ... &    4.6$\pm$0.2 &    1.0$\pm$0.1 & 194$\pm$13 \\
d56 &     &     &   59.8 &   66.0 & 00:54:38.782 & -37:41:10.52 &    ... &    1.5$\pm$0.3 &    3.2$\pm$0.2 &    0.6$\pm$0.1 & 200$\pm$12 \\
d57 &     &     &   56.7 &   54.4 & 00:54:38.835 & -37:41:12.84 &    ... &    ...0.1 &    1.4$\pm$0.1 &    0.6$\pm$0.1 & 172$\pm$4 \\
d58 &     &     &    8.4 &   31.1 & 00:54:39.649 & -37:41:17.49 &    2.3$\pm$0.3 &    ... &    5.6$\pm$0.3 &    1.8$\pm$0.2 & 174$\pm$11 \\
d60 &     &     &   30.3 &   41.0 & 00:54:39.279 & -37:41:25.52 &    0.5$\pm$0.1 &    ... &    1.7$\pm$0.1 &    0.6$\pm$0.1 & 184$\pm$11 \\
d61 &     &     &   69.2 &   36.9 & 00:54:38.628 & -37:41:16.32 &    2.3$\pm$0.3 &    ... &    9.4$\pm$0.5 &    1.6$\pm$0.1 & 187$\pm$5 \\
d65 &     &     &  186.8 &   18.6 & 00:54:36.643 & -37:41:19.99 &    6.4$\pm$0.4 &    ... &   29.4$\pm$1.5 &    3.3$\pm$0.2 & 193$\pm$8 \\
d66 &     &  7 &  238.7 &   72.3 & 00:54.35.769 & -37:41:09.26 &    5.2$\pm$0.4 &   14.1$\pm$0.8 &   14.1$\pm$0.7 &    6.1$\pm$0.3 & 190$\pm$7 \\
d68 &     &     &  241.8 &   39.6 & 00:54:35.716 & -37:41:15.79 &    2.2$\pm$0.2 &    1.3$\pm$0.2 &    9.3$\pm$0.5 &    2.5$\pm$0.2 & 187$\pm$5 \\
d78 &     &     &   16.0 &   61.5 & 00:54:39.521 & -37:41:11.41 &    0.4$\pm$0.1 &    1.0$\pm$0.1 &    2.1$\pm$0.1 &    0.7$\pm$0.1 & 133$\pm$20 \\
d82 &     &  6 &  246.7 &   36.5 & 00:54:35.633 & -37:41:16.42 &    1.1$\pm$0.1 &    3.0$\pm$0.2 &    4.2$\pm$0.2 &    2.1$\pm$0.1 & 186$\pm$6 \\
d83 &     &     &  273.6 &   24.0 & 00:54:35.181 & -37:41:18.92 &   75.3$\pm$4.0 &   14.1$\pm$0.9 &  266.0$\pm$13.3 &   58.3$\pm$3.0 & 186$\pm$7 \\
d97 &     &     &  238.2 &  104.9 & 00:54:35.777 & -37:41:02.74 &    1.6$\pm$0.2 &    2.5$\pm$0.2 &    4.8$\pm$0.3 &    2.0$\pm$0.1 & 193$\pm$8 \\
e12 &     &     &  113.7 &  132.6 & 00:54:40.231 & -37:39:55.22 &    1.3$\pm$0.2 &    1.1$\pm$0.2 &    4.5$\pm$0.3 &    1.6$\pm$0.1 & 184$\pm$7 \\
e13 &     &     &   44.9 &   57.1 & 00:54:41.390 & -37:40:10.33 &    ... &    3.2$\pm$0.3 &    2.5$\pm$0.2 &    1.6$\pm$0.1 & 190$\pm$13 \\
e29 &     &     &  124.0 &  133.1 & 00:54:40.057 & -37:39:55.13 &    ... &    ... &    1.2$\pm$0.1 &    ... & 172$\pm$20 \\
e48 &     &     &   78.9 &   85.2 & 00:54:48.818 & -37:40:04.69 &    1.0$\pm$0.2 &    0.1$\pm$0.2 &    3.1$\pm$0.2 &    1.3$\pm$0.1 & 176$\pm$9 \\
e49 &     &    & 106.6 &   95.1 & 00:54:40.351 & -37:40:02.73 &    1.0$\pm$0.2 &    ... &    3.5$\pm$0.2 &    1.4$\pm$0.1 & 188$\pm$7 \\
i10 &     &     &   25.5 &  276.6 & 00:54:44.638 & -37:41:40.80 &    1.3$\pm$0.3 &    5.4$\pm$0.4 &    4.7$\pm$0.3 &    ...0.1 & 149$\pm$4 \\
i29 &     &     &  121.7 &  218.7 & 00:54:43.016 & -27:41:52.37 &    5.3$\pm$0.9 &    ... &   16.7$\pm$0.9 &    0.8$\pm$0.3 & 191$\pm$6 \\
j02 &     &     &   39.1 &  274.0 & 00:54:50.323 & -37:41:50.62 &   13.1$\pm$0.7 &    5.2$\pm$0.4 &   42.3$\pm$2.1 &   13.3$\pm$0.7 & 147$\pm$11 \\
j03 &     &     &  123.9 &  294.1 & 00:54:48.893 & -37:41:46.58 &    3.0$\pm$0.4 &    8.2$\pm$0.6 &    7.3$\pm$0.4 &    4.7$\pm$0.3 & 156$\pm$5 \\
j11 &     &     &   59.7 &  209.3 & 00:54:49.975 & -37:42:03.55 &    2.3$\pm$0.3 &    ... &    5.9$\pm$0.4 &    1.8$\pm$0.2 & 168$\pm$10 \\
j20 &     &     &  114.1 &  179.2 & 00:54:49.059 & -37:42:09.57 &    1.7$\pm$0.2 &    0.5$\pm$0.1 &    5.1$\pm$0.3 &    1.4$\pm$0.1 & 154$\pm$12 \\
j21 &     &     &  121.2 &  177.9 & 00:54:48.938 & -37:42:09.84 &    1.0$\pm$0.2 &    3.4$\pm$0.3 &    2.4$\pm$0.2 &    0.9$\pm$0.1 & 146$\pm$6 \\
j22 &     &     &  149.1 &  180.1 & 00:54:48.469 & -37:42:09.39 &    1.4$\pm$0.2 &    ... &    6.3$\pm$0.4 &    1.3$\pm$0.1 & 158$\pm$2 \\
j24 &     &     &  244.7 &  175.6 & 00:54:46.857 & -37:42:10.29 &    8.5$\pm$0.6 &   33.6$\pm$1.7 &   26.7$\pm$1.4 &   10.0$\pm$0.5 & 172$\pm$3 \\
j43 &     &     &  185.0 &   49.9 & 00:54:47.864 & -37:42:35.44 &    9.9$\pm$0.6 &    ... &   27.3$\pm$1.4 &    6.1$\pm$0.4 & 163$\pm$4 \\
\hline                      
\end{tabular}
\end{small}
\tablefoot{Column~1, name; Col.~2, ID of PN in \citet{soffner1996}; Col.~3, ID of PN in  \citet{pena2012};
Col.~4, datacube spaxel x-coordinate; Col.~5, datacube spaxel y-coordinate; Col.~6, right ascension (J2000); Col.~7, declination (J2000); Col.~8, observed H$\beta$ flux in units of $10^{-17}$erg cm$^2$ s$^{-1}$; Col.~9, observed {\oiii} flux (5007~\AA) (same units); Col.~10, observed H$\alpha$ flux (same units); Col.~11, observed {\sii} flux (6716~\AA) (same units); Col~12, radial velocity [km~s$^{-1}$].  
} 
\end{table*}


{\noindent \bf Discussion:} One of the major objectives of our MUSE observations of NGC\,300 was the study of the PNLF down to very faint magnitudes. We have detected a total of 45 objects, reaching seeing-dependent completeness limits between m$_{5007}$=28.0 and m$_{5007}$=26.0 for a seeing FWHM of 0.6" and 1.2", respectively. PN candidates that were identified in field (e) reach magnitudes even fainter than m$_{5007}\approx29.0$. The latter may seem to be surprising in view of the fact that pointing (e) is not complete in the sense that only one third of the nominal exposure time of 1.5~h could initially be secured. However, with 9 firm PN detections and 4 more PN candidates, field (e) has the highest PN number density of all pointings. Even though our completeness limit estimates are supported by this result, another surprising finding is, that pointings (i) and (j) have yielded significantly fewer detections, although the excellent to good image quality would have suggested otherwise.

We have compared our detections with results from the literature, namely \citet{soffner1996} who used narrowband imaging at the 3.5m ESO-NTT, and \citet{pena2012} with data from narrowband imaging and spectroscopy using FORS at the VLT. Firstly, the MUSE observations have recovered all of the previous detections. We have classified, however, from the 18 objects in common with \citet{pena2012} as many as 5, i.e. almost one third, as cHII candidates. The same is true for 2 out of 8 PN candidates in common with \citet{soffner1996}.  As these objects are all fainter than m$_{5007}$=24.5, our reclassification should however have no major impact on the bright cutoff of the PNLF constructed from their samples. Our new detections that have no counterpart in the literature are typically fainter than m$_{5007}$=27.

 \begin{figure}[!t]
  \centering    
   \includegraphics[width=\hsize, bb=60 30 710 600,clip]{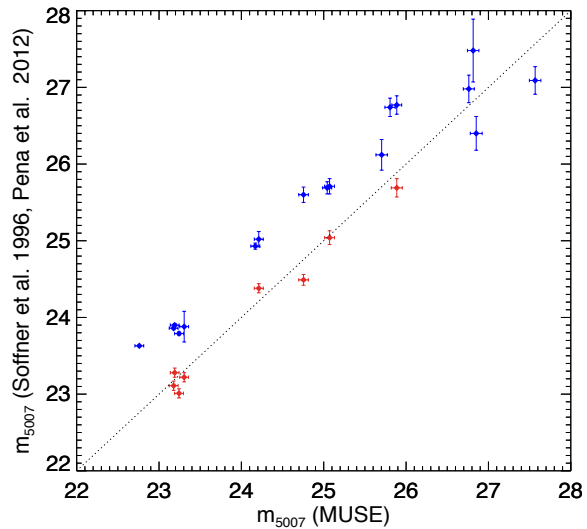}
    \caption{Comparison of PN and cHII magnitudes obtained with MUSE versus results from \citet{soffner1996} in red,  and \citet{pena2012} with blue plot symbols.}
    \label{PN_mag_comp}
 \end{figure}

We have also compared our m$_{5007}$ magnitudes with the samples from \citet{soffner1996} and \citet{pena2012} and plot the outcome in Fig.~\ref{PN_mag_comp}. We have 17 objects in common, of which 5 are cHII. The results obtained by \citet{soffner1996} are within the errors in agreement with our data, except for the very faintest magnitudes. However, there is a systematic offset with a median of $\sim$0.67 mag bet\-ween the FORS \citep{pena2012}  and our MUSE data. The radial velocities determined for our sample of PNe are discussed in $\S$\ref{discuss_kinematics} below.

In an attempt to understand the unequal detection rate over our pointings that cover the nucleus, parts of the central spiral arm to the NW, versus parts of the leading and trailing interarm regions around this spiral arm, we have complemented our detections with the PN candidates from \citet{pena2012} and plotted all objects for orientation over an H$\alpha$ map of NGC\,300, see Fig.~\ref{kinematics}. The coloured plot symbols refer to MUSE planetaries, while the filled white circles refer to the  \citet{pena2012} objects.

Although the latter do not reach the same faint magnitudes as our sample, there seems to be a hint of the surface density distribution to favour a higher concentration of PNe near the nucleus and along the spiral arms. Owing to the lack of deep coverage at the level of our MUSE data, this observation is not entirely compelling, however supported by the sparse population in field (j) and the interarm extension to the south-east (SE). Whether the latter is a generic feature of this region, or perhaps an effect of extinction owing to a dust lane extending along the interarm region, remains to be clarified. At any rate, a concentration along star forming regions of a spiral arm would not at all be expected from a theoretical point of view. According to \citet{renzini1986}, and more recently \citet{buzzoni2006}, the number of observable PNe at any time is given by the evolutionary flux that is tied to the underlying stellar population. The PNe that we observe stem from an intermediate age stellar population of $\sim$1~Gyr, so one would expect the PN surface density to correlate with the stellar mass surface density of the old stellar population, however not with young O stars and \hiiregs. As we found in 30\% of the cases of PN candidates from \citet{pena2012} evidence for contamination of their sample from cHII, it is not excluded that the provisional picture of PNe aligned with spiral arms is simply an artifact.

It remains unclear, however, why there is an extreme number of detections in field (e). More than half of the objects extend from the leading edge of the spiral arm into the interarm region to the NW. Two of the PNe in this area exhibit extremely low radial velocities (see below). The PN surface density pattern does not seem to continue into the adjacent field (d). Perhaps we are observing just a spurious density peak that would average out on larger scales. As the exposures in (d) and (e) are as yet incomplete, we consider the current census as preliminary. 

We note that this is not the first work to discover extragalactic PNe in MUSE datacubes. Recently, \citet{kreckel2017} published a paper on the PNLF of the face-on grand design spiral galaxy NGC\,628. The faintest PNe in their sample have a magnitude of $\sim$27.9, which is consistent with PNe in our field (e), where the exposure time was 30~min and the seeing 0.75". For NGC\,628,  \citet{kreckel2017} estimate PN surface densities of 1.6\ldots 3.1 PN/kpc$^2$ for two regions with different galactocentric distances. These numbers are significantly lower than the preliminary values that we find in our fields: 23.2 and 33.5  PN/kpc$^2$ for fields (a) and (e), and 16.6  PN/kpc$^2$ as an average of the total of 7 fields,  respectively, --- while NGC\,628 is $\sim$3 mag more luminous in the H band than NGC\,300, suggesting a larger parent stellar population for the observable PNe.

\subsection{Emission line stars}
\label{emStars}
{\bf Results: }The visual examination of the H$\alpha$ maps yielded point source candidates whose spectra are not compatible with PNe or cHII, however showing hints of a continuum. Blinking the H$\alpha$ map with the corresponding VRI image sometimes suggested coincidence with a star. Only a few of these stars could be associated with an object from the catalogue of {\textsc PampelMUSE}-extracted spectra, probably because the emission lines in H$\alpha$ and H$\beta$, filling in the absorption lines at the corresponding wavelengths, have prevented ULySS from converging to a reasonable fit, as emission line stars are not part of our empiricial library, nor would be supported by GLIB. In such cases, we have employed {\textsc P3D} to perform aperture spectro\-photometry on the stars in question, with the caveat of potentially subtle background subtraction issues at the wavelengths of emission lines. However, it was possible to extract spectra that in most cases presented broad, non-Gaussian emission line profiles with extended wings, distinct from the narrow nebular background emission, that are characteristic for hot massive stars with strong stellar winds, see e.g. \citet{klein1978}, \citet{leitherer1988}, \citet{lamers1993}, \citet{puls1996}, \citet{kudritzki1999}.  An example is shown in the bottom panel (i) of Fig.~\ref{NGC300_spectra}. For this object, a Gaussian fit to the H$\alpha$ emission line profile yields a FWHM of 6.6~\AA, however with a significant residual owing to extended wings to the red and the blue (the MUSE line-spread-function has a FWHM of 2.4\ldots3\AA). The linewidth and the extended wings point to stellar wind velocities of several hundreds up to 1000~km~s$^{-1}$, typical for hot massive stars. 


{\noindent \bf Discussion:} In many cases we could immediately associate a blue star from our VRI maps with the emission line object. However, there were also cases where we could not find an obvious bright optical counterpart, which could possibly be attributed to strong dust extinction: given that there is an immediately visible distinct dust lane at the trailing edge of the north-western (NW) spiral arm in NGC\,300, and another one at the leading edge of this spiral arm, and possibily significant intrinsic dust extinction in star forming regions as well, the broad line emission may act as  a beacon to detect the associated massive stars even in the event of high extinction. In some cases we also found objects with a narrow line width. At the distance of NGC\,300 it is impossible to disentangle compact nebulosities from intrinsic stellar emission on the basis of the MUSE data. Again, we visually inspected all of the spectra and decided which objects would qualify as emission line star candidates. The results of this ana\-ly\-sis are listed in Table~\ref{emStar-list}. 

Direct comparison with VRI colour maps has revealed emission line objects that most likely are associated with red giants rather than with hot massive stars as discussed above. As an example, i54 shows a broad H$\alpha$ line on top of a pronounced M star spectrum. Unfortunately, the object was found in a region where there is no coverage with HST images. Although we cannot rule out a chance alignment of the AGB star with a hot star that would be responsible for the emission, there is no indication from the MIUSCAT fit that this might be the case. There is, however, a striking similarity to the spectrum of the prototypical object R Aquarii, which is a well-studied case of a symbiotic M star, showing a hot companion, Roche lobe overflow, accretion onto the hot star, and jet activity (\citet{zhao-geisler2012}). The latter authors estimate the distance to R Aqr as 250~pc. \citet{schmid2017} used observations with HST and SPHERE ZIMPOL at the VLT to measure an H$\alpha$ flux of $4.6\times10^{-11}$ erg/cm$^2$/s.  Pushing this object to the distance of NGC\,300, this would translate to a flux of roughly $10^{-18}$ erg/cm$^2$/s, i.e. just detectable with MUSE. We therefore flagged red giant emission line stars as symbiotic star candidates.

A special case is the Wolf-Rayet star (WR) that was readily detected from the {\heii} narrowband image created for field (c). Inspection of the {\textsc PampelMUSE}-extracted spectrum confirmed that this object is indeed a hot star with the characteristic blue WR bump (see Fig.~\ref{WR_images}).
 
\begin{figure}[t]
  \centering
  \begin{minipage}[b]{0.24\textwidth}
     \includegraphics[width=120pt,bb=10 10 500 490,clip]{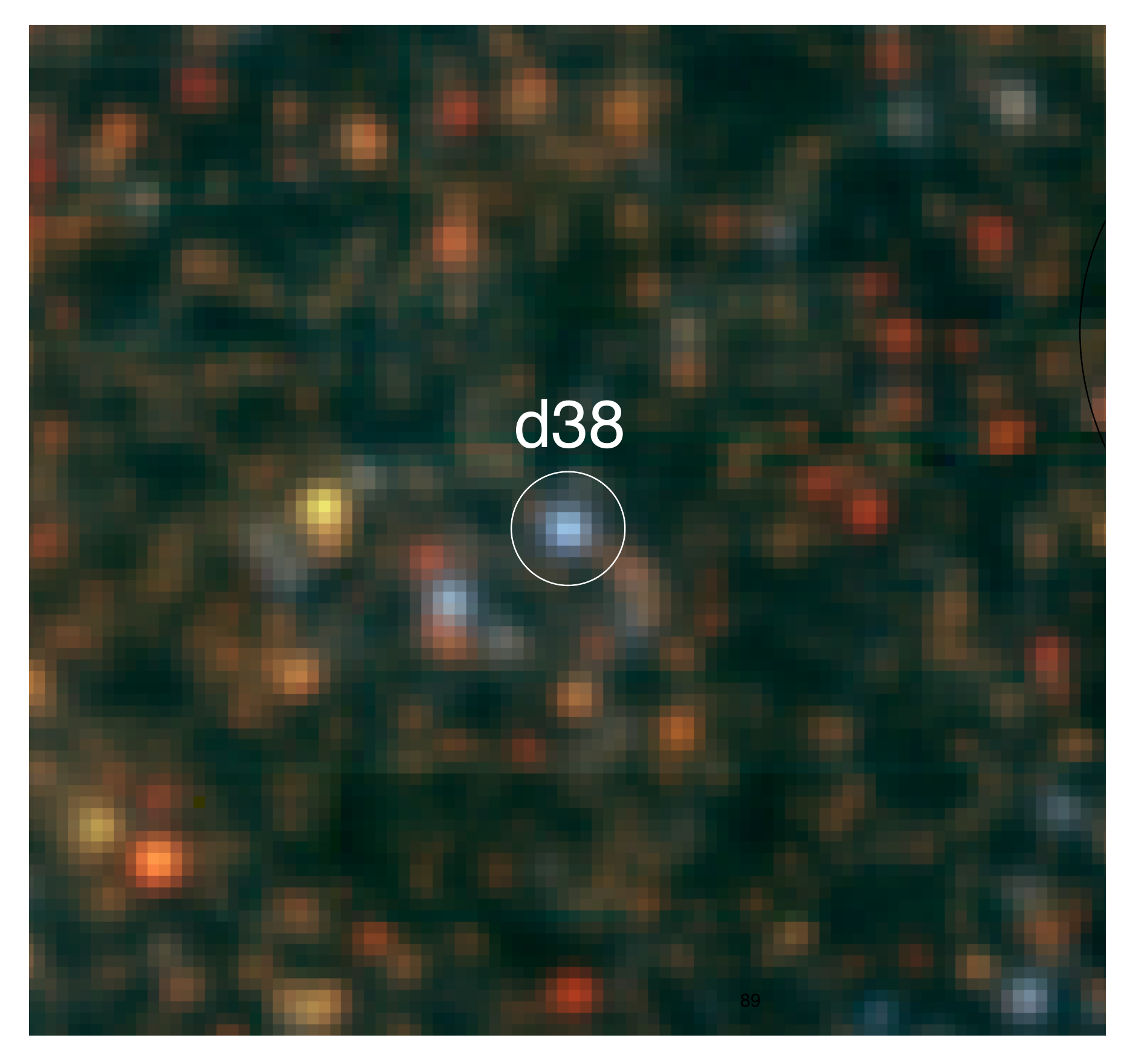}
  \end{minipage}
  \hfill
  \begin{minipage}[b]{0.24\textwidth}
    \includegraphics[width=120pt,bb=10 10 500 490,clip]{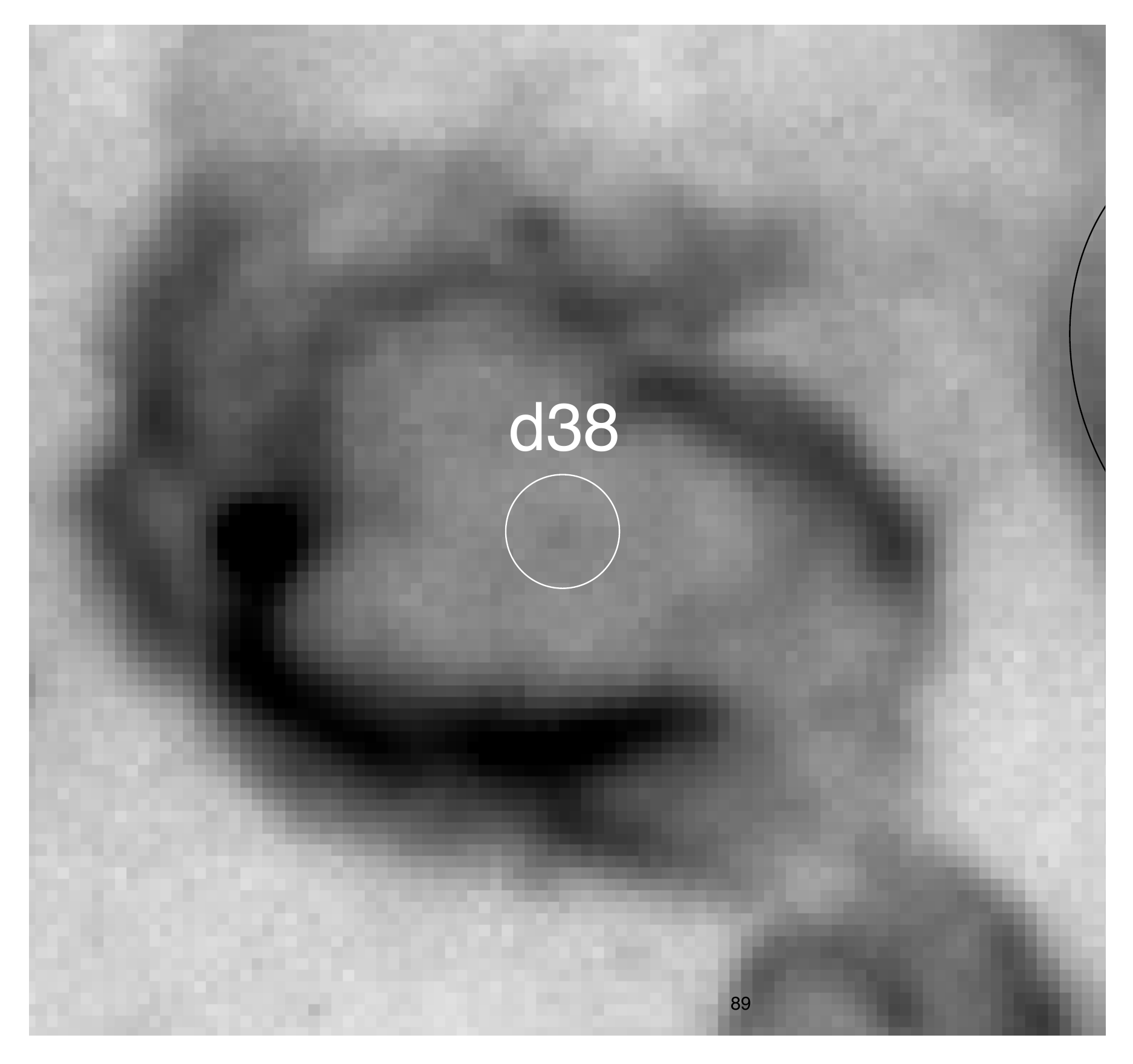}
  \end{minipage}

   \includegraphics[width=\hsize,bb=50 290 755 590,clip]{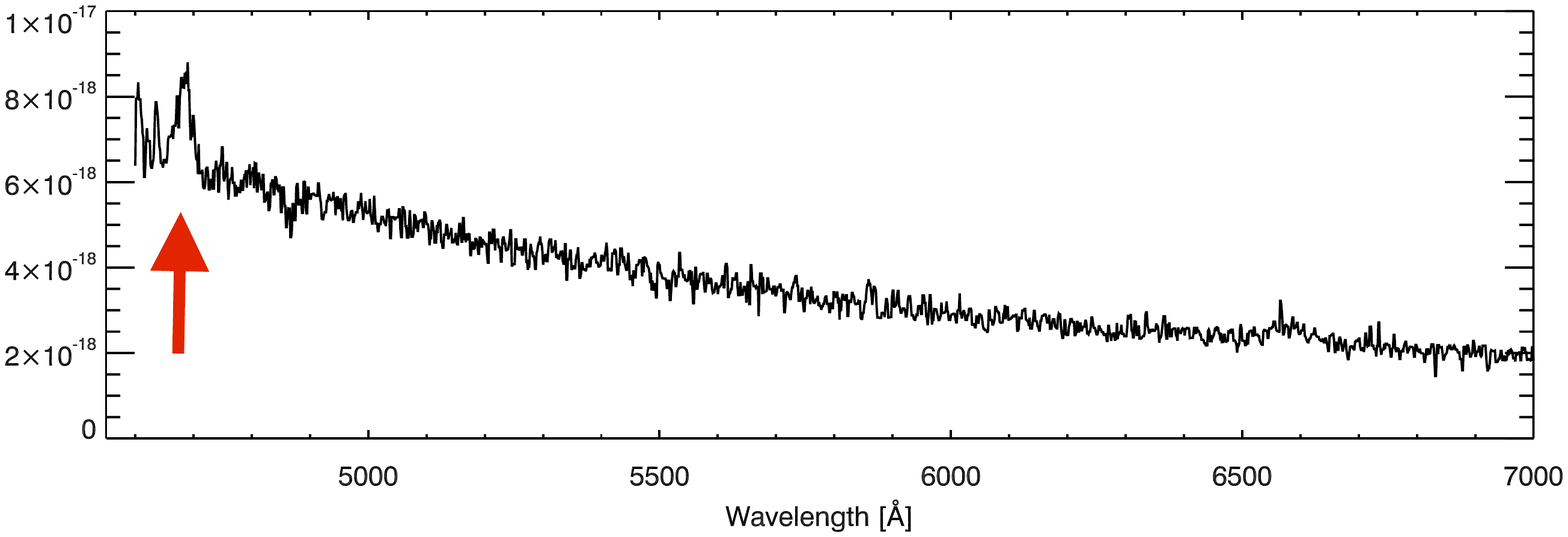}
   \caption{WR star in field (d). Upper left: VRI map, candidate star indicated as object d38. Upper right: H$\alpha$ map, showing surrounding shells.
   Bottom: spectrum showing the {\heii} feature (arrow),  in arbitrary units.
    }
    \label{WR_images}
 \end{figure}

 \pagebreak

 \begin{table*}[p]
 \caption{Excerpt of catalogue of emission line stars (in total 118 objects)}             
\label{emStar-list}      
\centering          
\begin{small}
\begin{tabular}{ c  r    r  r    c c        r  r         c c  c       l  }     
\hline\hline     
ID  &  Pm   &   x~~~~       &  y~~~~     & RA           &  Dec         &   F(H$\beta$)   &   F(H$\alpha$)   &  v$_\mathrm{rad}$ &FWHM & quality & colour \\
\hline
a05 &   409 &  85.58 & 281.07 & 00:54:54.936 & -37:40:39.93 &   8.8$\pm$0.5 &  25.8$\pm$1.3 & 164$\pm$3 &    6.0 & conf. &       blue \\
a07 &       & 277.07 & 275.74 & 00:54:51.710 & -37:40:41.00 &   1.2$\pm$0.1 &   1.2$\pm$0.1 & 141$\pm$12 &    4.5 & conf. &      red-f \\
a11 &       & 298.84 & 252.19 & 00:54:51.344 & -37:40:45.71 &   8.0$\pm$0.6 &  49.9$\pm$2.5 & 194$\pm$21 &   6.9w & conf. &        yellow \\
a18 &       &  38.93 & 203.32 & 00:54:55.722 & -37:40:55.48 &   ... &   0.4$\pm$0.1 & 152$\pm$20 &    3.1 & conf. &        red \\
a24 &       &  77.59 & 185.55 & 00:54:55.071 & -37:40:59.03 &   ... &   0.9$\pm$0.1 & 135$\pm$20 &    3.4 & ? &   red-s \\   
a26 &   266 & 159.78 & 177.99 & 00:54:53.686 & -37:41:00.55 &   ... &   6.7$\pm$0.5 & 161$\pm$20 &   6.3w & conf. &    blue \\   
a31 &   218 & 113.13 & 151.78 & 00:54:54.472 & -37:41:05.79 &   ... &   0.3$\pm$0.1 & 120$\pm$20 &    4.7 & conf. &   blue? \\   
a34 &   206 & 284.62 & 143.78 & 00:54:51.583 & -37:41:07.39 &   0.4$\pm$0.1 &   0.5$\pm$0.1 & 120$\pm$3 &    3.2 & conf. &        red \\
a35 &       & 280.18 & 135.34 & 00:54:51.658 & -37:41:09.08 &   0.6$\pm$0.1 &   0.9$\pm$0.1 & 136$\pm$1 &    4.8 & conf. &      red-f \\
a37 &   175 & 273.07 & 129.57 & 00:54:51.778 & -37:41:10.23 &   0.5$\pm$0.1 &   2.3$\pm$0.1 & 155$\pm$22 &    7.5 & conf. &       blue \\
a40 &    16 & 299.28 & 104.68 & 00:54:51.336 & -37:41:15.21 &   ... &   0.7$\pm$0.1 & 144$\pm$20 &    6.8 & conf. &       blue \\
a44 &       & 237.08 &  89.58 & 00:54:52.384 & -37:41:18.23 &   0.6$\pm$0.1 &   4.1$\pm$0.2 & 161$\pm$12 &    5.9 & conf. &       blue \\
a48 &    90 & 142.90 &  63.37 & 00:54:53.971 & -37:41:23.47 &   ... &   0.8$\pm$0.1 & 158$\pm$20 &    3.0 & conf. &      yel.-f \\
a49 &       & 202.43 &  65.14 & 00:54:52.968 & -37:41:23.12 &   ... &   0.6$\pm$0.1 & 165$\pm$20 &    3.2 & conf. & red-f-s \\   
a50 &    92 & 222.42 &  66.48 & 00:54:52.631 & -37:41:22.85 &   ... &   1.3$\pm$0.1 & 136$\pm$20 &    3.3 & conf. & red-f-s \\   
a52 &    10 &  52.26 &  51.81 & 00:54:55.498 & -37:41:25.78 &   ... &   3.0$\pm$0.2 & 150$\pm$20 &    6.2 & conf. &       blue \\
a54 &       & 142.45 &  50.04 & 00:54:53.978 & -37:41:26.14 &   0.5$\pm$0.1 &   0.7$\pm$0.1 & 135$\pm$12 &    3.3 & conf. &      yel.-f \\
a56 &       & 289.51 &  69.14 & 00:54:51.501 & -37:41:22.31 &   ... &   0.8$\pm$0.1 & 121$\pm$20 &   3.5 & conf. &     red-s \\   
b05 &       & 306.99 & 305.09 & 00:54:46.156 & -37:40:35.21 &   6.7$\pm$0.6 &  42.9$\pm$2.2 & 189$\pm$13 &   6.4 & conf. &     red-s \\    
b11 &       &  24.33 & 216.92 & 00:54:50.918 & -37:40:52.84 &   ... &   6.5$\pm$0.4 & 169$\pm$20 &    7.5 & conf. &        yel. \\
b18 &       & 196.23 & 180.59 & 00:54:48.022 & -37:41:00.11 &   9.3$\pm$0.7 &  37.1$\pm$1.9 & 153$\pm$1 &   6.3w & conf. &       blue \\
b31 &       & 263.57 &  67.18 & 00:54:46.887 & -37:41:22.79 &   2.4$\pm$0.2 &   9.4$\pm$0.5 & 146$\pm$1 &   5.9w & conf. &       blue \\
...   &       &              &            &                       &                       &                         &                         &                     &           &           &              \\
e72 &       & 306.44 &  12.35 & 00:54:36.985 & -37:40:19.27 &   ... &  12.4$\pm$0.7 & 165$\pm$20 &    5.0 & conf. &       blue \\
e77 &       & 155.32 & 235.01 & 00:54:39.530 & -37:39:34.74 &   ... &   0.8$\pm$0.1 & 207$\pm$20 &    7.5 & conf. & blue-f \\   
e79 &       & 220.15 &  25.31 & 00:54:38.438 & -37:40:16.68 &   ... &   0.7$\pm$0.1 & 166$\pm$20 &    7.0 & conf. &     blue-f \\
e82 &       & 208.97 & 162.58 & 00:54:38.627 & -37:39:49.22 &   ... &   1.0$\pm$0.1 & 184$\pm$20 &    7.5 & conf. & blue-f \\    
e83 &       &  81.10 & 268.09 & 00:54:40.780 & -37:39:28.12 &   ... &   1.4$\pm$0.1 & 183$\pm$20 &    7.5 & conf. &      red-c \\
e84 &       & 112.40 & 268.99 & 00:54:40.253 & -37:39:27.94 &   ... &   0.8$\pm$0.1 & 210$\pm$20 &    7.5 & --- &        --- \\
i03 &        &  42.17 & 300.98 & 00:54:44.356 & -37:41:35.92 &   ... &   1.5$\pm$0.1 & 185$\pm$20 &    7.4 & conf. &     blue-f \\
i07 &       & 217.48 & 289.68 & 00:54:41.402 & -37:41:38.18 &   ... &   0.6$\pm$0.7 & 114$\pm$20 &    5.1 & conf. & blue-f \\    
i09 &   379 & 283.45 & 296.46 & 00:54:40.290 & -37:41:36.82 &   ... &   2.9$\pm$0.2 & 166$\pm$20 &   5.3w & conf. &  blue \\   
i10 &       &  25.45 & 276.58 & 00:54:44.638 & -37:41:40.80 &   ... &   1.9$\pm$0.1 & 170$\pm$20 &   7.5w & conf. &  blue \\   
i12 &       &  59.34 & 273.41 & 00:54:44.067 & -37:41:41.43 &   ... &   0.4$\pm$0.1 & 172$\pm$20 &    5.6 & conf. &     blue-f \\
i13 &       &  62.50 & 265.28 & 00:54:44.013 & -37:41:43.06 &   ... &   0.7$\pm$0.1 & 110$\pm$20 &    3.5 & conf. &      red-c \\
i18 &       &  83.74 & 254.89 & 00:54:43.655 & -37:41:45.14 &   2.0$\pm$1.0 &  14.5$\pm$0.9 & 203$\pm$7 &   4.7w & conf. &   blue \\   
i21 &       & 132.97 & 236.81 & 00:54:42.826 & -37:41:48.75 &   ... &   0.6$\pm$0.1 & 167$\pm$20 &    5.6 & conf. &      red-c \\
i22 &  27 & 239.62 & 248.11 & 00.54:41.029 & -37:41:46.49 &   ... &   2.3$\pm$0.1 & 186$\pm$20 &    5.1 & conf. &        red \\
i25 &       & 313.72 & 273.86 & 00:54:39.780 & -37:41:41.34 &   ... &   1.0$\pm$0.1 & 163$\pm$20 &    3.9 & ? &   blue \\   
i30 &       & 151.06 & 211.06 & 00:54:42.521 & -37:41:54.90 &   ... &   0.6$\pm$0.1 & 167$\pm$20 &    7.0 & conf. &     blue-f \\
i33 &   197 & 146.99 & 188.01 & 00:54:42.590 & -37:41:58.51 &   ... &   4.0$\pm$0.2 & 202$\pm$20 &    7.5 & conf. &   blue \\   
i34 &    56 &  17.31 & 175.36 & 00:54:44.775 & -37:42:01.04 &   ... &   2.5$\pm$0.1 & 182$\pm$20 &    4.8 & conf. &       blue \\
i36 &       &  34.48 & 166.78 & 00:54:44.485 & -37:42:02.76 &   ... &   2.2$\pm$0.1 & 174$\pm$20 &   7.5w & conf. &   blue \\   
i39 &       &  70.63 & 167.23 & 00:54:43.876 & -37:42:02.67 &   ... &   1.7$\pm$0.1 & 186$\pm$20 &    7.2 & conf. &       blue \\
i40 &   244 &  73.79 & 169.94 & 00:54:43.823 & -37:42:02.13 &   ... &   0.3$\pm$0.1 & 185$\pm$20 &    5.3 & ? &        red \\
i41 &   237 & 109.04 & 164.97 & 00:54:43.229 & -37:42:03.12 &   ... &   1.5$\pm$0.1 & 165$\pm$20 &    5.0 & conf. &       blue \\
i42 &       & 176.81 & 157.74 & 00:54:42.087 & -37:42:04.57 &   ... &   1.6$\pm$0.1 & 162$\pm$20 &   7.5w & conf. &       blue \\
i43 &       & 209.80 & 159.55 & 00:54:41.531 & -37:42:04.21 &   ... &   1.5$\pm$0.1 & 189$\pm$20 &    6.8 & conf. &     blue-f \\
i44 &    45 & 222.00 & 150.96 & 00:54:41.326 & -37:42:05.92 &   ... &   2.8$\pm$0.2 & 175$\pm$20 &   6.1w & conf. &       blue \\
i49 &       &  24.54 & 135.60 & 00:54:44.653 & -37:42:08.99 &   ... &   1.2$\pm$0.1 & 167$\pm$20 &    7.5 & conf. &       blue \\
i51 &   212 &  63.40 & 143.73 & 00:54:43.998 & -37:42:07.37 &   ... &   0.5$\pm$0.1 & 253$\pm$20 &    3.8 & conf. &        yel. \\
i53 &       & 168.23 & 128.82 & 00:54:42.232 & -37:42:10.35 &   ... &   0.9$\pm$0.1 & 182$\pm$20 &    7.5 & conf. &       blue \\
i54 &   195 & 275.32 & 123.85 & 00:54:40.427 & -37:42:11.34 &   ... &   7.4$\pm$0.4 & 194$\pm$20 &   7.4 & conf. &  red-s\\   
i57 &       &  80.12 & 103.07 & 00:54:43.716 & -37:42.15.50 &   ... &   0.6$\pm$0.1 & 212$\pm$20 &    7.0 & conf. &        yel. \\
i58 &       &  90.96 & 112.11 & 00:54:43.534 & -37:42:13.69 &   ... &   0.8$\pm$0.1 & 152$\pm$20 &    5.8 & conf. &  blue \\   
i60 &     6 & 178.62 &  97.65 & 00:54:42.057 & -37:42:16.59 &   9.9$\pm$0.6 &  61.0$\pm$3.1 & 152$\pm$7 &   5.3w & conf. &   blue \\   
i63 &       & 180.43 &  83.19 & 00:54:42.026 & -37:42:19.48 &   ... &   0.8$\pm$0.1 & 152$\pm$20 &    4.8 & --- &        --- \\
i68 &    19 & 200.31 &  63.76 & 00:54:41.691 & -37:42:23.36 &   ... &   0.4$\pm$0.1 & 191$\pm$20 &    5.6 & conf. &       blue \\
i69 &       & 152.41 &  40.26 & 00:54:42.498 & -37:42:28.06 &   ... &   1.3$\pm$0.1 & 174$\pm$20 &   5.0w & conf. &       blue \\
i71 &       & 222.00 &  47.49 & 00:54:41.325 & -37:42:26.62 &   ... &   0.6$\pm$0.1 & 186$\pm$20 &    7.5 & conf. &     blue-f \\
...   &       &              &            &                       &                       &                         &                         &                     &           &           &              \\
\hline                      
\end{tabular}
\end{small}
\tablefoot{Excerpt (full table available online). Column~1, name; Col.~2, ID of {\textsc PampelMUSE} spectrum; Col.~3, datacube spaxel x-coordinate; Col.~4, datacube spaxel y-coordinate; Col.~5, right ascension (J2000); Col.~6, declination (J2000); Col.~7, H$\beta$ emission line flux; Col.~8, H$\alpha$ emission line flux; Col.~9, radial velocity [km~s$^{-1}$]; Col.~10, FWHM of H$\alpha$  emission line in [\AA], "w" indicates extended line wings; Col.~11, quality: confirmed, "?" uncertain, "---" no hint of a star, probably other type of emission line point source; Col.~11, apparent colour of star, "-f": faint, "-s": symbiotic star candidate, "-c": carbon star.
} 
\end{table*}


\begin{table*}[p]
 \caption{Catalogue of \hiiregs (61 objects)}             
\label{HII-list}      
\centering          
\begin{small}
\begin{tabular}{ l c   r  r     c  c      r  r      l  l  c  l  }     
\hline\hline 
ID   &  De88 &  x~~~~       &  y~~~    & RA                 &  Dec           & F(H$\beta$)~~~~~   & F(H$\alpha$) ~~~~~&   v$_\mathrm{rad}$    &   shape   &   size   & o.th. \\
\hline
 a10 &        & 128.5 &  252.5 & 00:54:54.213 & -37:40:45.74 &   66.0$\pm$3.9 &  226.0$\pm$11.4 & 150$\pm$6 &      oval       &  33$\times$36 & thick      \\
 a38 &        &  298.2 &  125.6 & 00:54:51.354 & -37:41:11.03 &  123.0$\pm$7.2 &  443.0$\pm$22.3 & 151$\pm$6 & irregular   &  41$\times$56 & thick      \\
 a58 &        &   31.9 &   25.9 & 00:54:55.840 & -37:41:30.97 &   19.8$\pm$1.7 &   63.7$\pm$3.3 & 137$\pm$3 &     round         &  15$\times$15 &  thin     \\
 a65 &  80  &  302.0 &  172.7 & 00:54:51.294 & -37:41:01.64 &  378.0$\pm$21.4 & 1260.0$\pm$63.4 & 162$\pm$7 & round   & 100$\times$100$^*$ &  thin    \\
 a66 & 100 &   23.8 &  125.6 & 00:54:55.977 & -37:41:11.03 & 1000.0$\pm$57.7 & 3410.0$\pm$171.6 & 138$\pm$8 & round/irr. & 136$\times$136$^*$ &  thin    \\  
 a67 &        &  205.4 & 24.7 & 00:54:52.918 & -37:41:31.21 &   33.3$\pm$3.7 &  107.0$\pm$5.6 & 145$\pm$6 &     round         & 47$\times$47 &  thin    \\  
 a68 &        &  264.6 &  145.6 & 00:54:51.920 & -37:41:07.03 &   42.9$\pm$3.3 &  136.0$\pm$7.0 & 151$\pm$3 &      oval        & 47$\times$33 &  thin    \\  
 a69 &        &  213.8 & 273.6 & 00:54:52.777 & -37:40:41.43 &  126.0$\pm$6.7 &  428.0$\pm$21.5 & 149$\pm$6 &     round    &  34$\times$34 &  thin    \\  
 a70 &  86  &  233.4 &  294.1 & 00:54:52.446 & -37:40:37.33 &  308.0$\pm$16.6 & 1040.0$\pm$52.2 & 147$\pm$5 & elliptical &  98$\times$76$^*$ &  thin  \\  
 a73 &  85  &  265.5 &    7.0 & 00:54:51.904 & -37:41:34.74 &  167.0$\pm$9.0 &  507.0$\pm$25.5 & 140$\pm$8 &         ...        &   ... $^*$           &     ...    \\
 b20 &        &  278.6 &  194.8 & 00:54:46.634 & -37:40:57.27 &   30.5$\pm$2.3 &   87.3$\pm$4.5 & 162$\pm$6 &     round        &  25$\times$25 &  thin    \\  
 b29 &        &  258.2 &   82.7 & 00:54:46.977 & -37:41:19.69 &   36.3$\pm$2.3 &  124.0$\pm$6.3 & 159$\pm$6 &      oval          &  31$\times$29 &  thin    \\  
 b32 &        &   21.2 &   31.3 & 00:54:50.971 & -37:41:29.97 &   40.1$\pm$2.7 &  130.0$\pm$6.6 & 147$\pm$5 &     round         &  21$\times$21 &  thin    \\  
 b39 &        &   50.9 &  277.2 & 00:54:50.470 & -37:40:40.79 &  194.0$\pm$12.1 &  631.0$\pm$31.9 & 155$\pm$7 &      oval     &  75$\times$50 &  thin    \\  
 b49 &        &  141.3 &   60.5 & 00:54:48.948 & -37:41:24.12 &  284.0$\pm$14.8 &  908.0$\pm$45.5 & 148$\pm$5 &      oval     &  38$\times$62 &  thin    \\  
 b51 &  80  &   20.3 &  183.2 & 00:54:50.985 & -37:40:59.57 &  369.0$\pm$18.9 & 1310.0$\pm$65.6 & 162$\pm$8 & elliptical   &  34$\times$52$^*$ & thick  \\
 b52 &  80  &   46.9 &  194.3 & 00:54:50.538 & -37:40:57.36 &  243.0$\pm$13.5 &  793.0$\pm$39.8 & 161$\pm$6 &      oval     &  47$\times$78 & thick    \\  
 b56 &  74  &   96.5 &  129.2 & 00:54:49.702 & -37:41:10.39 &  632.0$\pm$32.5 & 2030.0$\pm$101.8 & 154$\pm$5 &      oval  &  36$\times$63 &  thin    \\  
 b57 &  74  &   85.5 &  120.8 & 00:54:49.888 & -37:41:12.07 &  302.0$\pm$16.6 &  961.0$\pm$48.3 & 152$\pm$5 &      oval     &  76$\times$94 &  thin    \\  
 b59 &        &   89.0 &  197.0 & 00:54:49.828 & -37:40:56.83 &   51.7$\pm$3.7 &  174.0$\pm$8.8 & 163$\pm$1 & irregular        &  42$\times$32 &  thin    \\  
 b60 &        &   85.0 &  161.1 & 00:54:49.896 & -37:41:04.01 &  168.0$\pm$10.1 &  567.0$\pm$28.6 & 162$\pm$9 & irregular   &  52$\times$42 &  thin    \\  
 c06 &        &  296.8 &  286.0 & 00:54:41.210 & -37:40:39.35 &  126.0$\pm$7.8 &  402.0$\pm$21.5 & 182$\pm$5 & elliptical     &  33$\times$44 &  thin    \\  
 c44 &  60  &   84.0 &  140.2 & 00:54:44.795 & -37:41:08.50 & 1170.0$\pm$71.6 & 3680.0$\pm$197.0 & 161$\pm$4 & complex & 103$\times$154 & thick    \\  
 c45 &        &  168.0 &  236.6 & 00:54:43.380 & -37:40:49.22 &  431.0$\pm$48.6 & 1180.0$\pm$83.8 & 167$\pm$4 & complex$^a$ & 236$\times$163 &  thin    \\
 c46 &        &  288.0 &  251.3 & 00:54:41.359 & -37:40:46.29 &   90.6$\pm$6.5 &  315.0$\pm$17.5 & 169$\pm$5 &  open ring$^b$ &  54$\times$59 &  thin    \\
 c48 &        &  221.3 &   39.8 & 00:54:42.481 & -37:41:28.58 &   13.7$\pm$2.5 &   49.2$\pm$4.1 & 164$\pm$8 &     round          &  36$\times$36 &  thin    \\ 
 d08 &        &  102.8 &  275.2 & 00:54:38.059 & -37:40:28.67 &   14.8$\pm$2.2 &   47.7$\pm$2.5 & 184$\pm$7 & irregular         &  36$\times$36 &  thin    \\ 
 d21 &        &  244.9 &  215.3 & 00:54:35.664 & -37:40:40.65 &    5.5$\pm$0.6 &   20.7$\pm$1.1 & 189$\pm$9 &      oval            &  25$\times$19 &  thin    \\ 
 d22 &        &  267.3 &  204.6 & 00:54.35.287 & -37:40:42.79 &   48.3$\pm$4.1 &  153.0$\pm$7.8 & 193$\pm$5 & elliptical         &  54$\times$77 &  thin    \\  
 d49 &        &  300.0 &  101.3 & 00:54:34.737 & -37:41:03.45 &   19.4$\pm$1.8 &   69.2$\pm$3.5 & 196$\pm$6 & irregular         &  44$\times$36 &  thin    \\
 d70 &        &   75.5 &  277.0 & 00:54:38.518 & -37:40:28.31 &   10.6$\pm$3.5 &   63.3$\pm$3.4 & 195$\pm$8 & irregular          & 111$\times$49 &  thin    \\  
 d72 &        &  156.4 &  229.6 & 00:54:37.152 & -37:40:37.83 &  292.0$\pm$15.3 &  952.0$\pm$47.7 & 194$\pm$6 &     round   &  62$\times$62 & thick    \\  
 d74 &  42  &   37.9 &  173.3 & 00:54:39.151 & -37:40:49.05 &  409.0$\pm$17.4 & 1360.0$\pm$54.6 & 181$\pm$4 &      oval     &  53$\times$64 & thick    \\ 
 d77 &        &  278.9 &  124.1 & 00:54:35.091 & -37:40:58.89 &  126.0$\pm$2.7 &  419.0$\pm$8.4 & 193$\pm$5 &      oval         &  36$\times$27 & thick    \\  
 d81 &        &  183.2 &   52.6 & 00:54:36.705 & -37:41:13.20 &   38.5$\pm$2.4 &  132.0$\pm$6.7 & 186$\pm$7 &      oval           &  24$\times$36 &  thin    \\  
 d91 &        &  244.1 &  153.2 & 00:54:35.679 & -37:40:53.08 &  248.0$\pm$14.0 &  778.0$\pm$39.0 & 194$\pm$5 & irregular   &  69$\times$76 &  thin    \\  
 d92 &        &  266.9 &   98.2 & 00:54:35.295 & -37:41:04.08 &   23.2$\pm$1.5 &   70.9$\pm$3.6 & 189$\pm$5 &     round          &  23$\times$23 &  thin    \\  
 d93 &        &  283.9 &   85.2 & 00:54:35.008 & -37:41:06.67 &   37.3$\pm$2.1 &  128.0$\pm$6.4 & 192$\pm$6 &     round         &  27$\times$27 &  thin    \\  
 d95 &        &  181.9 &  141.1 & 00:54:36.726 & -37:40:55.49 &  722.0$\pm$40.7 & 2430.0$\pm$121.9 & 188$\pm$5 & open ring$^c$ & 163$\times$103 &  thin   \\
 e19 &        &  149.1 &   97.3 & 00:54:39.636 & -37:40:02.27 &   77.9$\pm$3.6 &  301.0$\pm$13.6 & 184$\pm$5 &     round       &  24$\times$24 & thick    \\  
 e43 &        &  293.9 &  123.7 & 00:54:37.196 & -37:39:57.00 &    2.2$\pm$0.4 &    7.0$\pm$0.4 & 195$\pm$7 &      oval             &  17$\times$23 &  thin    \\  
 e90 &        &  117.3 &  121.4 & 00:54:40.170 & -37:39:57.45 &   53.7$\pm$3.0 &  199.0$\pm$10.0 & 188$\pm$6 &      oval        &  24$\times$32 &  thin    \\  
 e91 &        &   12.7 &  129.0 & 00:54:41.933 & -37:39:55.93 &  349.0$\pm$21.0 & 1030.0$\pm$51.8 & 187$\pm$6 & complex   &  73$\times$97$^*$ &  thin  \\ 
 i26 &         &   41.3 &  196.6 & 00:54:44.371 & -37:41:56.80 &    1.0$\pm$0.2 &    4.7$\pm$0.2 & 178$\pm$4 & irregular            &  22$\times$11 &  thin    \\  
 i62 &         &  260.4 &   91.3 & 00:54:40.678 & -37:42:17.85 &   33.0$\pm$2.9 &  107.0$\pm$5.5 & 166$\pm$5 &     round         &  29$\times$29 &  thin    \\  
 i67 &         &  190.4 &   62.4 & 00:54:41.859 & -37:42:23.64 &    9.4$\pm$1.4 &   29.9$\pm$1.7 & 162$\pm$8 & elliptical           &  24$\times$20 &  thin    \\  
 i81 &         &  216.6 &  231.4 & 00:54:41.417 & -37:41:49.84 &   89.1$\pm$5.5 &  285.0$\pm$14.4 & 175$\pm$6 & elliptical      &  14$\times$13 &  thin    \\   
 i83 &         &  169.6 &   78.7 & 00:54:42.209 & -37:42:20.38 &    0.5$\pm$0.2 &    2.4$\pm$0.2 & 167$\pm$6 &     round            &  16$\times$16 &  thin    \\  
 i91 &         &  290.2 &  221.9 & 00:54:40.176 & -37:41:51.73 &  132.0$\pm$7.0 &  464.0$\pm$23.3 & 174$\pm$7 &     round     &  54$\times$54 &  thin    \\   
 i92 &         &  277.1 &   58.3 & 00:54:40.396 & -37:42:24.45 &    1.6$\pm$1.0 &    6.5$\pm$0.6 & 175$\pm$5 &     round            &  20$\times$20 &  thin    \\  
 i94 &         &  228.3 &  315.9 & 00:54:41.219 & -37:41:32.94 &  174.0$\pm$13.1 &  607.0$\pm$31.0 & 169$\pm$8 & complex$^d$ & 207$\times$98$^*$ &  thin   \\
 i96 &         &    9.2 &  240.0 & 00:54:44.912 & -37:41:48.12 &  197.0$\pm$10.2 &  608.0$\pm$30.5 & 172$\pm$7 &      oval      &  27$\times$102$^*$ &  thin    \\
 i97 &         &   29.5 &  206.1 & 00:54:44.569 & -37:41:54.90 &  277.0$\pm$14.6 &  864.0$\pm$43.3 & 177$\pm$5 & elliptical    &  45$\times$53 & thick    \\ 
 i98 &         &  279.4 &   11.3 & 00:54:40.358 & -37:42.33.85 &  133.0$\pm$7.1 &  412.0$\pm$20.7 & 176$\pm$8 & complex$^d$ & 136$\times$25$^*$ &  thin  \\
 i99 &         &  313.3 &   62.4 & 00:54:39.787 & -37:42:23.63 &  184.0$\pm$11.8 &  596.0$\pm$30.1 & 174$\pm$6 & complex$^d$ & 245$\times$187$^*$ &  thin  \\
i100 &        &  150.6 &    9.1 & 00:54:42.529 & -37:42:34.30 &  121.0$\pm$7.5 &  372.0$\pm$18.8 & 177$\pm$4 & complex$^d$ & 117$\times$109$^*$ &  thin   \\
 j04 &         &   41.8 &  294.1 & 00:54:50.277 & -37:41:46.58 &  271.0$\pm$15.5 &  748.0$\pm$37.7 & 146$\pm$4 & complex    & 132$\times$63$^*$ &  thin    \\ 
 j07 &         &  269.4 &  288.3 & 00:54:46.441 & -37:41:47.74 &   35.2$\pm$2.3 &  131.0$\pm$6.6 & 163$\pm$6 & irregular        &  88$\times$52 &  thin    \\  
 j33 &         &  165.7 &   66.1 & 00:54:48.189 & -37:42:32.20 &  323.0$\pm$16.4 & 1020.0$\pm$51.2 & 162$\pm$7 &      oval     &  63$\times$51 & thick   \\  
 j52 &         &  116.3 &  160.8 & 00:54:49.021 & -37:42:13.25 &   83.4$\pm$4.4 &  297.0$\pm$14.9 & 156$\pm$5 &     round       &  33$\times$33 & thick   \\  
 j57 &        &   24.3 &  250.6 & 00:54:50.572 & -37:41:55.29 &   22.2$\pm$1.3 &   75.4$\pm$3.8 & 143$\pm$3 & irregular           &  36$\times$29 &  thin    \\  
\hline                      
\end{tabular}
\end{small}
\tablefoot{Column~1, name; Col~2, \citet{deharveng1988}; Col.~3, datacube spaxel x-coordinate; Col.~4, spaxel y-coordinate; Col.~5, right ascension (J2000); Col.~6, declination (J2000);  Col.~7, H$\beta$ flux [$10^{-17}$erg cm$^2$ s$^{-1}$]; Col.~8, H$\alpha$ flux, same units; Col.~9, radial velocity [km~s$^{-1}$], Col.~10, morphological description,  remarks: (a) supershell consisting of serveral bubbles, (b) classified SNR by \citet{blair1997},  (c) giant double-shell around WR star d38, (d) diffuse giant shell; Col.~11, apparent size in projection [pc], $^*$: truncated at the edge of the field, Col.~12, optical thickness.   
} 
\end{table*}
 \clearpage



\subsection{\hiiregs}
\label{HII-SNR-ISM}
{\bf Results: }Besides emission line point sources, we have catalogued extended objects from the visual examination of emission line maps, appearing bright in H$\alpha$ and/or in {\oiii} $5007~\AA$, with a variety of morphologies such as contiguous high surface brightness regions with circular or elliptical geometry, bright regions with irregular geometries, contiguous patches of low surface brightness, ring-like shells, diffuse and filamentary low surface brightness emission, and large scale arcs or closed shells. 

For example, the \hiireg labelled De100 in Fig.~\ref{NGC300_Ha} bears some similarity with the 30~Dor region in the LMC. There is quite a number of prominent shells of diameters up to 100~pc visible, e.g. quite prominently in field (d). Besides the point sources that were catalogued as cHII, we also found many compact, only slightly extended \hiiregs. As one can appreciate from Fig.~\ref{NGC300_Ha}, the earlier compilation of \hiiregs from \citet{deharveng1988} mainly comprise bright objects, while our new data reveal a wealth of newly discovered, fainter objects. 

We have made an attempt to distinguish between classical photoionized, optically thick \hiiregs, optically thin \hiiregs, and diffuse ISM using the technique of ionization para\-meter mapping (IPM) that is based on the extent of Str\"omgren spheres in different ions \citep{koeppen1979a,koeppen1979b}, as excercised extensively by \citet{pellegrini2012} in the Magellanic Clouds, and also recently applied for NGC\,628 with MUSE data by \citet{kreckel2016}. Fig.~\ref{IPM} shows two examples of optically thick and thin \hiiregs, catalogued as i97 and i81, respectively. i97 has a compact appearance with a peak H$\alpha$ surface brightness of $9.72\times10^{-16}$~erg~cm$^{-2}$~s$^{-1}$~arcsec$^{-2}$ (left). The IPM greyscale image, with low values of {\sii}/{\oiii} in black and high values in white, shows the typical picture of an optically thick region with a high excitation core and an enhanced line ratio as a ring, as presented by \citet{pellegrini2012} in their Fig. ~3 for DEM S38 in the LMC. Conversely, i81 has a peak H$\alpha$ surface brightness of 
$1.54\times10^{-16}$~erg~cm$^{-2}$~s$^{-1}$~arcsec$^{-2}$  and shows qualitatively the same appearance as DEM S159 (their Fig.~4.)
with areas of both high and low optical depths, namely a distinct rim to the E, complemented by an open lobe to the W, and again highly ionized gas in the central part. 

We inspected the IPM images for all fields (a)$\ldots$(j) and marked the discovered \hiiregs accordingly in the catalogue. 
The {\sii}/{\oiii} line ratio map is also sensitive to reveal other interesting features. Besides the brighter PNe appearing as black blotches that were already discovered otherwise as described above, the map brings out high excitation emission, e.g. a peculiar structure seen as extending radially from the rim of i97 (arrow in Fig.~\ref{IPM}, top row, middle). It it not visible  in any other emission line rather than [O$\,$III] and is roughly, but not quite aligned with a blue star. The nature of this strange feature is unclear. 

The catalogue of \hiiregs  is given in Table~\ref{HII-list}. Emission line fluxes, radial velocities, positions, as well as sizes and morphology were determined as described in Section~\ref{extract-gas}.

Also the compact \hiireg a32 that is embedded in a high surface brightness area of the \hiireg a65, impossible to discern on the basis of an H$\alpha$ image, was discovered this way. Altogether, we discovered 114 \hiiregs, measured their fluxes, sizes, radial velocities, and recorded any luminous blue stars visible with the nebulae that could be responsible for photoionization.

For the ISM, \citet{chu2008} distinguishes interstellar bubbles with diameters up to 30 pc that are powered by the stellar winds of individual massive stars,  superbubbles with sizes of $\sim$100~pc, dynamical ages of $\sim10^6$ years that require only one episode of star formation, and supergiant shells that have sizes of $\sim$1~kpc, dynamical ages of $\sim10^7$ years and require multiple episodes of star formation. Furthermore, there are SNR, that can be distinguished from the latter class of objects on the basis of the line intensity ratio F([S\,II)]/F(H$\alpha$) > 0.4 \citep{mathewson1973}. A detailed investigation of the nature of all of these nebulae is beyond the scope of this paper, so we have restricted ourselves to list the objects by characterizing them as compact \hiiregs, classical optically thick/thin \hiiregs, or giant \hiiregs with sizes larger than 100~pc, following the scheme of \citet{franco2003}, furthermore as SNR candidates, shells, or diffuse/filamentary ISM. 
SNR and the DIG are discussed further below.

\begin{figure}[!t]
  \centering
  \begin{minipage}[b]{0.15\textwidth}
     \includegraphics[width=85pt,bb=10 10 550 500,clip]{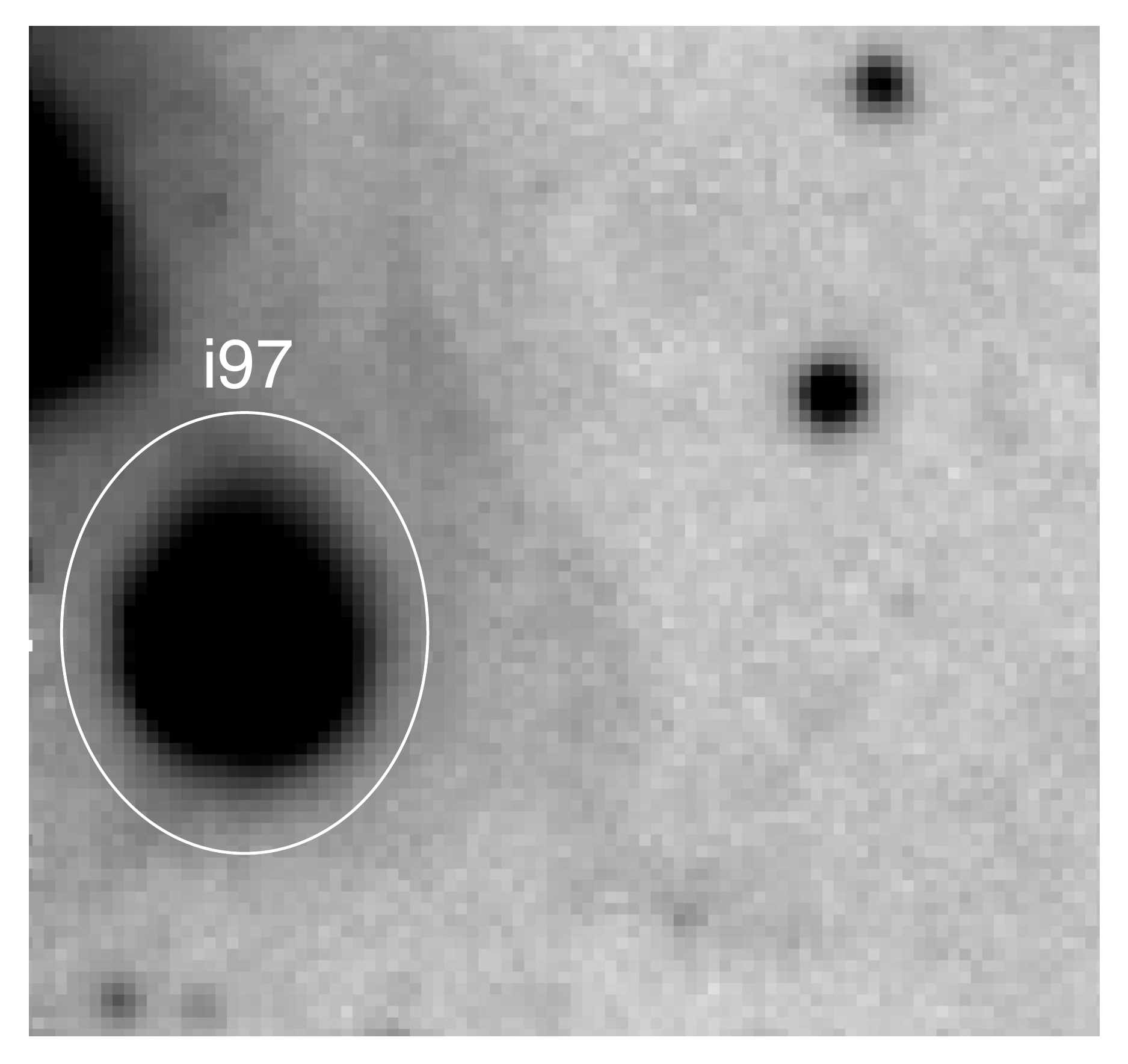}
  \end{minipage}
  \hfill
  \begin{minipage}[b]{0.15\textwidth}
    \includegraphics[width=85pt,bb=10 10 550 500,clip]{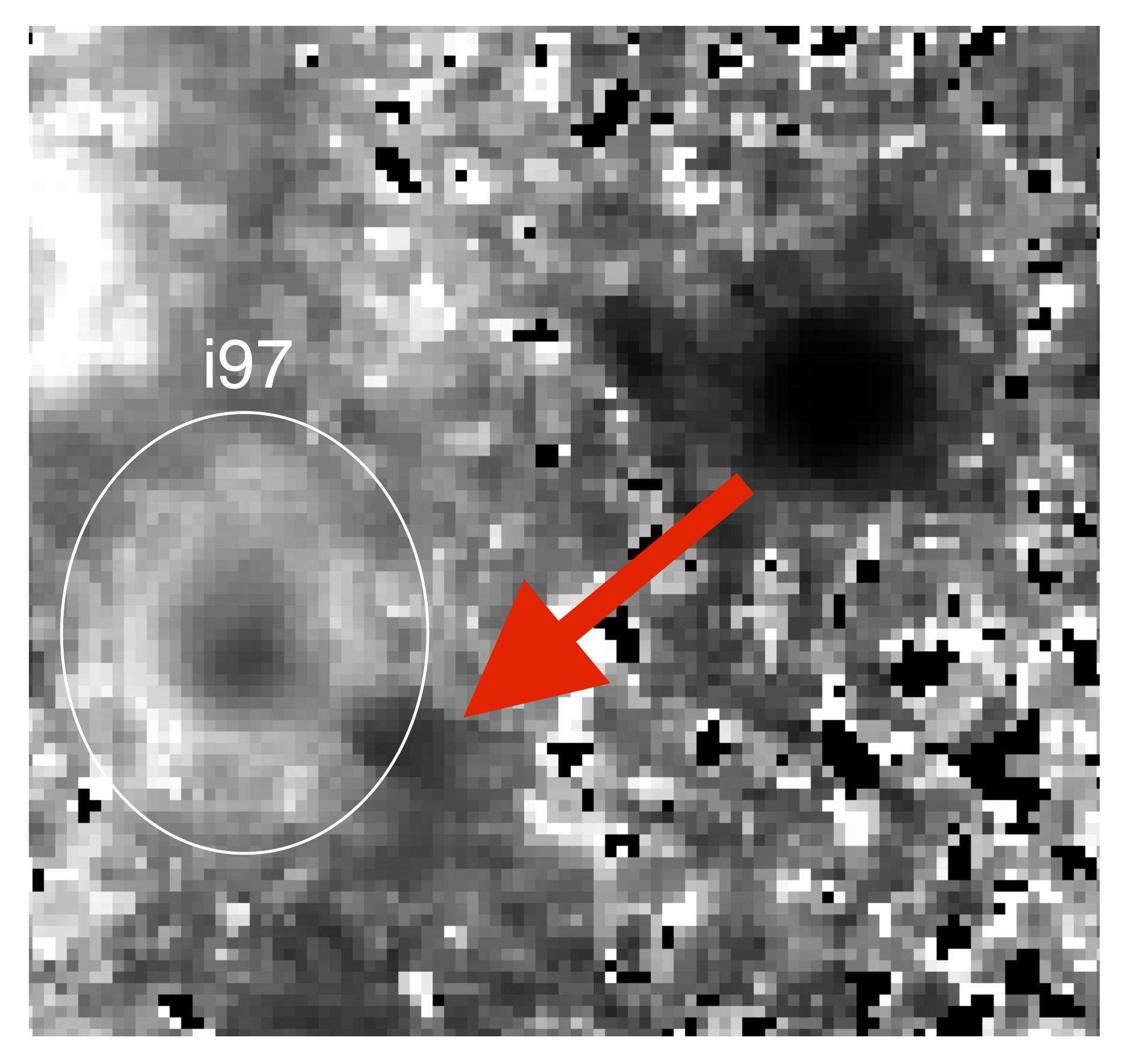}
  \end{minipage}
    \hfill
  \begin{minipage}[b]{0.15\textwidth}
    \includegraphics[width=85pt,bb=10 10 550 500,clip]{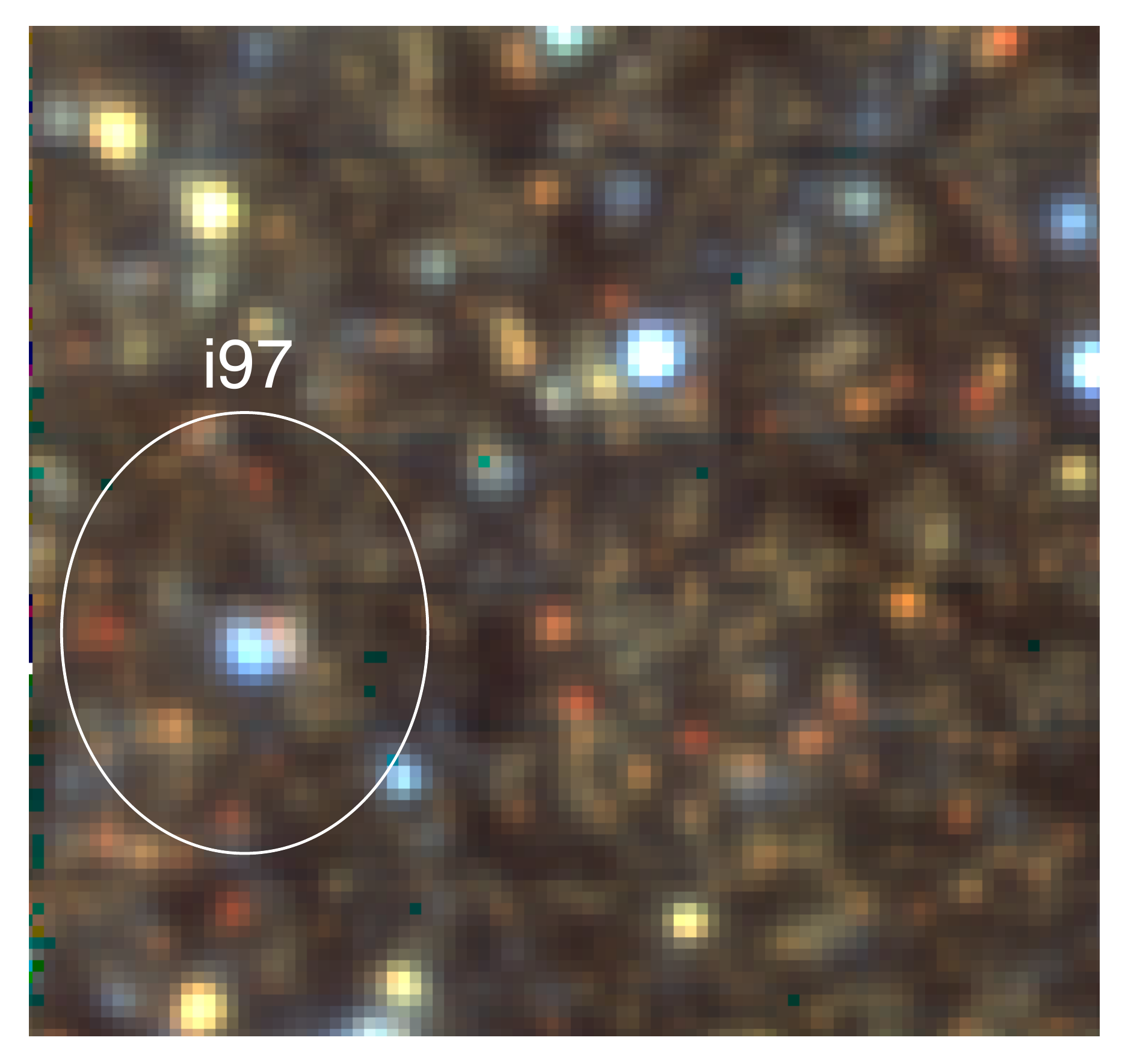}
  \end{minipage}

 \begin{minipage}[b]{0.15\textwidth}
     \includegraphics[width=85pt,bb=10 10 550 550,clip]{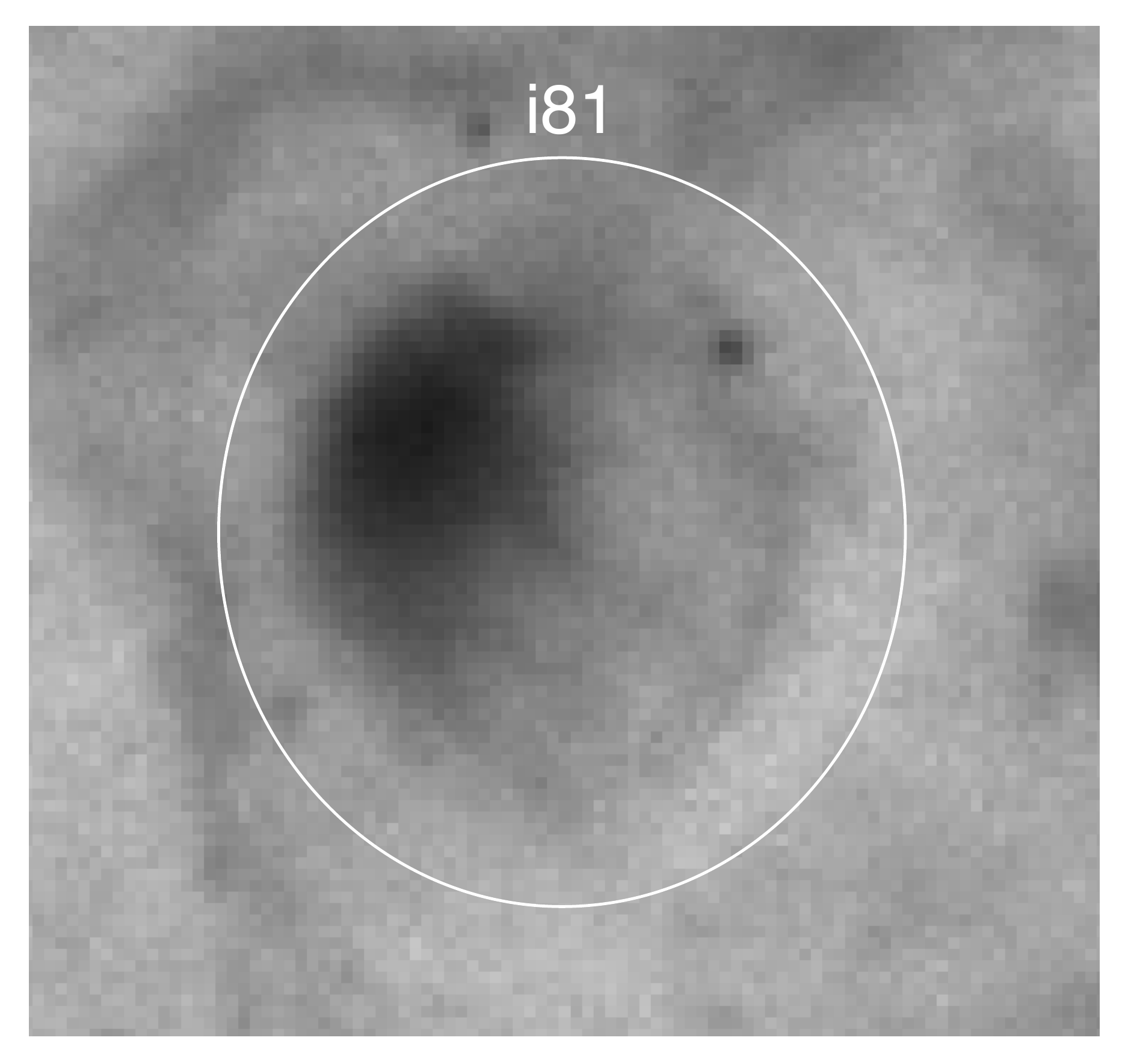}
  \end{minipage}
  \hfill
  \begin{minipage}[b]{0.15\textwidth}
    \includegraphics[width=85pt,bb=10 10 550 550,clip]{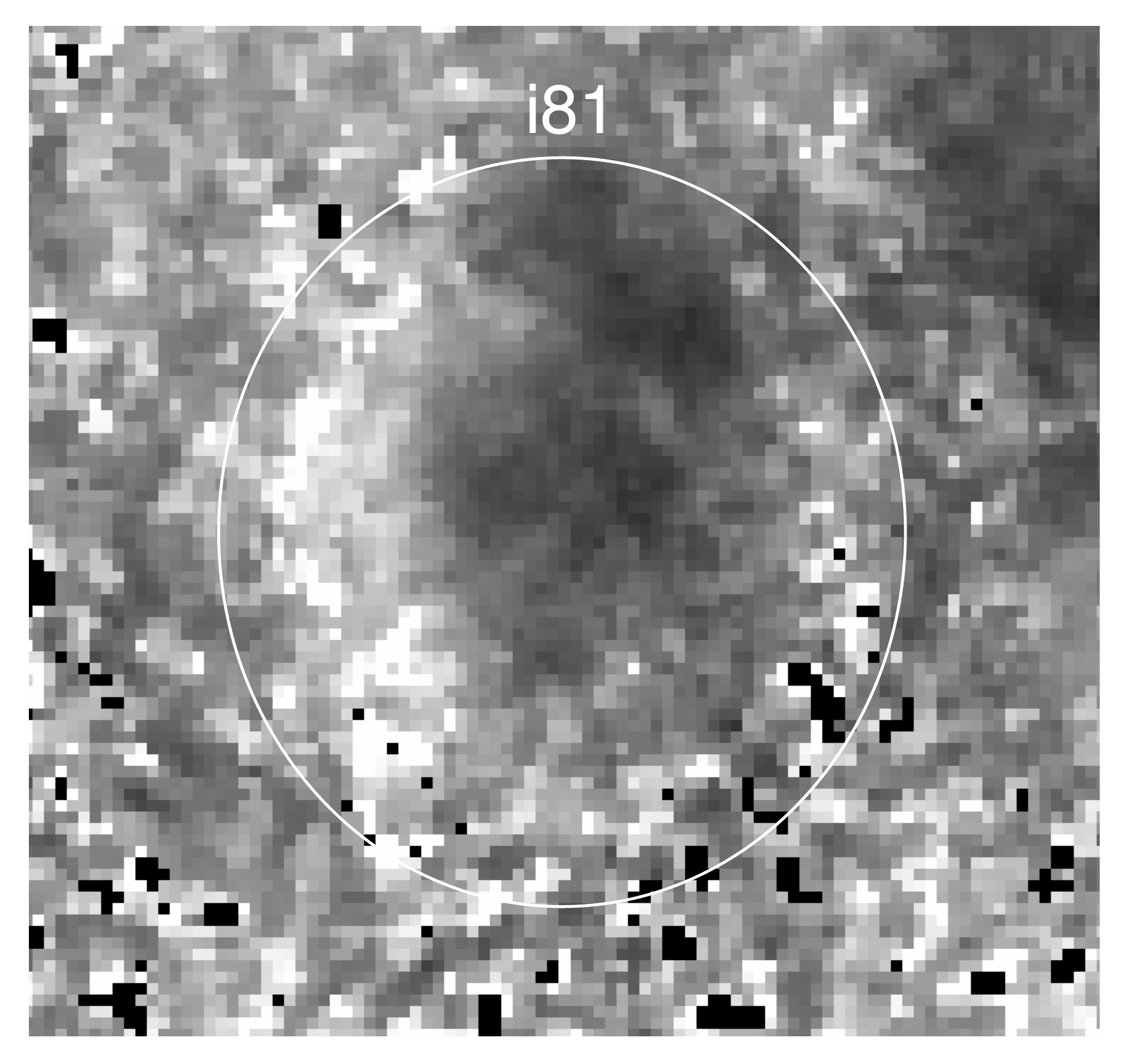}
  \end{minipage}
    \hfill
  \begin{minipage}[b]{0.15\textwidth}
    \includegraphics[width=85pt,bb=10 10 550 550,clip]{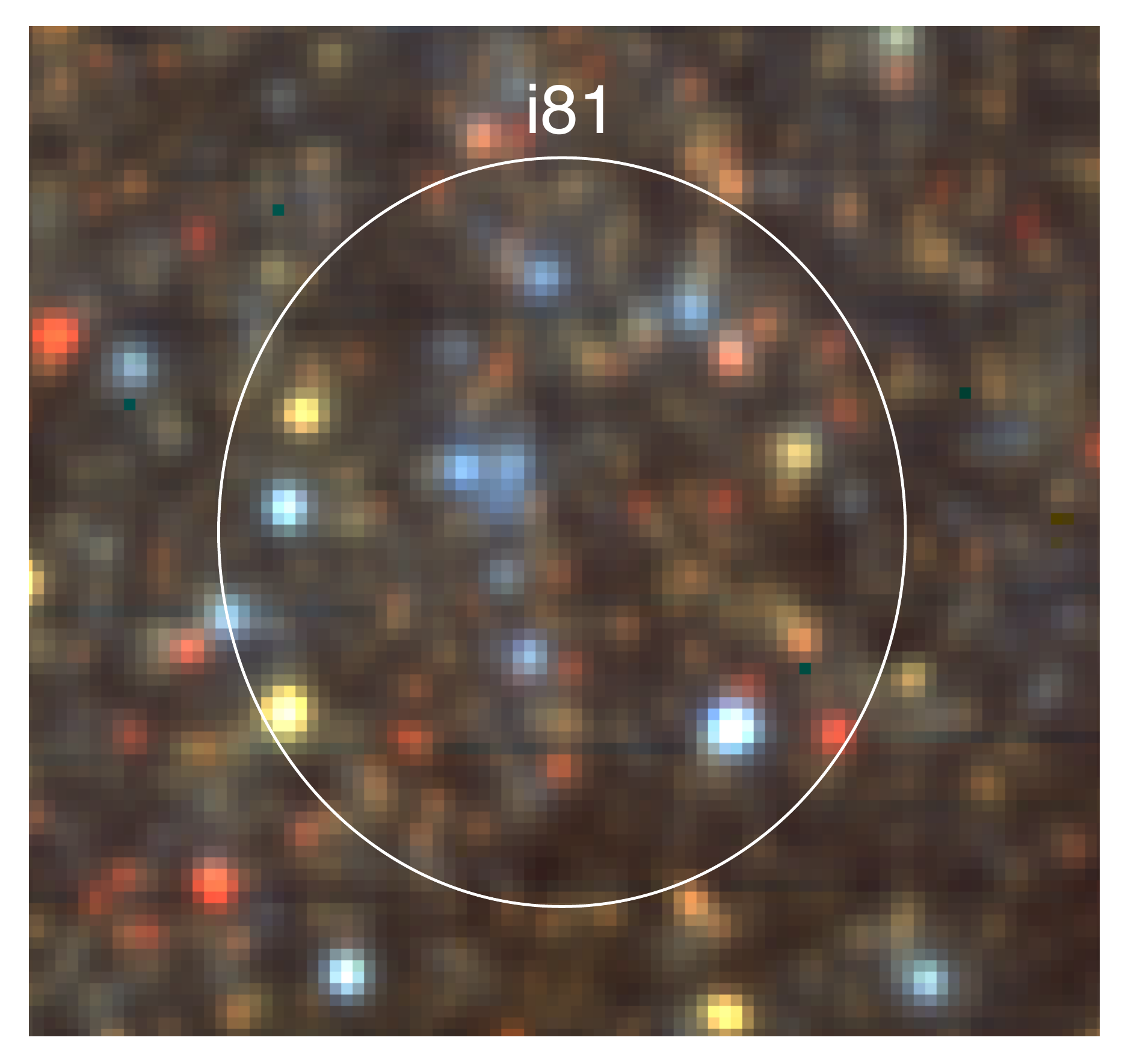}
  \end{minipage}
  
  \caption{Ionization parameter mapping of two \hiiregs, from left to right: H$\alpha$, [S$\,$II]/[O$\,$III], VRI. The greyscale for the H$\alpha$ maps shows high intensity in black, whereas the IPM maps feature high ratios in white. The region i97 (top row, 4.8"x5.8") is optically thick, while i81 (bottom, 10.4"x11.8") is apparently leaking ionizing radiation, a "blister" in the nomenclature of \citet{pellegrini2012}. A strange, {\oiii}-bright feature near the rim of i97 is marked by an arrow (see text). The stars that are likely responsible for the ionization of the nebulae are visible as blue objects in the VRI images.}
    \label{IPM}
\end{figure}

{\noindent \bf Discussion:} \label{discuss_HII} The most comprehensive catalogue of \hiiregs in NGC\,300 is still the one of \citet{deharveng1988}. It was complemented with more recent data from \citet{bresolin2009}, whose 28 \hiiregs, however, do not overlap with our MUSE observations. The red circles in Fig.~\ref{NGC300_Ha} indicate 8 objects from \citet{deharveng1988}, that coincide partially or fully with fields (a)\ldots(j). 
It is not surprising that with a total of 61 \hiiregs our MUSE observations go significantly deeper than their photographic data from 30 years ago. Since we were  able to sample low emission line intensities down to levels below 10$^{-17}$erg~cm$^{-2}$~s$^{-1}$~arcsec$^{-2}$, we could use the IPM technique to categorize \hiiregs as optically thick or thin, and to identify superbubbles and giant shells. The fluxes measured by \citet{deharveng1988} are in accord with our measurements within the errors, as far as their objects are entirely contained in our fields. As a convenient feature of the MUSE datacubes, we are able to immediately identify blue stars that are likely to be the ionizing sources  of the \hiiregs. As a caveat, it is important to realize that the spatial resolution of MUSE per se is no guarantee for having resolved individual stars. With a sampling of 1.8~pc per spaxel,  even in the event of seeing at the Nyquist limit, the image of the PSF projected into NGC\,300 corresponds to 3.6~pc, which is a size too large to safely resolve e.g. OB associations. 

\begin{figure*}[th!]
   \centering
    \includegraphics[width=\hsize,bb=35 20 750 570,clip]{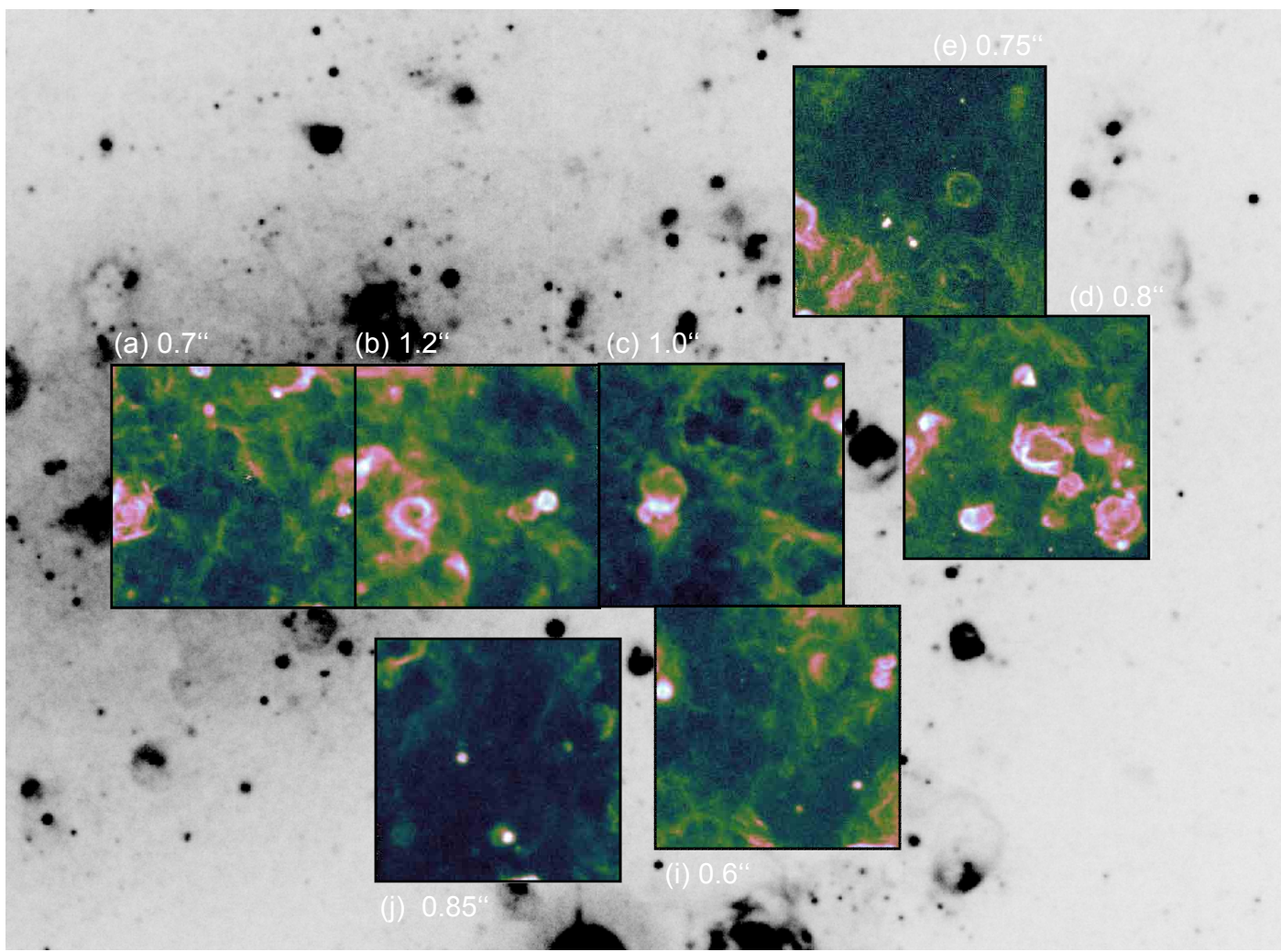}
    \caption{{\sii} images reconstructed from MUSE datacubes for pointings (a), (b), $\ldots$, (j). The colour scale brings out the DIG more prominently than the corresponding H$\alpha$ map, also   
     suppressing many point sources.}
     \label{NGC300_SII}
\end{figure*}

Therefore, the use of HST images and catalogue stars as input to {\textsc PampelMUSE} is an important prerequisite for the identification and possibly deblending of the ionizing stars. Even with HST at a sampling of 0.048"/pixel, i.e. an image scale of 0.44~pc/pixel, very compact clusters or binaries may not be resolved.

The decomposition of blended spectra with our library fitting technique may provide a solution to this problem. In order to make this work, we are planning to increase the scope of our stellar library with more O and B stars. The goal of this exercise would be to determine the energy budget of \hiiregs and obtained a unified picture concerning leaking Lyman continuum radiation and the ionization of the DIG (see below). The well-known fact that photometry in the optical cannot constrain the number of ionizing photons of hot stars, because the Rayleigh-Jeans tail of the spectral energy distribution is insensitive to the effective temperature \citep{hummer1988}, was realized in the outcome of an experiment by  \citet{niederhofer2016}, that consisted in the attempt of predicting the content of O stars in clusters within \hiiregs from isochrone fitting to HST photometry. In a follow-up project, similar to the work of, e.g. \citet{bresolin2002}, or \citet{Kudritzki2008}, we are planning to experiment with MUSE spectra to constrain the luminosity of the ionizing stars. In addition, we are also planning to further investigate the chemical (oxygen) content of our data in order to better constrain the distribution of the metal within this galaxy using different metallicity indicators as well combining our data with the literature ones.

\begin{figure*}[th]
  \centering
  \begin{minipage}[b]{0.49\textwidth}
     \includegraphics[width=300pt,bb=55 30 600 680,clip]{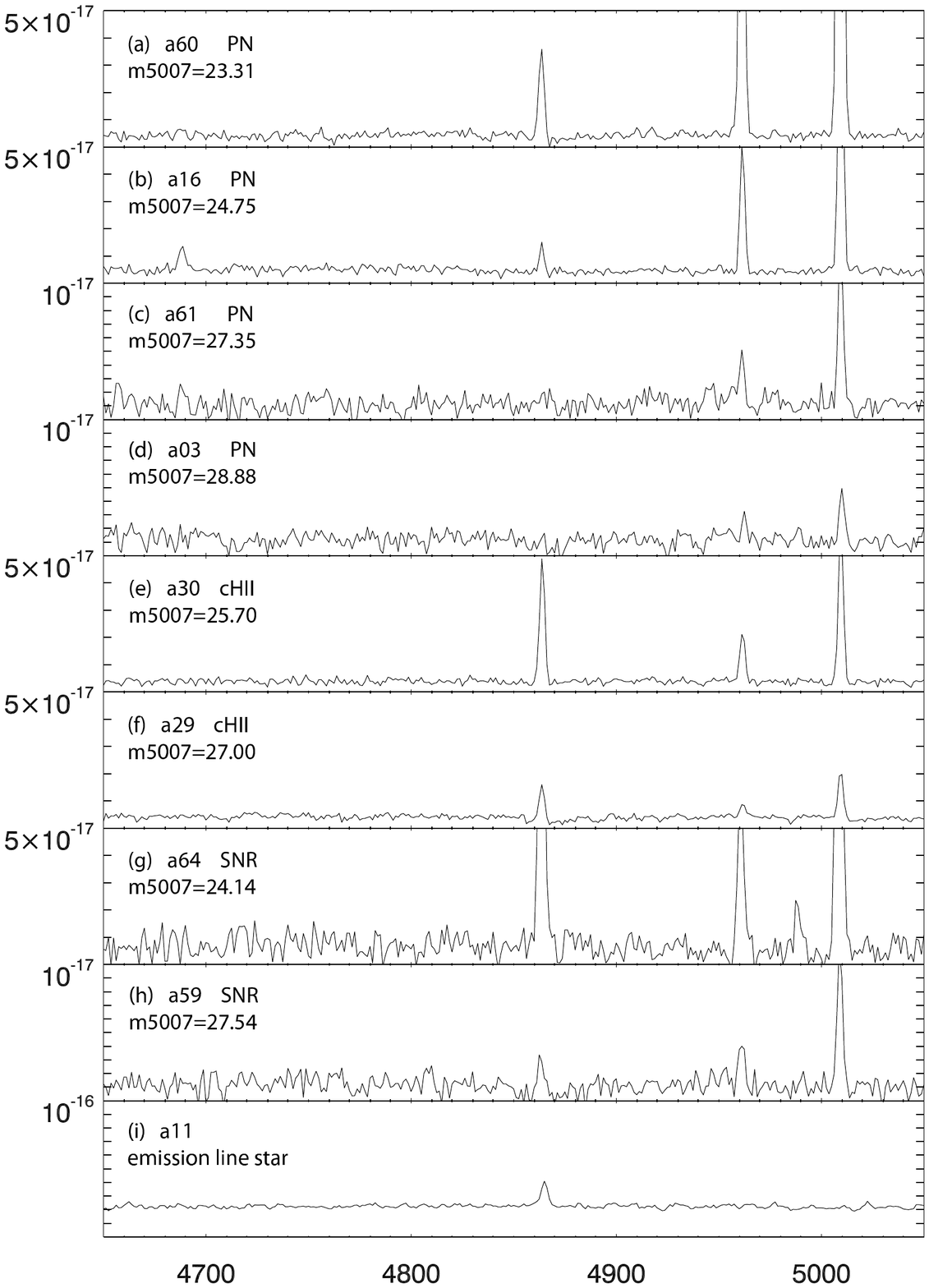}
  \end{minipage}
  \hfill
  \begin{minipage}[b]{0.49\textwidth}
    \includegraphics[width=300pt,bb=55 30 600 680,clip]{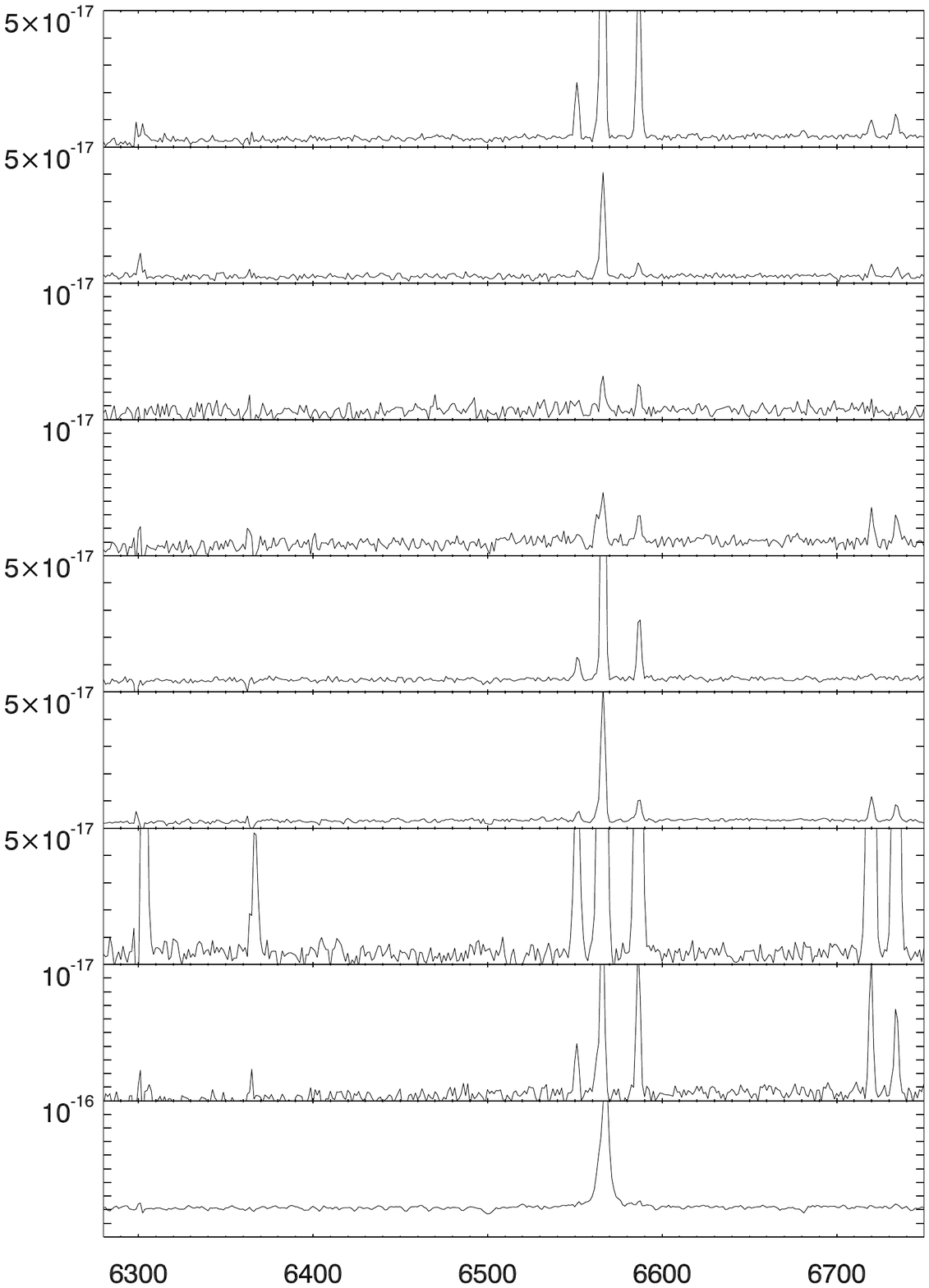}
  \end{minipage}
  \caption{Example spectra of PNe, cHII, SNR, and an emission line star in units of erg/cm$^2$/s per 1.25~{\AA} spectral bin in the H$\beta$-{\oiii} and H$\alpha$ regions, plotted against wavelength in \AA. The spectra are offset by some arbitrary bias in order to better show the noise level. Panels (a)\ldots(d) present PNe over the full range from the brightest to the faintest objects. The very faint object a03 with m$_{5007}$=28.88 is an uncertain PN candidate, given that the [SII] line ratio, although critically depending on diffuse nebular background subtraction, could also be indicative of a cHII, or even a SNR. Panels (e), (f) present a bright and a faint cHII, and panels (g), (h) a bright and a faint SNR. In panel (i) an emission line star is shown, presenting an H$\alpha$ emission line width of 6.6~\AA. Wavelengths in units of \AA, flux in units of erg cm$^{-2}$ s$^{-1}$ $\AA^{-1}$.}
    \label{NGC300_spectra}
\end{figure*}

\subsection{ Supernova remnants}
\label{SNR}
{\noindent \bf Results:} We inspected the {\oiii} maps for extended emission to find SNR candidates and subsequently used the spectra to select only those objects that show evidence for shock excitation on the basis of the canonical line ratio criterion {\sii}/H${\alpha} > 0.4$ \citep{mathewson1973}. We found prominent, high surface brightness examples, e.g. Fig.~\ref{SNR_images}, as well as low surface brightness objects that barely stand out from the noise. In our dataset, we have reached limiting surface brightness levels of 2$\times10^{-17}$erg~cm$^{-2}$~s$^{-1}$~arcsec$^{-2}$ in [O$\,$III]. In total, we detected 38 SNR and measured their fluxes, sizes, and radial velocities. 
Nebulae that have a similar appearance as SNR and spectra pointing to shock excitation rather than photoionization, however formally failing the $[$S II$]$/H${\alpha}$ line ratio criterion, were classified as shells to distinguish them from \hiiregs. 

\begin{table*}[!h]
 \caption{Catalogue of SNR candidates (38 objects)}             
\label{SNR-list}      
\centering          
\begin{small}
\begin{tabular}{ l  c    r r   c c      r r r     l  l     c  }     
\hline\hline     
ID    &  BL97 &  x      &     y      & RA                 &  Dec            &   F(${H\beta}$)   &   F(${H\alpha}$)   &   F{\sii}  &   v$_\mathrm{rad}$   &   shape   &   size    \\
\hline
 a02 &           & 261.1 &  301.1 & 00:54:51.980 & -37:40:35.93 &    4.1$\pm$0.3 &   10.9$\pm$0.6 &    3.9$\pm$0.2 & 152$\pm$7 & elliptical &  18$\times$15  \\
 a04 &           &  200.6 &  285.5 & 00:54:52.998 & -37:40:39.04 &    ... &    1.1$\pm$0.1 &    0.6$\pm$0.1 & 141$\pm$10 &     round &  18$\times$18  \\
 a13 &           &  131.8 &  233.1 & 00:54:54.158 & -37:40:49.53 &    3.7$\pm$0.4 &   12.0$\pm$0.6 &    3.0$\pm$0.2 & 146$\pm$8 &      oval &  18$\times$15  \\
 a15 &           &   86.9 &  219.3 & 00:54:54.914 & -37:40:52.28 &    2.0$\pm$0.3 &    6.8$\pm$0.4 &    3.8$\pm$0.2 & 138$\pm$8 &     round &  18$\times$18  \\
 a17 &           &  249.5 &  220.6 & 00:54:52.174 & -37:40:52.02 &    2.1$\pm$0.6 &    6.1$\pm$0.4 &    2.6$\pm$0.2 & 149$\pm$5 &     round &  18$\times$18  \\
 a25 &           &  106.9 &  182.9 & 00:54:54.577 & -37:40:59.57 &    0.8$\pm$0.4 &    2.2$\pm$0.2 &    0.8$\pm$0.1 & 139$\pm$5 &     round &  18$\times$18  \\
 a41 &           &   81.1 &   98.9 & 00:54:55.011 & -37:41:16:36 &    ... &    4.8$\pm$0.4 &    1.8$\pm$0.2 & 137$\pm$6 &     round &  18$\times$18  \\
 a42 &           &  242.9 &   96.2 & 00:54:52.287 & -37:41:16.90 &    ... &    2.7$\pm$0.2 &    1.8$\pm$0.1 & 154$\pm$6 & elliptical &  18$\times$20  \\
 a45 &           &   25.2 &   74.5 & 00:54:55.955 & -37:41:21.25 &    ... &    1.0$\pm$0.1 &    0.5$\pm$0.1 & 142$\pm$10 &     round &  18$\times$18  \\
 a47 &           &  129.1 &   69.6 & 00:54:54.203 & -37:41:22.23 &    ... &    0.9$\pm$0.1 &    0.7$\pm$0.1 & 155$\pm$5 & elliptical &  20$\times$13  \\
 a55 &           &  248.6 &   57.6 & 00:54:52.189 & -37:41:24.63 &    0.6$\pm$0.3 &    2.1$\pm$0.2 &    1.9$\pm$0.2 & 156$\pm$6 &     round &  18$\times$18  \\
 a59 &           &   75.8 &   19.4 & 00:54:55.101 & -37:41:32.27 &    1.5$\pm$0.2 &    3.7$\pm$0.2 &    2.2$\pm$0.1 & 134$\pm$11 &     round &  18$\times$18  \\
 a64 &   16    &  118.4 &  304.6 & 00:54:54.383 & -37:40:35.22 &   53.6$\pm$3.7 &  184.0$\pm$9.3 &  143.0$\pm$7.2 & 138$\pm$4 & elliptical &  51$\times$40$^*$   \\
 a71 &           &  297.5 &   35.4 & 00:54:51.366 & -37:41:29.07 &    1.0$\pm$0.2 &    1.7$\pm$0.1 &    0.7$\pm$0.1 & 160$\pm$5 &     round &  18$\times$18  \\
 b22 &           &  229.0 &  131.0 & 00:54:47.470 & -37:41:10.03 &   32.6$\pm$2.7 &  110.0$\pm$5.6 &   75.7$\pm$3.8 & 156$\pm$3 &      ring &  29$\times$29  \\
 b23 &   14    &  251.6 &  140.7 & 00:54:47.089 & -37:41:08.08 &  132.0$\pm$7.2 &  462.0$\pm$23.2 &  338.0$\pm$17.0 & 152$\pm$4 &      ring &  48$\times$48  \\
 b44 &           &  206.0 &  127.9 & 00:54:47.858 & -37:41:10.65 &   16.4$\pm$1.6 &   60.5$\pm$0.7 &   36.9$\pm$2.2 & 161$\pm$4 & irregular &  29$\times$29  \\
 b48 &           &   44.3 &   69.4 & 00:54:50.582 & -37:41:22.35 &   27.4$\pm$2.7 &   95.8$\pm$5.0 &   48.6$\pm$2.5 & 459$\pm$20 & elliptical &  34$\times$24  \\
 b58 &           &  301.2 &  233.8 & 00:54:46.253 & -37:40:49.47 &    ... &    0.4$\pm$0.1 &    0.5$\pm$0.1 & 166$\pm$7 &     round &  19$\times$19  \\
 c31 &           &  244.9 &  185.1 & 00:54:42.085 & -37:40:59.52 &    1.6$\pm$0.5 &    5.5$\pm$0.6 &    2.6$\pm$0.3 & 187$\pm$10 &     round &  19$\times$19  \\
 c47 &   10    &  309.7 &  244.6 & 00:54:40.992 & -37:40:47.61 &  197.0$\pm$11.1 &  710.0$\pm$37.1 &  195.0$\pm$10.3 & 174$\pm$2 & elliptical &  40$\times$87$^*$  \\
 c49 &           &   49.4  &  302.8 & 00:54:45.378 & -37:40:35.98 &    ... &   15.4$\pm$2.5 &    8.2$\pm$1.0 & 167$\pm$9 &      oval &  59$\times$36$^*$  \\
 d80 &     8    &   97.0 &   56.2 & 00:54:38.157 & -37:41:12:48 &  257.0$\pm$14.9 &  856.0$\pm$43.0 &  315.0$\pm$15.9 & 186$\pm$3 &      oval &  75$\times$57  \\
 d88 &           &  211.4 &   97.8 & 00:54:36.229 & -37:41:04.17 &   85.3$\pm$5.0 &  289.0$\pm$14.5 &  101.0$\pm$5.1 & 185$\pm$6 & double-ri &  49$\times$45  \\
 d89 &           &  194.0 &   71.8 & 00:54:36.522 & -37:41:09.35 &   25.6$\pm$3.1 &   92.7$\pm$4.8 &   35.5$\pm$1.9 & 190$\pm$5 & elliptical &  56$\times$85  \\
 d90 &           &  270.0 &   53.0 & 00:54:35.242 & -37:41:13.11 &  192.0$\pm$12.9 &  683.0$\pm$34.3 &  342.0$\pm$17.2 & 187$\pm$4 &     round & 109$\times$109  \\
 d96 &           &  220.8 &  194.3 & 00:54:36.070 & -37:40:44.85 &   34.6$\pm$3.1 &  117.0$\pm$5.9 &   53.4$\pm$2.7 & 192$\pm$4 & irregular &  43$\times$83  \\
 e03 &           &  206.7 &  269.0 & 00:54:38.664 & -37:39:27.94 &    2.7$\pm$0.4 &   11.1$\pm$0.6 &    4.8$\pm$0.3 & 184$\pm$7 &     round &  18$\times$18  \\
 e21 &           &  209.0 &  162.6 & 00:54:38.627 & -37:39:49.22 &   32.1$\pm$5.8 &   88.9$\pm$5.0 &   55.8$\pm$3.2 & 210$\pm$9 &     round &  85$\times$85  \\
 e40 &           &  272.0 &  269.9 & 00:54:37.565 & -37:39:27.76 &   29.5$\pm$3.8 &   89.6$\pm$4.7 &   26.4$\pm$1.6 & 185$\pm$7 & elliptical &  44$\times$65  \\
 i14 &            & 165.5 &  255.8 & 00:54:42.277 & -37:41:44.96 &    1.8$\pm$0.2 &    6.1$\pm$0.3 &    3.0$\pm$0.2 & 171$\pm$7 &     round &  18$\times$18  \\
 i23 &            & 295.6 &  243.1 & 00:54:40.085 & -37:41:47.49 &   30.2$\pm$1.9 &  109.0$\pm$5.5 &   53.1$\pm$2.7 & 177$\pm$11 &      ring &  38$\times$39  \\
 i59 &            & 175.5 &  110.8 & 00:54:42.110 & -37:42:13.97 &    2.9$\pm$0.5 &   12.6$\pm$0.7 &    3.8$\pm$0.3 & 160$\pm$3 &     round &  31$\times$31  \\
 i72 &            & 258.1 &   31.2 & 00:54:40.716 & -37:42:29.87 &   17.1$\pm$1.8 &   52.2$\pm$2.7 &   22.6$\pm$1.2 & 172$\pm$6 &     round &  63$\times$63  \\
 i80 &            & 210.7 &  140.1 & 00:54:41.516 & -37:42:08.09 &    0.2$\pm$0.1 &    0.6$\pm$0.0 &    0.3$\pm$0.0 & 165$\pm$4 &     round &  18$\times$18  \\
 i82 &            & 120.8 &   61.5 & 00:54:43.031 & -37:42:23.82 &    2.9$\pm$0.3 &    6.8$\pm$0.4 &    2.2$\pm$0.2 & 182$\pm$7 &     round &  18$\times$18  \\
 j38 &            &  44.0 &   64.7 & 00:54:59.240 & -37:42:32.47 &   20.5$\pm$1.6 &   69.1$\pm$3.6 &   30.8$\pm$1.7 & 156$\pm$5 &     round &  63$\times$63  \\
 j58 &            &  290.5 &  205.2 & 00:54:46.086 & -37:42:04.36 &   31.6$\pm$2.3 &  128.0$\pm$6.6 &   84.5$\pm$4.3 & 147$\pm$5 & irregular & 102$\times$78  \\
\hline                      
\end{tabular}
\end{small}
\tablefoot{Column~1, name; Col.~2, Objects in common with \citet{blair1997}: S16$^{(1)}$, S14$^{(2)}$, S10$^{(3)}$,  S8$^{(4)}$; Col.~3, datacube spaxel x-coordinate; Col.~4, datacube spaxel y-coordinate; Col.~5, right ascension (J2000); Col.~6, declination (J2000);
Col.~7, H$\beta$ flux in units of $10^{-17}$erg cm$^2$ s$^{-1}$; Col.~8, H$\alpha$ flux, same units; Col.~9, [SII] flux (6717~\AA), same units; Col~10, radial velocity [km~s$^{-1}$]; Col.~11, morphological description;  Col.~12, apparent size in projection [pc], $^*$ indicates objects truncated at the edge of the field.
} 
\end{table*}

~~\\
{\noindent \bf Discussion:} SNR in NGC\,300 were first recorded by \citet{d'dorico1980} who made photographic observations with narrowband filters at the UK Schmidt and found 8 SNR candidates. \citet{blair1997} performed a systematic narrowband filter search in H$\alpha$, [S$\,$II], and continuum using the 2.5m telescope at Las Campanas Observatory with a CCD imager and identified 28 SNR candidates in NGC\,300. With follow-up spectroscopy at the same telescope they investigated in detail the empirical classification parameter of the [SII]/H$\alpha$ line ratio that is required to be larger than or equal to 0.4 to qualify for a SNR. The extra effort of spectroscopy was necessary to eliminate the additional flux contribution from the [N$\,$II] lines adjacent to H$\alpha$ that would normally fall within the transmission band of a narrowband filter. Their survey reached a limiting H$\alpha$ surface brightness of 1$\times10^{-16}$ erg~cm$^{-2}$~s$^{-1}$~arcsec$^{-2}$. For comparison, the H$\alpha$ surface brightness limit of our data is below 1$\times10^{-17}$ erg/cm2/s/arcsec$^2$, i.e. an order of magnitude fainter, see [S$\,$II] map in Fig.~\ref{DIG_diagnostic} for comparison. The total number of SNR detected in our MUSE datacubes is 38, with diameters in the range of  20$\ldots$110~pc. The size distribution of SNR in M33, a galaxy similar to NGC\,300, was modeled and inferred observationally by \citet{asvarov2014}. Assuming a similar distribution for NGC\,300, we expect  that  the median of the distribution is found at SNR diameters near 40~pc, with maximum diameters between 150 and 200~pc, depending on different values of the filling factor of the warm phase of the ISM. However, as pointed out  by \citet{asvarov2014}, the predicted size distribution is subject to observational selection effects, diminishing the detection rate for extended low surface brightness SNR in particular. We found one example of an extremely nitrogen-rich SNR (a71) similar to the one reported for the galactic SNR Pup~A by \citet{dopita1977}, or several such objects in M81, M101, and NGC\,6946 \citet{matonick1997}.

The dichotomy of filter spectrophotometry and the associated problem of [N$\,$II] contamination of H$\alpha$ fluxes  vs. slit spectroscopy, as discussed by the latter authors,  is lifted when using IFS, which is an advantage in terms of efficiency, because no extra observations are necessary. Moreover, the fact  that all derived quantities were obtained under identical observing conditions within a given data\-cube is an asset.

\begin{figure}[bh!]
  \centering
  \begin{minipage}[b]{0.24\textwidth}
     \includegraphics[width=120pt,bb=10 10 500 490,clip]{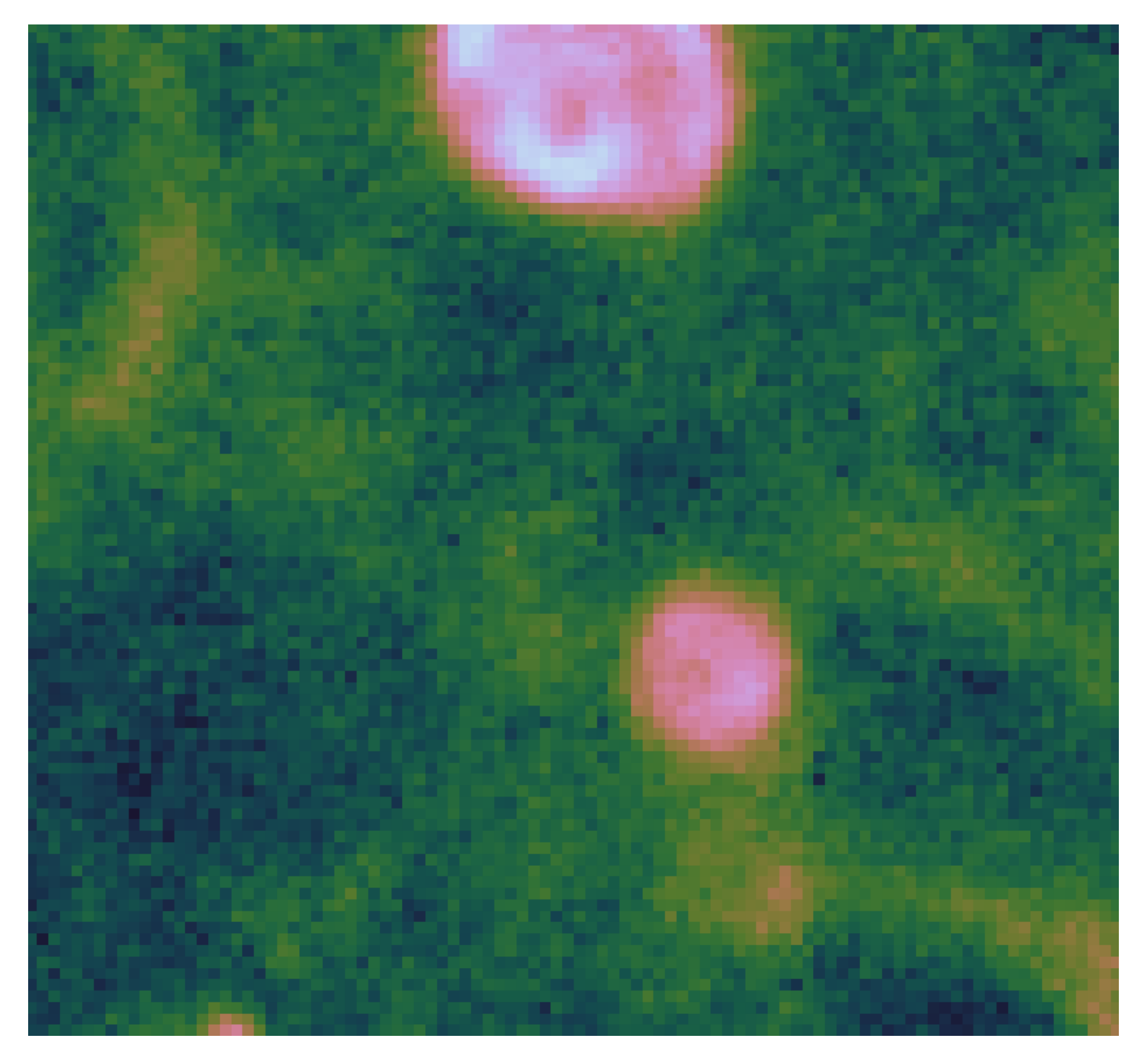}
  \end{minipage}
  \hfill
  \begin{minipage}[b]{0.24\textwidth}
    \includegraphics[width=120pt,bb=10 10 500 490,clip]{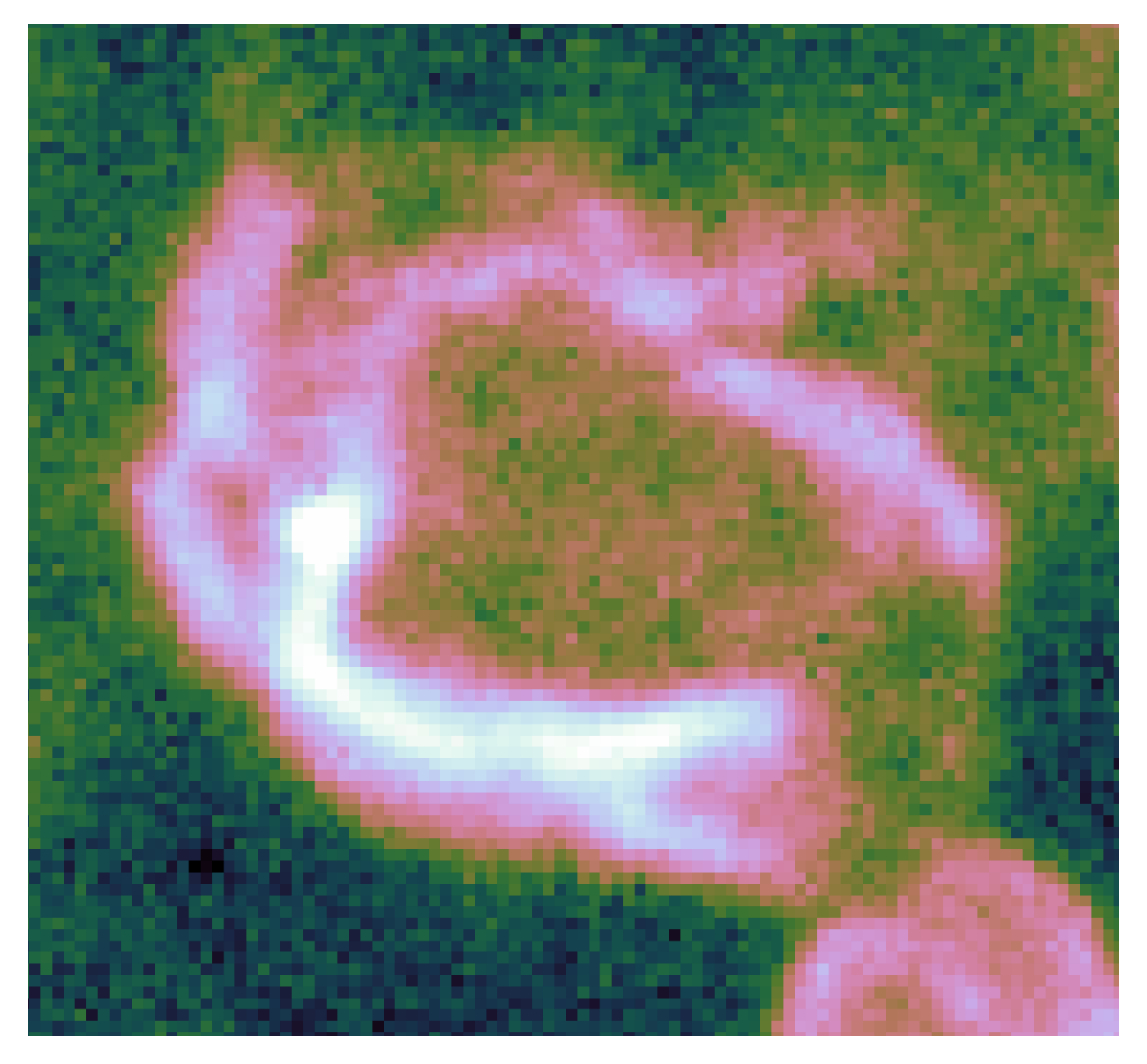}
  \end{minipage}
 \begin{minipage}[b]{0.24\textwidth}
     \includegraphics[width=120pt,bb=10 10 500 530,clip]{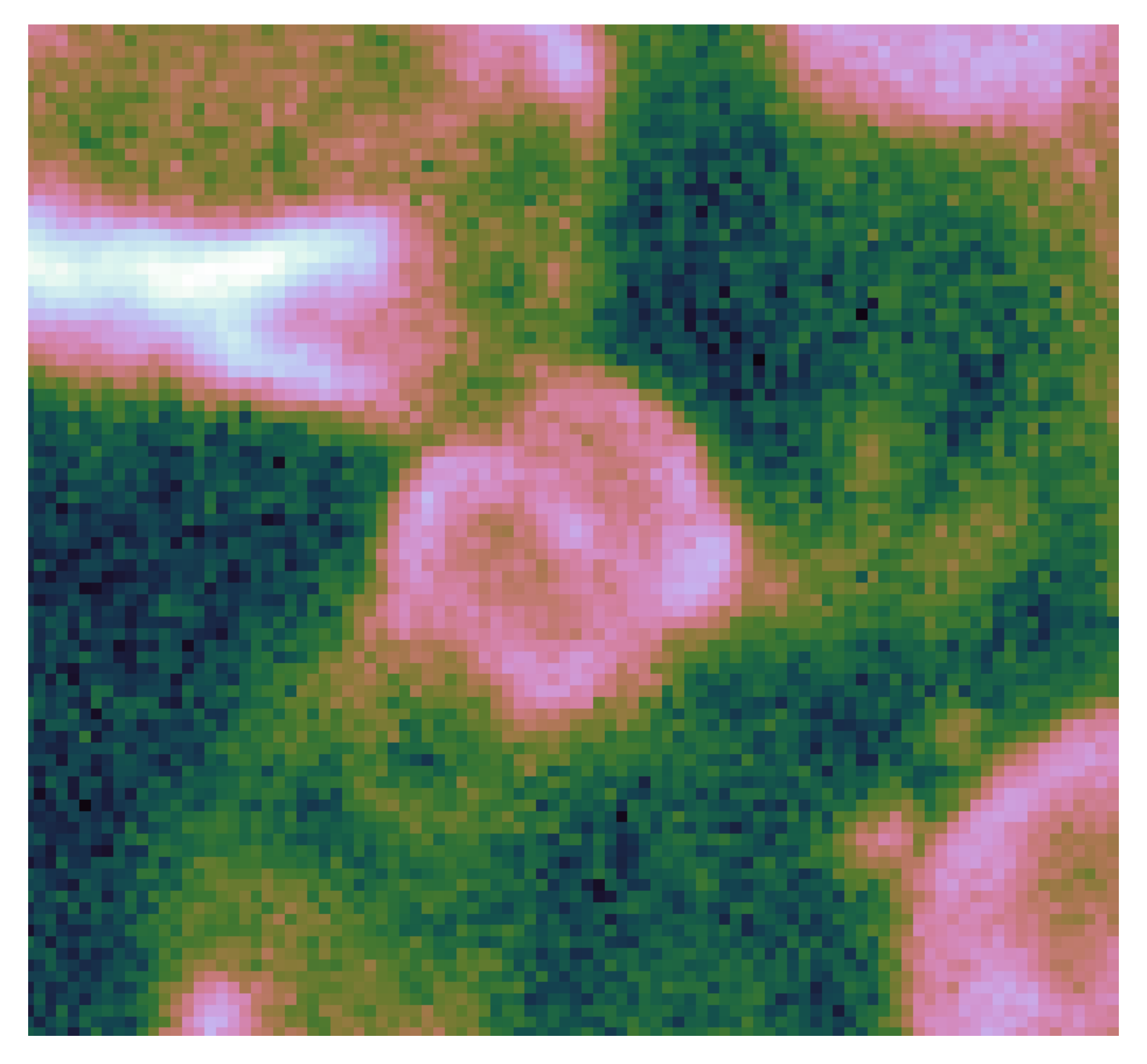}
  \end{minipage}
  \hfill
  \begin{minipage}[b]{0.24\textwidth}
    \includegraphics[width=120pt,bb=10 10 500 490,clip]{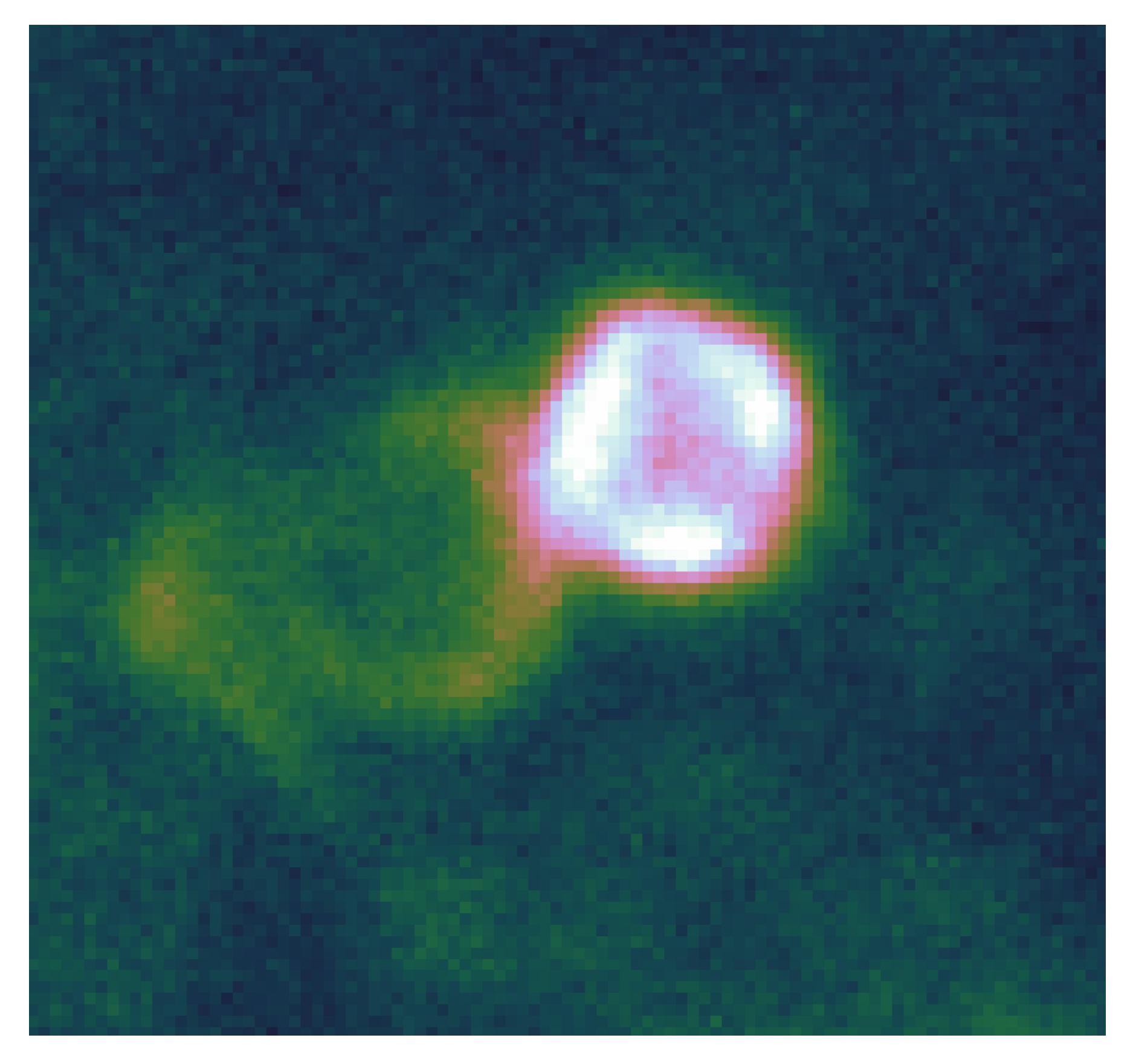}
  \end{minipage}
   \caption{[S$\,$II] images of a supernova remnant and shells. UL: SNR a64, truncated at the northern edge of field (a); the object to the SW from the center is the optically thick \hiireg a10. UR: Double-lobe shell d95/d96, hosting WR star. LL: Double-ring shell d87/d88, reminiscent of the rings of SN1987A -- however with a diameter of $\sim$10 pc. LR: shell b23, suspiciously aligned with the somewhat fainter shell B22.
    }
    \label{SNR_images}
 \end{figure}

 \begin{figure*}[th!]
   \centering
  \begin{minipage}[b]{0.487\textwidth}
        \includegraphics[width=\hsize,bb=35 60 510 535,clip]{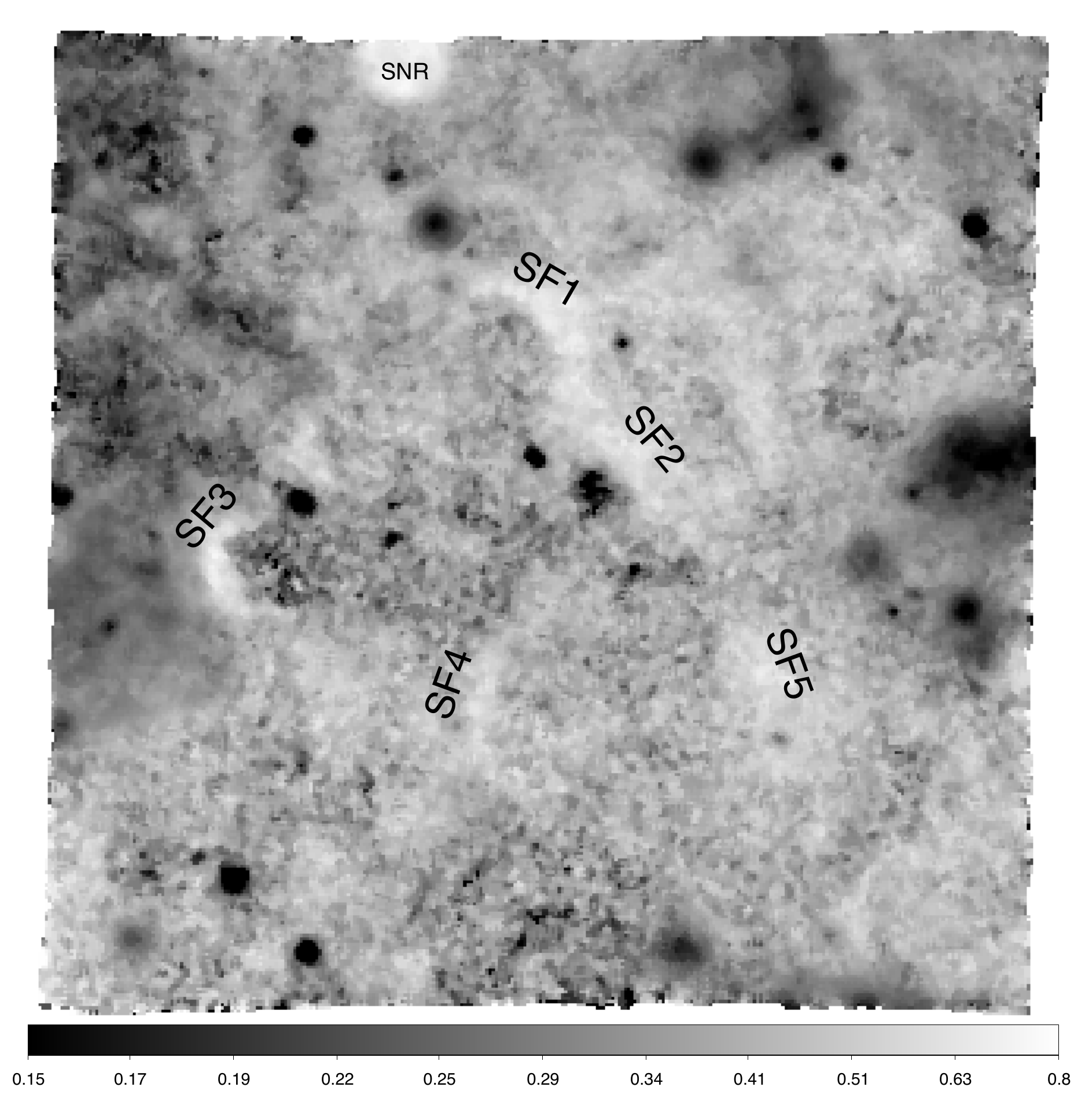}
  \end{minipage}
    \hfill
  \begin{minipage}[b]{0.487\textwidth}
        \includegraphics[width=\hsize,bb=35 60 510 535,clip]{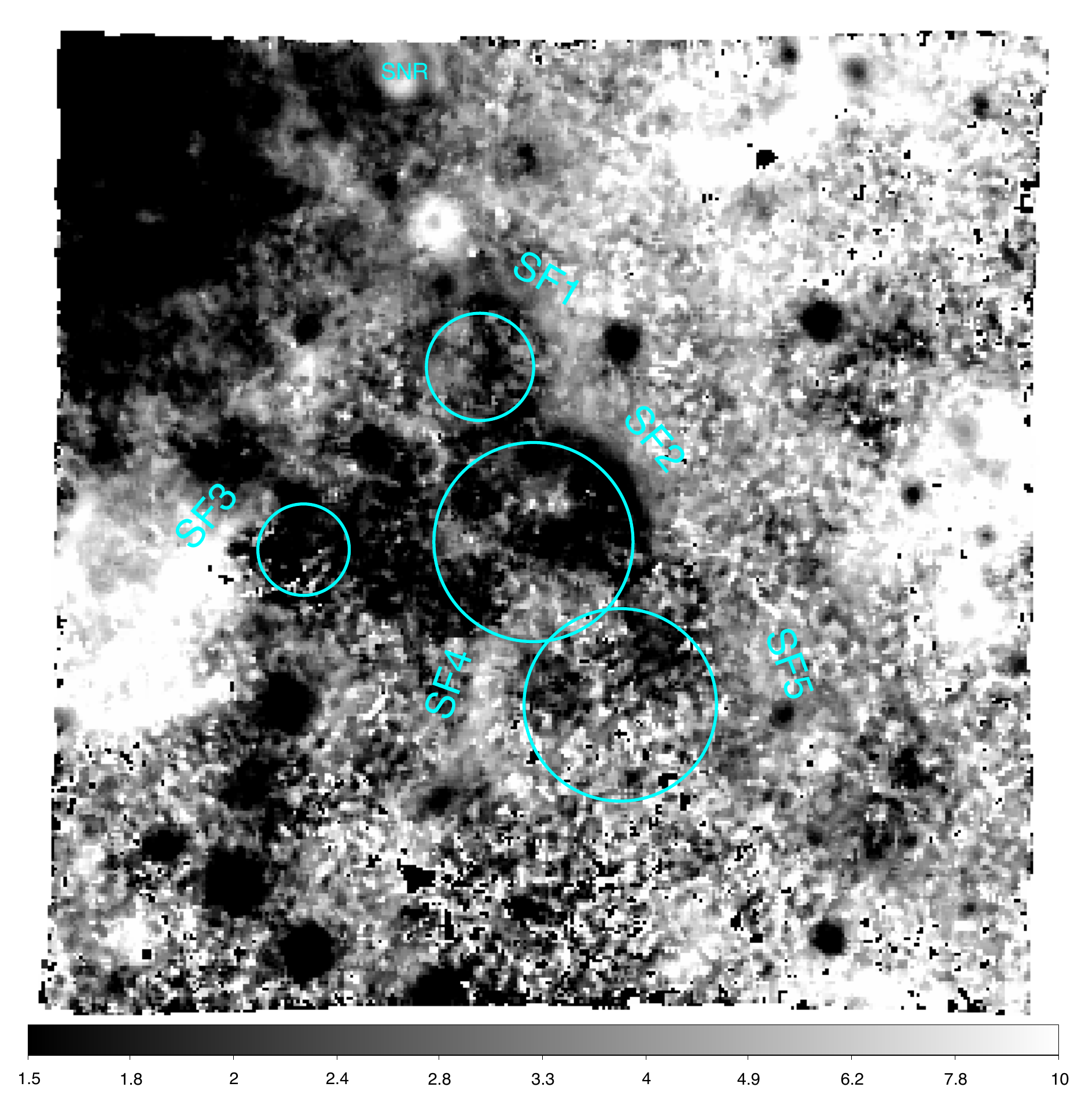}
  \end{minipage}
    \caption{Ionization parameter mapping used to analyze the DIG in field (a). Left: [S$\,$II]/H$\alpha$, greyscale $0.15\ldots0.8$ (black--white). Right: [S$\,$II]/[O$\,$III], greyscale $1.5\ldots10$ (black--white).
    }
    \label{DIG_maps}
 \end{figure*}

From the observational point of view in the optical, there exists no sharp distinction between SNR, superbubbles, or giant shells as all of these objects are characterized by shock excitation rather than photoionization and a  high [SII]/H$\alpha$ line ratio. SNR are defined by a line ratio larger than 0.4, altough \citet{blair1997} found that this criterion is washed out in some galaxies, e.g. NGC7793. We have chosen to assign a classification "shell" if blue stars can be seen within the nebula whose stellar winds are likely shaping the structure, and "SNR" if this is not the case. Fig.~\ref{SNR_images} shows examples of a SNR and several superbubbles/shells. Although we recorded the presence of luminous blue stars within the nebulae whose stellar winds might contribute to the expansion of the shell, e.g. the WR star d38 in shell d95 (Fig.~\ref{WR_images}), we note that we are not able at this stage on the basis of the optical data alone to assess with certainty whether the shells are driven by stellar winds, by expanding SNR, or a combination of the two. As an interesting feature, we show three objects that exhibit a double-ring structure whose nature is not exactly clear --- perhaps only a by-chance alignment. 

Detailed studies of extragalactic SNR, including Xray data, were for a long time restricted to the Magellanic Clouds. Recent advances, e.g. the study of \citet{long2018} in M33, show that more distant galaxies have come into reach. Considering the data quality and sensitivity obtained with MUSE, as well as the advantages of imaging spectroscopy, we are now in the position to collect larger samples within different environments for the purpose of testing numerical simulations, e.g. the SNR models from \citet{li2015}, wind-driven superbubble models from \citet{ntormousi2011}, or the more recent results from the SILCC project \citep{peters2017}.

\subsection{ Diffuse ionized gas}
\label{DIG}
{\noindent \bf Results:} Diffuse ionized gas, that has been thought for some time to be due to photoionization from field stars and Lyman continuum radiation leaking from \hiiregs \citep{oey1997}, is ubiquitously found in our dataset. Fig.~\ref{NGC300_SII} illustrates the distribution of DIG emission, whose visibility is enhanced by choosing the [S$\,$II] emission line map as opposed to H$\alpha$, as well as a suitable color table. Apart from bright \hiiregs, SNR, and shells that appear in pink and white hues, the greenish structure can mostly be attributed to the ISM. Typical [S$\,$II] surface brightness values corresponding to this color coding are from a few up to  50$\times10^{-18}$~erg~cm$^{-2}$~s$^{-1}$~arcsec$^{-2}$. The presence of DIG is very prominent along the NW spiral arm, and even near the nucleus of the galaxy, with a particularly enhanced level over a roughly 500~pc wide strip extending SW-NW in the lower left quadrant of field (b) to the upper right quadrant of field (a). The interarm regions in the northern parts of field (d12) and in field (j) are comparatively quiet. Still  the minimum [S$\,$II] surface brightness levels in these areas amount to 2$\times10^{-18}$~erg~cm$^{-2}$~s$^{-1}$~arcsec$^{-2}$ and 2$\times10^{-17}$~erg~cm$^{-2}$~s$^{-1}$~arcsec$^{-2}$  in fields (d) and (j), respectively. 

We have investigated in more detail field (a) with the expectation that the old stellar population in the center should exhibit little star formation, hence providing a more undisturbed view on the diffuse emission of the ISM. Within a radius of roughly 200~pc around the nucleus we find indeed no major star forming regions. From Figs.~\ref{NGC300_SII} and \ref{DIG_diagnostic} it is apparent that the distribution of [S$\,$II] emission is not smooth and homogenous, but shows a distinct fuzzy and filamentary structure that has as yet not been reported in the literature. In particular, there are several narrow ridges of enhanced surface brightness up to $2\ldots5\times10^{-17}$~erg~cm$^{-2}$~s$^{-1}$~arcsec$^{-2}$ that extend over lengths of 100\ldots200~pc. In order to obtain alternative surface brightness estimates independently from any unknown systematic errors of our continuum-subtracted narrowband images, and also with the goal to measure radial velocities of the gas, we performed spectrophotometry with {\textsc P3D} over apertures of $5\times5$ spaxels, i.e. 1~arcsec$^2$, sampling a total of 20 positions aD01$\ldots$aD20 along filaments, some other high surface brightness areas, and three areas of the lowest apparent surface brightness (Fig.~\ref{DIG_diagnostic}). 

A correction for the underlying stellar continuum was applied using a ULySS fit to the local unresolved stellar population, that takes into account the LOSV and velocity dispersion. This was subtracted from the flux measured within the aperture defined with the {\textsc P3D} tool. The background spectrum needed to be scaled individually to the actual continuum level near each line for the wavelength regions around H$\beta$/{\oiii} and H$\alpha$/{\nii}/{\sii} since ULySS applies a polynomial fit to accomodate variations of the overall spectral energy distribution, that however does not necessarily match the exact shape of the background continuum at the percent level of accuracy. Since the scaling was iteratively performed under visual control, it was also reassuring to see that the model background spectrum reproduced not only the strong absorption lines, but also numerous weak absorption features. The background correction turned out to be instrumental for the elimination of systematic errors that would otherwise arise strongly at stellar absorption lines.  
The results from these measurements are listed in Table~\ref{DIG-list}.

 \begin{table*}[h]
 \caption{Diffuse ionized gas}             
\label{DIG-list}      
\centering          
\begin{small}
\begin{tabular}{ c  r    c  c     c c      c  c      c    c    c    l  }     
\hline\hline     
ID  & x       &  y     & RA           & Dec     &   F(5007) & F(H$\alpha$) & F(6716) &  v$_\mathrm{rad}$   &  comment  \\
\hline
aD01 &  407.00 & 309.00 & 00:54:55.580 & -37:40:34.33 &  8.23 &  1.59 &  6.07  &  150$\pm$9 &  arbitrary area \\
aD02 &   40.00 & 272.00 & 00:54:55.707 & -37:40:41.62 &  3.42  &  4.95 &  0.80 &  163$\pm$11 & local surface brightness min. \\
aD03 &  156.00 &  23.00 & 00:54:53.750 & -37:40:50.16 &  2.64   &  4.89 &  3.35  &  146$\pm$6 &  filament SF1 \\
aD04 &  165.00 & 222.00 & 00:54:53.597 & -37:40:51.77 &  1.95  &  5.57 &  4.25  &  143$\pm$6 &  filament SF1 \\
aD05 &  170.00 & 214.00 & 00:54:53.514 & -37:40:53.39 &  2.10  &  4.66 &  3.60  &  149$\pm$4 &  filament SF1 \\
aD06 &  175.00 & 198.00 & 00:54:53.425 & -37:40:56.57 &  1.72  &  4.89 &  3.44  &  145$\pm$6 &  filament SF1 \\
aD07 &  180.00 & 185.00 & 00:54:53.345 & -37:40:59.15 &  3.92  &  6.25 &  4.72  &  152$\pm$7 &  filament SF2 \\
aD08 &  186.00 & 176.00 & 00:54:53.243 & -37:41:00.98 &  6.38  &  6.02 &  4.25  &  159$\pm$9 &  filament SF2 \\
aD09 &  196.00 & 167.00 & 00:54:53.075 & -37:41:02.77 &  2.38  &  4.42 &  2.91  &  150$\pm$11 & filament SF2\\
aD10 &  158.00 & 275.00 & 00:54:53.716 & -37:40:41.18 &  1.87  &  5.69 &  1.55  &  153$\pm$7 &  arbitrary area \\
aD11 &  184.00 & 267.00 & 00:54:53.275 & -37:40:42.77 &  1.10  &  5.98 &  3.13  &  155$\pm$3 &   arbitrary area \\
aD12 &  223.00 & 231.00 & 00:54:52.620 & -37:40:49.95 &  1.27  &  6.12 &  3.19  &  153$\pm$9 &  arbitrary area \\
aD13 &  268.00 & 191.00 & 00:54:51.864 & -37:40:57.92 &  \ldots      &  2.01 &  0.74 &  146$\pm$20 & arbitrary area \\
aD14 &   79.00 & 141.00 & 00:54:55.046 & -37:41:07.94 &   \ldots      &  2.00 &   \ldots      &  143$\pm$20 & local surface brightness min. \\
aD15 &  146.00 & 115.00 & 00:54:53.919 & -37:41:13.07 &  0.85 &  4.05 &  2.25  &  155$\pm$7 &   filament SF4 \\
aD16 &  142.00 & 101.00 & 00:54:53.986 & -37:41:16.01 &  0.88 &  4.70 &  2.45  &  148$\pm$1 &   filament SF4 \\
aD17 &  139.00 &  85.00 & 00:54:54.038 & -37:41:19.12 &  1.05   &  4.42 &  2.31  &  146$\pm$2 &   filament SF4 \\
aD18 &  160.00 &  31.00 & 00:54:53.686 & -37:41:29.78 &   \ldots      &  1.98 &  0.28 &  144$\pm$11 & local surface brightness min. \\
aD19 &  232.00 & 110.00 & 00:54:52.474 & -37:41:14.14 &  1.52  &  4.26 &  2.76  &  150$\pm$5 &   filament SF5 \\
aD20 &  242.00 &  94.00 & 00:54:52.300 & -37:41:17.36 &  2.21   &  5.21 &  3.18  &  157$\pm$9 &   filament SF5 \\
\hline                      
\end{tabular}
\end{small}
\tablefoot{Spectrophotometry of DIG at selected spots in field (a). Column~1, name; Col.~2, datacube spaxel x-coordinate; Col.~3, datacube spaxel y-coordinate; Col.~4, right ascension (J2000); Col.~5, declination (J2000); Col.~6, [OIII] flux (5007~\AA) in units of $10^{-17}$erg cm$^2$ s$^{-1}$; Col.~7, H$\alpha$ flux, same units; Col.~8, [SII] flux (6716~\AA), same units; Col~9, radial velocity [km~s$^{-1}$]; Col~9, comment.} 
\end{table*}

~\\

\citet{pellegrini2012} have stressed that ionization parameter mapping is effective at highlighting large, extremely faint structures. We have applied the IPM technique exemplarily in field (a). As reported in $\S$\ref{HII-SNR-ISM}, the [S$\,$II] surface brightness recorded in the narrowband images agrees generally well with spectro\-photometry that was performed  in selected areas. We have chosen to measure with {\textsc P3D} specifically  (a) prominent filamentary DIG features, and (b) local minima of low surface brightness regions, in order to shed light on the physical nature of those regions. The results of these distinct surface brightness measurements are tabulated in Table~\ref{DIG-list}, presenting the flux in {\oiii}, H$\alpha$, and {\sii} across an aperture of 1~arcsec$^2$, and the LOSV of the measured spots.  The radial velocities for the filament SF1 north of the nucleus, sampled by aD03$\ldots$aD06 are 145$\pm$2 km~s$^{-1}$, which is within the error identical with the systemic velocity of the galaxy, i.e. quite reasonably close to the nucleus. Interestingly, the branch of the filament SF2 sampled by aD07$\ldots$aD09 exhibits a slightly more positive value of 151$\pm$2 km~s$^{-1}$, which can be understood if SF1 and SF2 are aligned in projection, however are not physically associated with each other, each having their own kinematic history. The [S$\,$II]/H$\alpha$ line ratio map in Fig.~\ref{DIG_maps} shows that filaments SF1 and SF2 stand out from their environment with a value of $\sim$0.6. 


{\noindent \bf Discussion:}  Following \citet{mathewson1973}, we interpret the enhanced line ratio as shock fronts from ancient supernova remnants, rather than due to photoionization from leaking \hiiregs. There are more features suggesting the same interpretation, namely filaments SF4 and SF5. SF3 is especially prominent, marking the transition from the extremely low surface brightness area around aD14 towards the bright giant \hiireg  De100. For comparison, the SNR a64 (marked "S16" in Fig.~\ref{NGC300_Ha}, "SNR" in Fig.~\ref{DIG_maps}) appears prominently at the northern edge of the field. The [S$\,$II]/H$\alpha$ map further supports this interpretation, e.g. delineating a region of high excitation gas by an almost perfectly circular arc (SF2),  as indicated in Fig.~\ref{DIG_maps} with a circle. The same argument can be made for the arc of SF1, as well as for the pair of SF4/SF5, and SF3. The four examples are somehow  similar to the enigmatic large ionized bubble discovered by \citet{pellegrini2012} in the LMC (their Fig.~10), although the latter is lacking a rim.\citet{ogden1985} have studied a similar, kinematically distinct filament in the Milky Way, concluding that most likely the object is not photoionized by a nearby O star, but rather shock ionized.

The emission line intensity measurements of aD01$\ldots$aD020, that were  chosen specifically along filaments bright in H$\alpha$, and then also in regions of local surface brightness minima, were used to determine line ratios and place the resulting values in diagnostic diagrams, such as the one of \citet{sabbadin1977} that distinguishes shock-excitation gas in SNR from photoionized gas in \hiiregs. The resulting values of log(H$\alpha$/[SII]) vs. log(H$\alpha$/[NII]) are plotted in Fig.~\ref{DIG_diagnostic}, along with the envelopes for SNR and \hiiregs that were determined empirically by the latter authors. The red plot symbols refer to samples aD03$\ldots$aD09, aD15$\ldots$aD17, and aD19$\ldots$aD20, that are associated with filaments of enhanced H$\alpha$ and [S$\,$II] emission, whereas the blue squares refer to the local minima of emission line surface brightness aD02, aD14, aD18, see Fig.~\ref{DIG_diagnostic}. There is a clear separation between shock-excitation in filaments (red circles), and photoionization in the undisturbed, low surface brightness regions (blue squares). It is interesting to note that aD14 is located within 20~pc of the giant \hiireg a66, i.e. relatively close to the photoionizing star cluster.

 \begin{figure*}[!h]
   \centering
  \begin{minipage}[b]{0.487\textwidth}
       \includegraphics[width=\hsize,bb=45 30 500 480,clip]{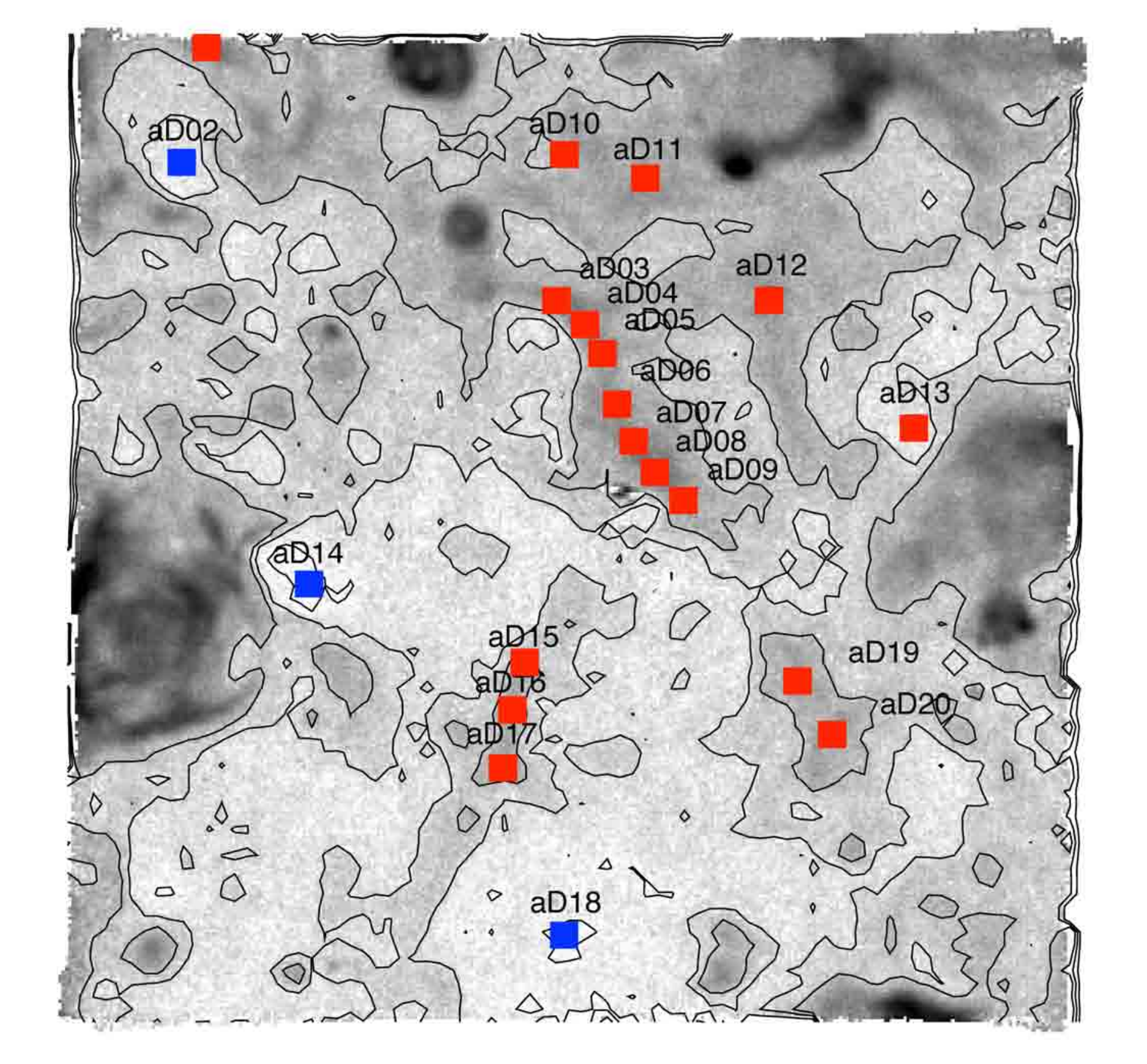}
   \end{minipage}
    \hfill
  \begin{minipage}[b]{0.487\textwidth}
        \includegraphics[width=\hsize,bb=70 360 420 685,clip]{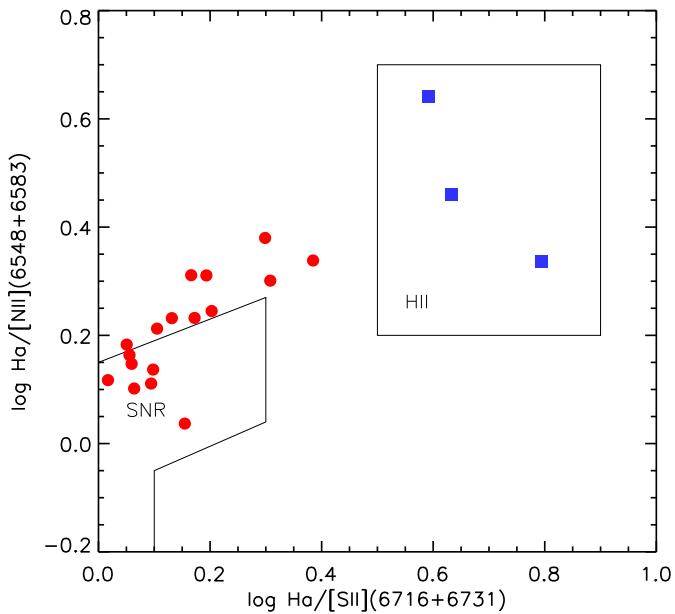}
  \end{minipage}
    \caption{\bf Left: Continuum-subtracted [S$\,$II] image of diffuse ionized gas in field (a). Contours are set to surface brightness levels of 0.5, 1.0, and $1.5\times10^{-17}$erg~cm$^{-2}$~s$^{-1}$~arcsec$^{-2}$. Right: Diagnostic diagram to distinguish photoionization from shock excitation according to \citet{sabbadin1977}. Red plot symbols correspond to shock ionized filaments aD03...aD09, aD15...aD17, etc., and blue symbols to photoionized local surface brightness mimima aD02, aD14, aD18 (see text). 
    }
    \label{DIG_diagnostic}
 \end{figure*}

Therefore the question arises, whether, perhaps, the emission of DIG is not dominated by shock excitation owing to SNR and stellar winds of massive stars, rather than photoionization. We felt compelled to study this more closely, because already  \citet{reynolds1985a} who had observed the DIG in the Milky Way using the Wisconsin Fabry-Perot Spectrometer, pointed out a high [S$\,$II]/H$\alpha$ ratio, typical for shock excitation, and concluded that shock ionization was likely to be at work,  or perhaps a combination of photo- and shock ionization. The model of \citet{wood2010}, however, demonstrated, that also photoionization alone could possibly be the relevant mechanism, owing to the patchy and turbulent nature of the ISM. 

As a first test towards a more thorough future quantitative analysis, we have used the region-of-interest feature of the {\textsc P3D} tool to measure the flux emerging from filaments likely to emit shock-excited features in proportion to flux from the remaining areas in field (a) outside of \hiiregs to obtain an order of magnitude estimate of the fraction of DIG emission that might be due to Lyman continuum photons leaking from \hiiregs, rather than due to shock excitation. As a result, we find a total H$\alpha$ flux of 1.1$\times10^{-14}$erg/cm$^2$/s for the filaments (5342 spaxels coadded), and a total of  8.8$\times10^{-15}$erg/cm$^2$/s for the remaining area (9934 spaxels). In essence, the two contributions are of the same order of magnitude, while it is worth mentioning that one cannot exclude that even the latter may be, to some fraction, also due to shocked gas.

   \begin{figure}[!b]
   \centering
    \includegraphics[width=\hsize,bb=45 30 500 480,clip]{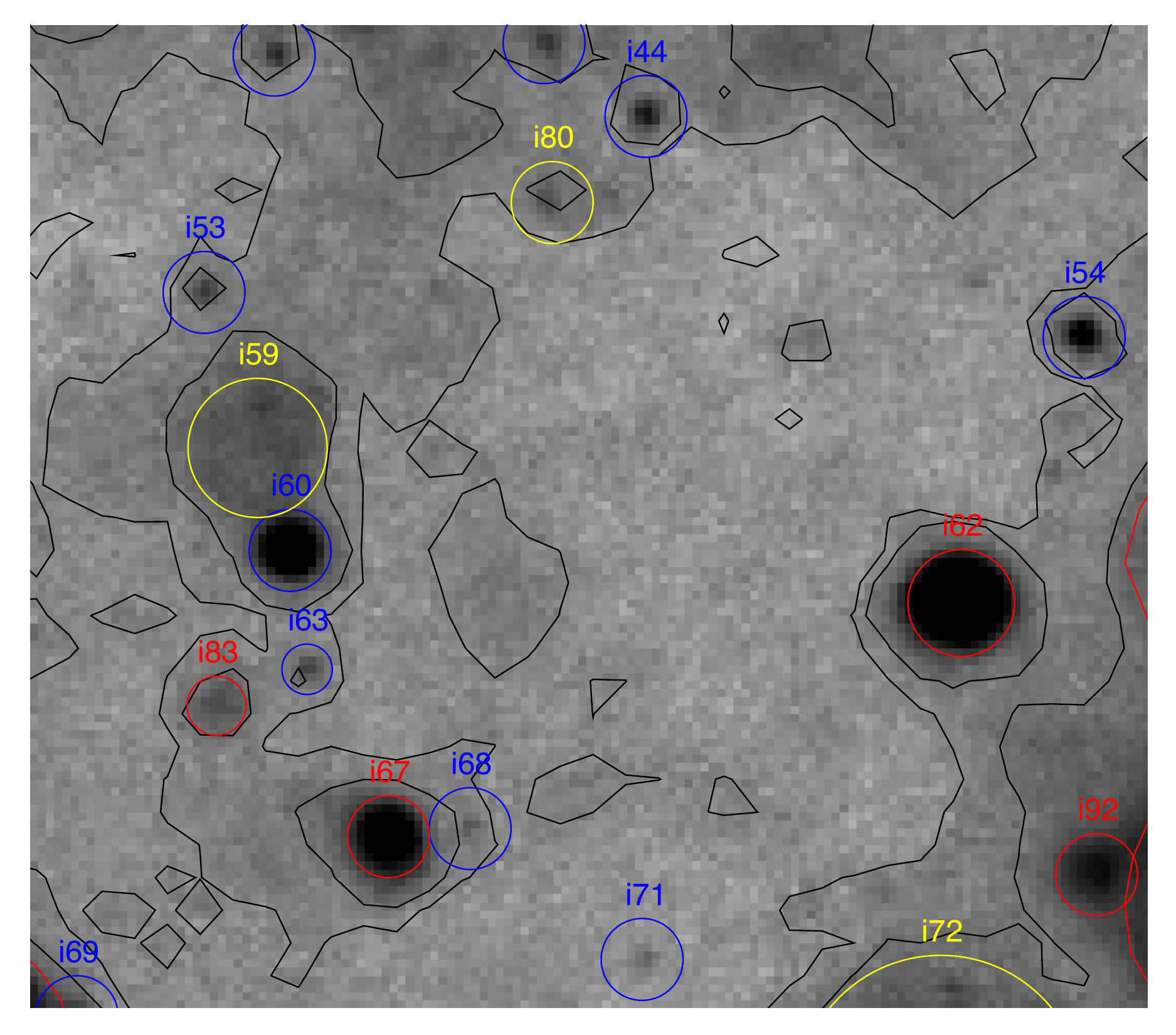}
    \caption{H$\alpha$ image around the LBV candidate i60 (blue circle to the left) in field (i), covering $245\times320$~pc$^2$. Red and yellow circles indicate \hiiregs and SNR, respectively. Contours at surface brightness levels of 1.5 and $2\times10^{-17}$erg~cm$^{-2}$~s$^{-1}$~arcsec$^{-2}$. }
    \label{LBV}
 \end{figure}

We note in passing that bright blue field stars without a noticeable \hiireg, also emission line stars that possibly represent massive luminous stars, show no prominent enhancement of DIG emission in their vicinity that would be comparable to the intensity of the filaments. An example is shown in Fig.~\ref{LBV}, where the LBV candidate star i60 is seen in H$\alpha$ in comparison to \hiiregs and SNR within a radius of $\sim$200~pc. This star emits only in H$\alpha$ a flux of $6\times10^{-16}$erg/cm$^2$/s, which is more than half of the total H$\alpha$ a flux of the entire optically thick \hiireg i62 (to the right in Fig.~\ref{LBV}). If the large width of the H$\alpha$ line (5.3~\AA~FWHM, extended wings) is indeed an indication of a strong stellar wind, the star is likely extremely hot and luminous, probably outshining any of the O stars that are ionizing the  \hiiregs in this field. Since there is no indication of an \hiireg around i60, one would expect significant leakage of Lyman continuum photons that should ionize the neutral hydrogen known to exist in NGC\,300 \citep{westmeier2011}. The contour levels in Fig.~\ref{LBV} show no evidence that this would be the case (the object i59 to the north is a shock-ionized SNR).


Given the interest in understanding the nature of DIG, we are planning a future paper to extend this analysis to a study of the entire set of pointings in NGC\,300.

 \begin{figure*}[!th]
   \centering
  \begin{minipage}[b]{0.487\textwidth}
           \includegraphics[width=\hsize,bb=28 0 500 370,clip]{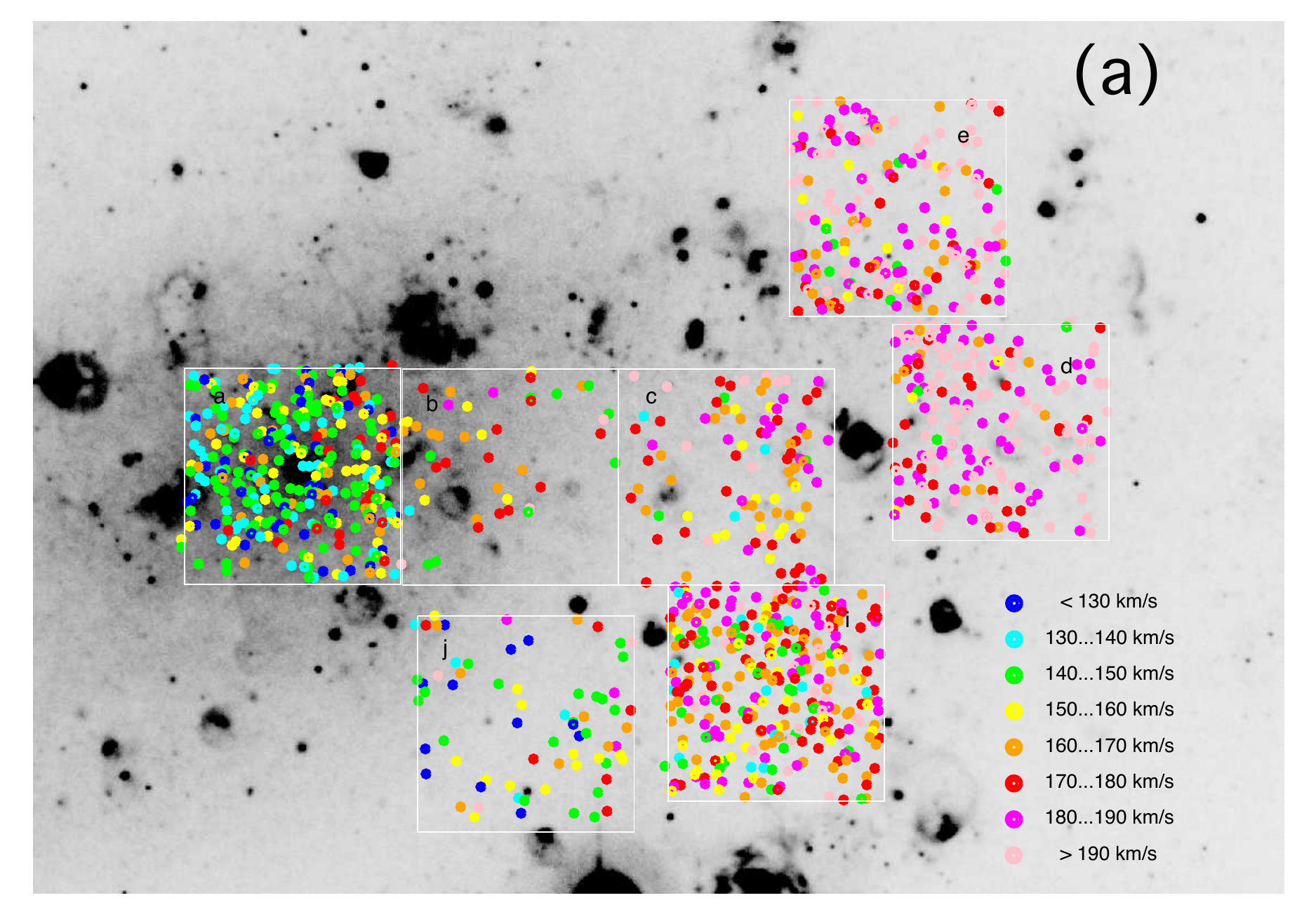}
  \end{minipage}
    \hfill
  \begin{minipage}[b]{0.487\textwidth}
            \includegraphics[width=\hsize,bb=35 20 750 580,clip]{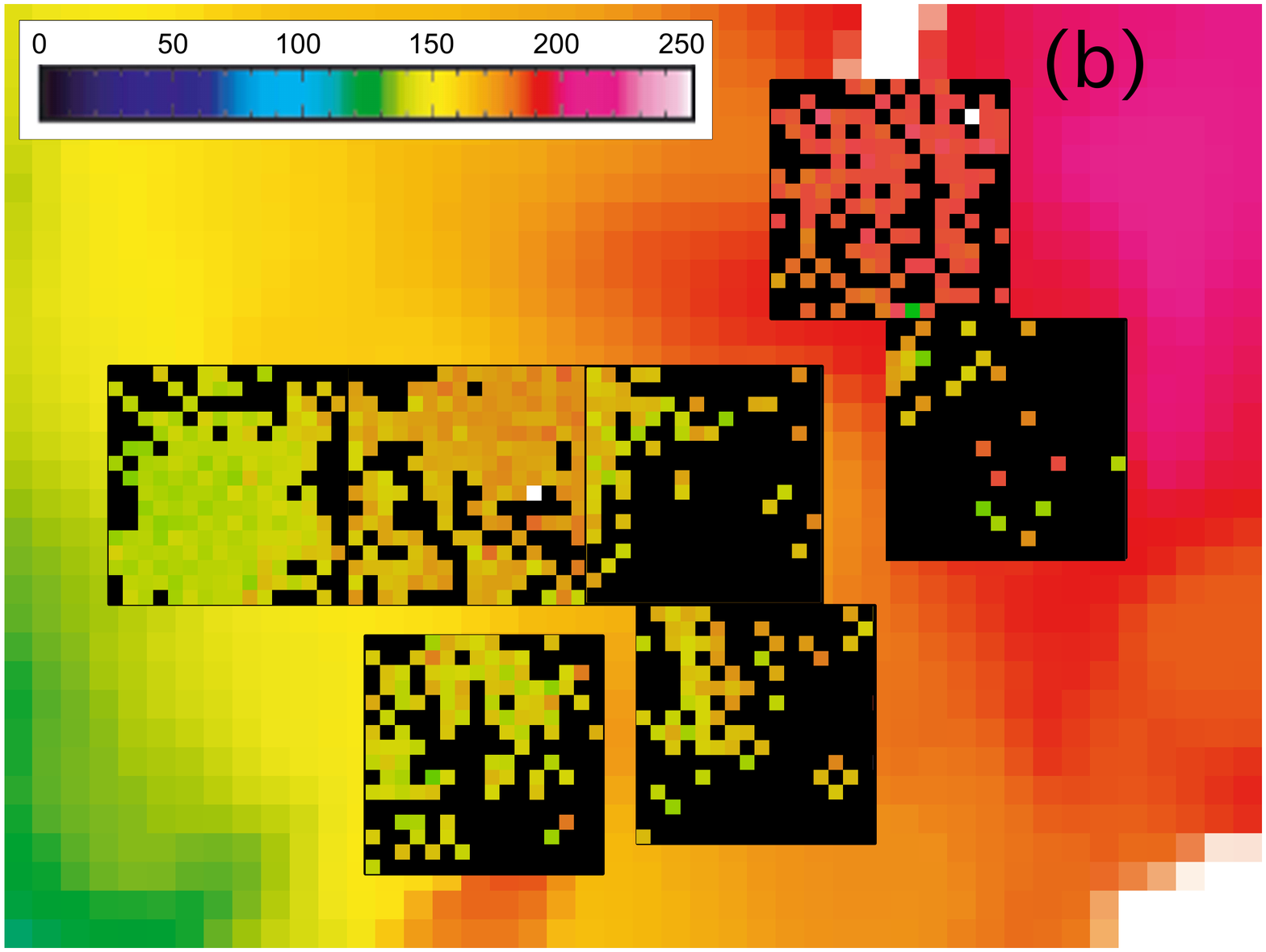}
  \end{minipage}
    \begin{minipage}[b]{0.487\textwidth}
           \includegraphics[width=\hsize,bb=28 0 500 370,clip]{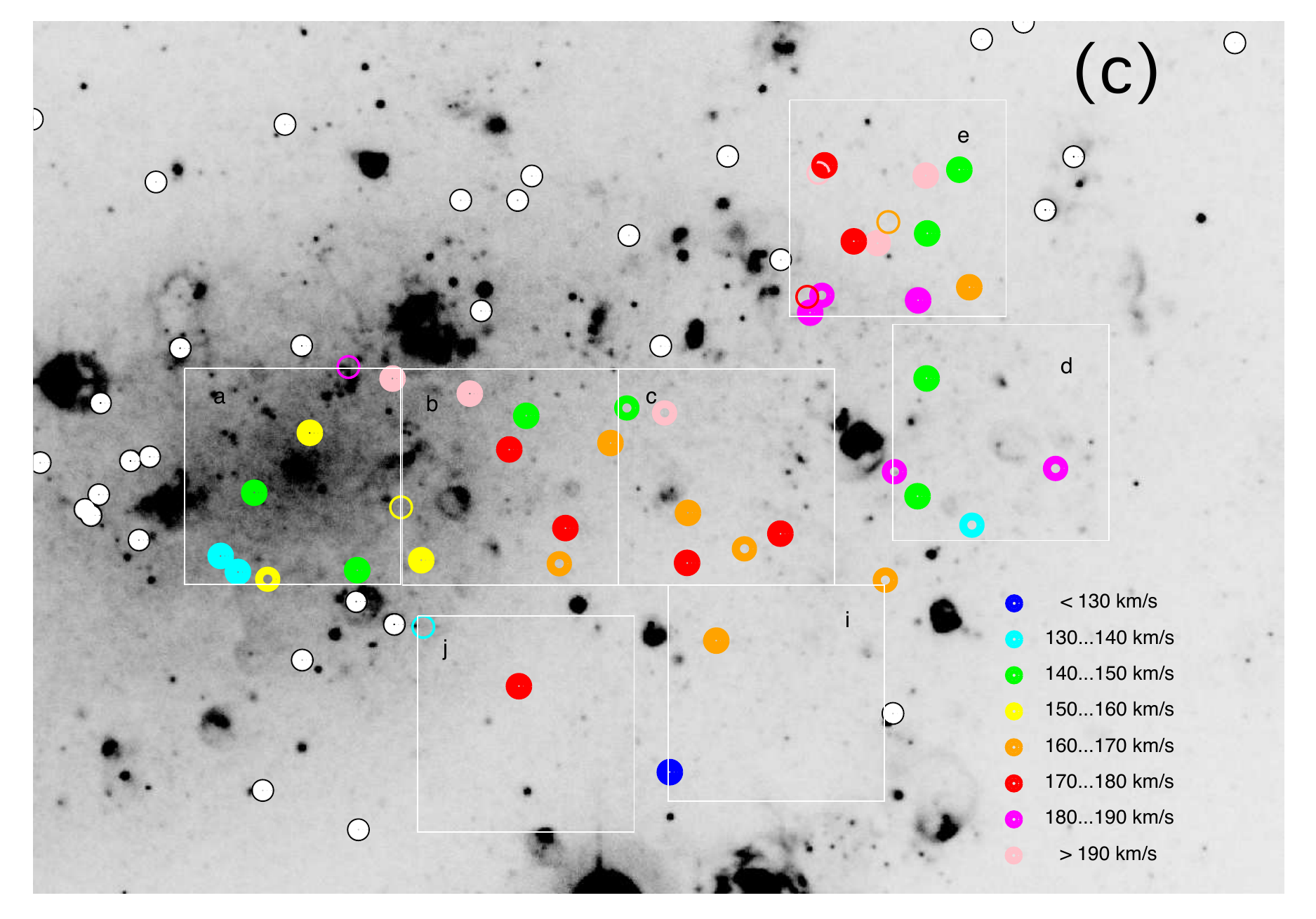}
  \end{minipage}
    \hfill
  \begin{minipage}[b]{0.487\textwidth}
            \includegraphics[width=\hsize,bb=35 20 750 580,clip]{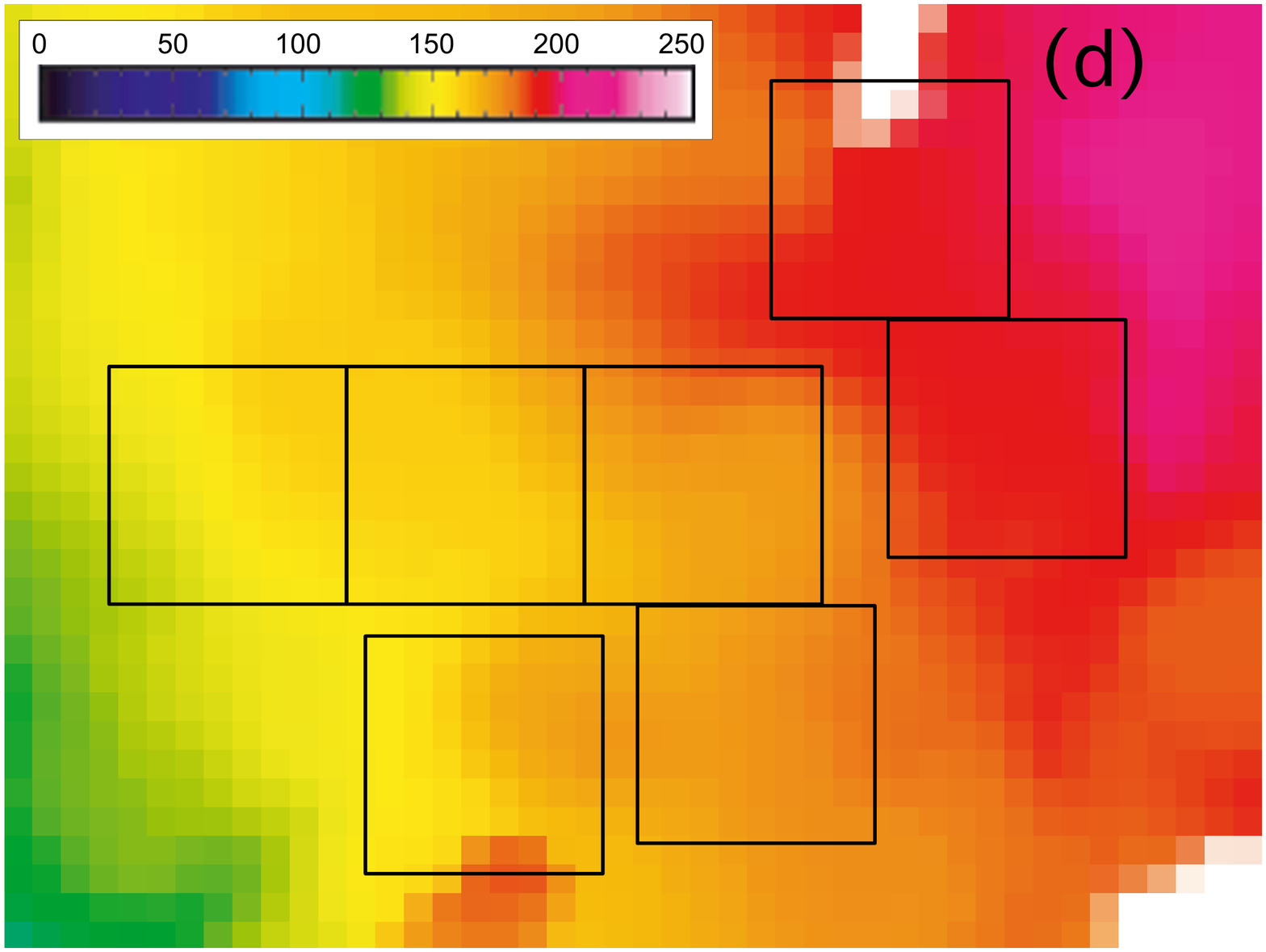}
  \end{minipage}  
    \caption{Kinematics in NGC\,300. Panel (a): radial velocities determined for individual stars; panel (b): radial velocities obtained    
    from unresolved background tiles produced by {\textsc PampelMUSE}; panel (c): radial velocities determined for PNe, errors coded
    with plot symbols (full circles  $<$10~km~s$^{-1}$, thick rings  $<$20~km~s$^{-1}$, thin rings $>$20~km~s$^{-1}$); 
    panel (d): reference velocity field reproduced from \citet{hlavacek2011}. Note that the velocity field colour coding is similar to, but 
    differs slightly from the colour code for point sources in panels (a) and (c). 
    }
    \label{kinematics}
 \end{figure*}

\subsection{Kinematics}
\label{discuss_kinematics}
As highlighted in $\S$\ref{subsection_stars}, the analysis with ULySS and the {\textsc P3D} tool has provided us with LOSV information for stars and emission line sources. The results from our measurements are plotted in Fig.~\ref{kinematics}. The panel (a) shows the LOSV of individual stars that were extracted with {\textsc PampelMUSE} with a colour code that is explained in the legend, stretching from below 130~km~s$^{-1}$ to above 190~km~s$^{-1}$ in bins of 10~km~s$^{-1}$. For better orientation, the symbols are plotted over the WFI H$\alpha$ image shown already in Fig.~\ref{NGC300_Ha}. One can immediately see that on average the stellar LOSV increases steadily with galactocentric distance from a median at 144~km~s$^{-1}$ near the nucleus to around 200~km~s$^{-1}$ at the extreme north-west. We have compared our results with the literature and reproduced to this end the  velocity field from \citet{hlavacek2011} as projected onto our fields (panel (d) in Fig.~\ref{kinematics}). This data was obtained from observations with a scanning Fabry-Perot instrument that has sampled the H$\alpha$ line in emission and a post-processing that produced a smoothly varying velocity field without gaps. They found their results to agree with H$\,$I data from \citet{puche1990} and \citet{westmeier2011} concerning the systemic velocity of $144\pm2$km~s$^{-1}$, and the distribution of the velocity field. As the sampling of the H$\,$I maps is rather coarse (10~arcsec and 30~arcsec grids, respectively), we have preferred a direct comparison to the  H$\alpha$ velocity map. The stellar LOSV follows well the H$\alpha$  velocity field, with values around the systemic velocity close to the nucleus in field (a), out to 200 km~s$^{-1}$ in fields (d) and (e). This result should be taken with some caution because only the stars in field (i) have been fully analyzed including visual inspection and quality parameter assignment at this stage, such that the output of ULySS fits were taken at face value for the remaining fields, only excluding obvious outliers from failed fits. From our experience with the exercise in field (i) that has shown a robust behaviour with regard to radial velocity determinations, we are however expecting no fundamental change of the picture after full analysis of the remaining fields. 

Since we have obtained ULySS fits also for the local background spectra as an output from {\textsc PampelMUSE} for each data\-cube, there are also radial velocity maps  for the unresolved stellar population. For the parameter setting of our {\textsc PampelMUSE} runs, we have obtained unresolved background spectra for tiles with sizes of $20\times20$ spaxels, i.e. areas of 4"$\times$4" on the sky. However, these measurements are hampered by contamination from nebular lines for tiles at or close to the location of bright \hiiregs, superbubbles, shells, etc. that mostly rendered a fit impossible. For the scope of this paper, we have not attempted to remove the disturbing emission, e.g. following the discussion of \citet{falcon2004} for the measurement of emission-free stellar kinematics. We reserve this step to the future analysis of the complete dataset with the pPXF tool \citep{cappellari2004}, that would also address a stellar population analysis of the background component.
The present results are plotted in panel (b) of  Fig.~\ref{kinematics} with the same colour-code as the H$\alpha$ velocity map, where black areas indicate that no ULySS fit was obtained due to bright nebular emission lines. Despite the fact that significant areas are not covered with LOSV information, for the tiles that do carry information the velocities are almost indistinguishable from the H$\alpha$ map. The predominance of 
G type main sequence star spectra of the ULySS fits for the background in field (a), changing to predominantly B type spectra for the fields associated with the spiral arm is suggestive of a population age $\ge$ 10 Gyr  near the nucleus \citep{bertelli2008} and ongoing star formation elsewhere, prompting us to perform a detailed analysis in a forthcoming paper.

Finally, the radial velocities measured for PNe are plotted in panel (c) of Fig.~\ref{kinematics} using the same colour code as for the stars. 
We have also added the PN candidates from \citet{pena2012} as white circles, for which we have, however, no velocity information. The first striking feature has already been addressed above, namely the unequal distribution of PN detections across our seven fields, despite the fact that we have estimated completeness limits of m$_{5007} = 27\ldots28$, depending on the seeing attained in each field. The next observation is the apparent scatter of velocities, which for most objects track the general distribution of the H$\alpha$ velocity map. However, there is a number of remarkable outliers: two high velocity PNe in the nuclear region: (a03) with v$_\mathrm{rad}$=197$\pm$9~km~s$^{-1}$, (b06) with v$_\mathrm{rad}$=244$\pm$3~km~s$^{-1}$, and at larger galactocentric distances four low velocity PNe: v$_\mathrm{rad}$=139$\pm$11~km~s$^{-1}$ (d63), 144$\pm$7~km~s$^{-1}$ (d79), v$_\mathrm{rad}$=145$\pm$4~km~s$^{-1}$ (e02), and 143$\pm$3~km~s$^{-1}$ (e20), respectively, as well as the extreme case of PN i84 with v$_\mathrm{rad}$=108$\pm$8~km~s$^{-1}$. It is going to be interesting to see after the final analysis of our full dataset whether a similar behavior can be found for AGB stars as the progenitors of PNe. We also speculate that we may be able to find evidence for the existence of migration along the leading and trailing edges of spiral arms as suggested by the predictions from numerical simulations by \citet{grand2016}, although it would be premature at this stage to draw any conclusions from our as yet provisional dataset.

\subsection{Serendipitous discoveries}
As mentioned in ${\S}$\ref{section_analysis}, the process of data analysis has involved significant human interaction and visual control. As an asset, the detailed inspection of our data has enabled the discovery of objects that were not initially targeted, or expected. One such example is the WR star found in field (d) as reported in ${\S}$\ref{emStars} that immediately became apparent upon inspection of the He$\,$II image of that region. This star was not specifically classified or ear-marked by ULySS. Other examples are the detection of blue emission line stars representing the rare class of hot massive stars whose discovery conventionally would have involved the two-step procedure of imaging and follow-up spectroscopy, the discovery of symbiotic star candidates, and the detection of carbon stars.

\subsubsection*{Variable sources}
Although the layout of our observations was not directly addressing photometric or spectroscopic variability, multi-epoch MUSE surveys inherently offer the capability of detecting variable sources. Incidentally, by comparison of MUSE data with HST images, we have serendipitously made such a discovery:
upon the investigation of the apparent foreground star ID 32224 as described in $\S$\ref{subsection_stars}, the visual inspection of the stellar images shown in Fig.~\ref{foreground} revealed that the star denoted B in the HST image (epoch 2005) is in fact absent from the MUSE image (epoch 2015). Apparently, we have picked up a transient or variable star (perhaps a Mira?), whose brightness has faded over a decade by at least 2 orders of magnitude, reminiscent of a similar case of a red supergiant that was described as a failed supernova by \citet{adams2017}.  A possible alternative is that star B has moved closer to the position of star A. In that case B would be the actual foreground star with a rather high proper motion. Unlike an ongoing MUSE GTO program on globular clusters \citep{giesers2018}, we have as yet not attempted to obtain multi-epoch observations in NGC\,300 with the purpose of detecting radial velocity or spectrophotometric variability. However, such an observing strategy of future MUSE surveys in nearby galaxies would open unique opportunities, e.g., to discover O star binaries. 

 \begin{figure}[t!]
   \centering
    \includegraphics[width=\hsize,bb=45 200 755 570,clip]{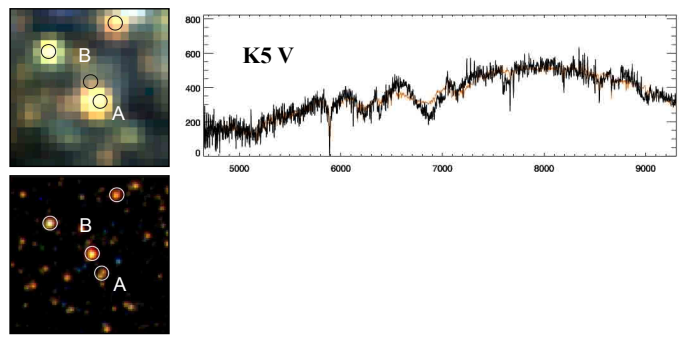}
    \caption{Foreground star ID~32224 denoted A, from field (i). The nearby peculiar red star B (ID~1003, F475W=24.45, F606W=22.82,
    see Table~\ref{stars-i}), that in the HST    
    image appears even 1.4~mag brighter than A,  is not detectable in our MUSE image. The less than perfect fit of the K5V spectrum from A is 
    understood as an effect of blending.
    }
    \label{foreground}
 \end{figure}

   \begin{figure}[h!]
   \centering
    \includegraphics[width=\hsize,bb=45 30 500 480,clip]{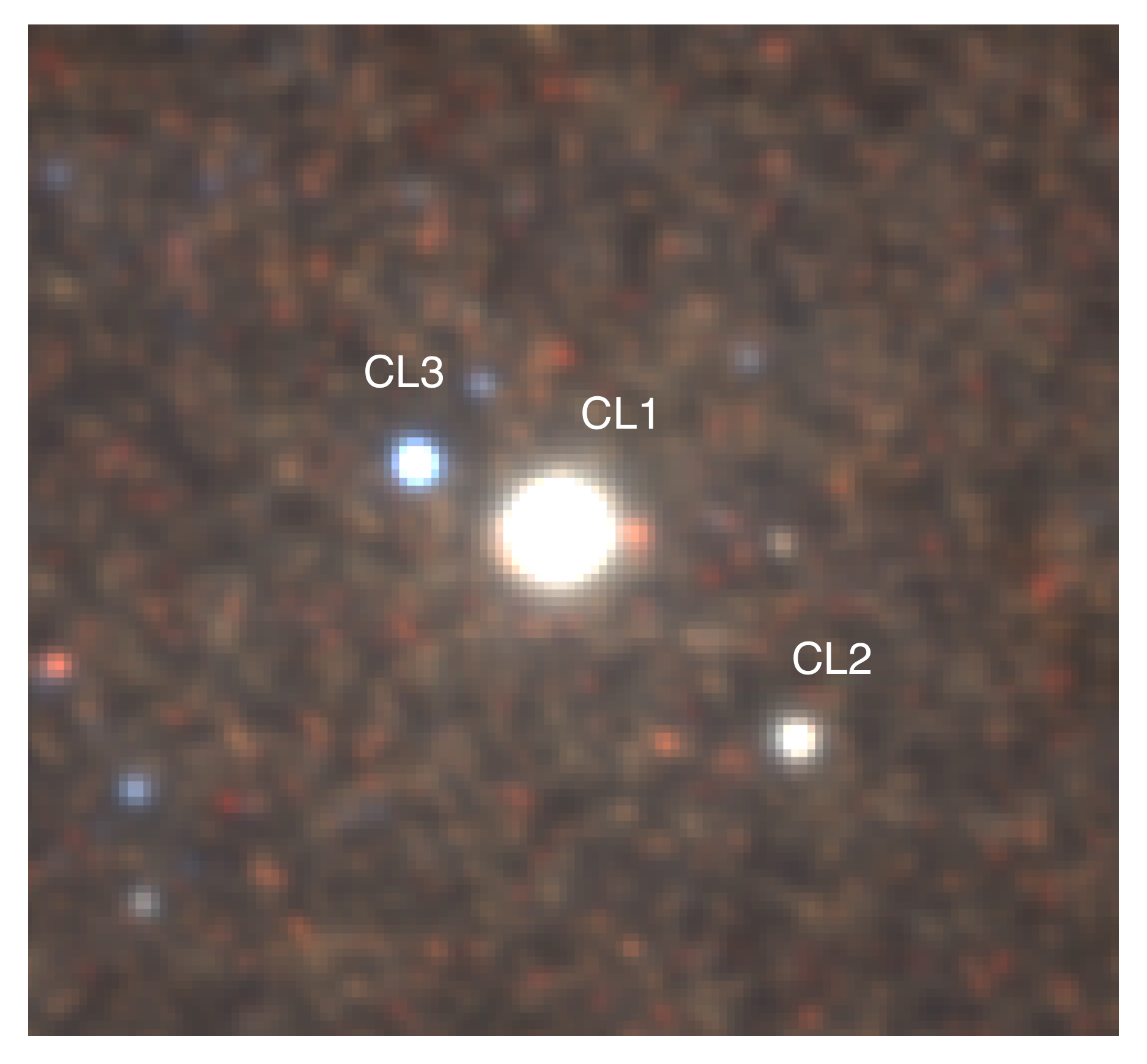}
    \caption{Reconstructed VRI image from MUSE datacube, showing the nuclear cluster CL1 and two newly discovered clusters in its vicinity. The FoV measures 245$\times$245~pc$^2$ (in projection).}
    \label{Clusters}
 \end{figure}

\subsubsection*{Nuclear star cluster}
Like most Scd galaxies, NGC\,300 has a nuclear star cluster at its center \citep{seth2008}, CL1 in Fig.~\ref{Clusters}. Nuclear star clusters are thought to grow dissipatively from in situ star formation \citep{milosavljevic2004} and/or by dissipationless accretion of migrating star cluster \citep{tremaine1975,antonini2013}. The at first glance point-like object CL2 in field~(a) attracted our attention, in that the spectrum is similar to the spectrum of the nuclear star cluster CL1 that is known to harbour a young stellar population \citep{walcher2005,carson2015}. Inspection of the corresponding HST F606W image revealed that the image of CL2 has a FWHM of 4.0~pixels, that is significantly wider than the PSF with a FWHM $\sim$2.2~pixels.  This confirms that we have found another, less luminous cluster at a projected distance of $\sim$71~pc to the SW of the nucleus.




The decomposition of the spectrum of the nuclear cluster  CL1 with ULySS yields major fractional contributions from spectral types G8V (66.1\%), A2IV (10.8\%), K2IIIb  (9.9\%), and v$_\mathrm{rad}$=139$\pm2$km~s$^{-1}$,  whereas the decomposition of the new cluster CL2 is reporting G8V (64\%), B0.5IV (19\%),  B9III (12\%), and AIa (30\%) as the major components, indicating an even younger population than the one of the nuclear cluster. The LOSV is determined as  v$_\mathrm{rad}$=119$\pm3$km~s$^{-1}$. 

There is another bright object with a distinct blue color in the vicinity of the nucleus, whose HST F606W image is resolved with a FWHM of 3.4~pixels. This new cluster CL3 is located to the NE of CL1 with a projected distance of 35~pc. The spectral decomposition in this case yielded B0.5IV (45\%),  G8V (32\%), B3Ib (22\%). The LOSV is measured to v$_\mathrm{rad}$=114$\pm10$~km~s$^{-1}$, a value that is significantly lower than the average of 144~km~s$^{-1}$ for the stellar population in this region, which is also the case for CL2. The presence of of a blue supergiant component, as well as a broad H$\alpha$ emission line in the composite spectrum, points to a cluster age of only a few million years. With a LOSV differential of 20~km~s$^{-1}$, the cluster would have travelled within a timespan of 1~Myr over a distance of no more than 20~pc on an inclined trajectory through the disk of NGC\,300. There is also no evidence for a recent merger or an interacting companion \citep{blandhawthorn2005}. 
Clusters C1, C2, and C3 are also visible in HST I band images of NGC\,300 by \citet{boeker2002}. The latest compilation of nuclear star clusters in late type galaxies is published in \citet{georgiev2014}. The stellar clusters orbiting the nucleus together with the presence of young stars motivates a more detailed study of the assembly of the nuclear star cluster in NGC\,300.

\begin{table}[t!]
 \caption{Background galaxies (redshifts with a colon are uncertain).}             
\label{bg_galaxies}      
\centering
\begin{small}          
\begin{tabular}{ l  r r   c c   l  }     
\hline\hline
ID          &    X             &    Y          &    RA(J2000)    &    Dec(J2000)  &      ~~~z       \\
\hline
aZ01   &  123 &   156 &  00:54:54.303 &  -37:41:04.8  &  0.3567 \\    
aZ02a  &  204 &    19 &  00:54:52.934 &  -37:41:32.3  &  1.0066 \\    
aZ02b  &  206 &    22 &  00:54:52.903 &  -37:41:31.8  &  1.0071 \\    
aZ03a  &  118 &    48 &  00:54:54.399 &  -37:41:26.6  &  0.2730 \\    
aZ03b  &  123 &    44 &  00:54:54.315 &  -37:41:27.4  &  0.2737 \\    
aZ04a  &  262 &  226 &  00:54:51.966 &  -37:40:50.9  &  0.7468 \\   
aZ04b  &  259 &  229 &  00:54:52.023 &  -37:40:50.3  &  0.7465 \\   

bZ01    &  196 &  308 &  00:54:48.027 &  -37:40:34.6  &  0.4996 \\   
bZ02a  &  129 &  268 &  00:54:49.159 &  -37:40:42.6  &  0.5737 \\   
bZ02b  &  123 &  272 &  00:54:49.256 &  -37:40:41.8  &  0.5737 \\   
bZ03a  &  281 &  242 &  00:54:46.591 &  -37:40:47.8  &  1.1555 \\   
bZ03b  &  284 &  244 &  00:54:46.548 &  -37:40:47.3  &  1.1544 \\   

cZ01    &  198 &   63 &  00:54:44.150 &  -37:41:24.0  &  0.4543 \\  

dZ01   &  297 &  258 &  00:54:34.785 &  -37:40:32.0  &  1.1885 \\   
dZ02   &  255 &  149 &  00:54:35.495 &  -37:40:53.9  &  1.0623 \\   
dZ03   &  274 &   49 &  00:54:35.180 &  -37:41:13.9  &  0.7449 \\    
dZ04   &  264 &  207 &  00:54:35.344 &  -37:40:42.3  &  1.0611 \\   
dZ05   &  298 &   57 &  00:54:34.765 &  -37:41:12.3  &  1.1890 \\    
dZ06   &   24 &  140 &  00:54:39.386 &  -37:40:55.7  &  0.7990 \\    
dZ07   &  171 &   24 &  00:54:36.911 &  -37:41:18.9  &  0.7046 \\    
dZ08   &  297 &  258 &  00:54:34.794 &  -37:40:32.0  &  1.1885 \\   

eZ01   &  161 &  138 &  00:54:39.442 &  -37:39:54.1  &  4.4718: \\  
eZ02   &    31 &  145 &  00:54:41.617 &  -37:39:52.8  &  0.9186: \\   

iZ01    &  255 &  280 &  00:54:40.776 &  -37:41:40.2  &  0.2548 \\   
iZ02    &  246 &  168 &  00:54:40.921 &  -37:42:02.4  &  0.1275 \\   
iZ03a  &  163 &  108 &  00:54:42.321 &  -37:42:14.4  &  1.0346 \\   
iZ03b  &  162 &  107 &  00:54:42.337 &  -37:42:14.8  &  1.0344 \\   
iZ04a  &    61 &    84 &  00:54:44.033 &  -37:42:19.2  &  1.0946 \\   
iZ04b  &    67 &    92 &  00:54:43.939 &  -37:42:17.7  &  1.0926 \\   
iZ05a  &    10 &    58 &  00:54:44.898 &  -37:42:24.5  &  0.9558 \\   
iZ05b  &    16 &    53 &  00:54:44.798 &  -37:42:25.5  &  0.9563 \\   
iZ05c  &    22 &    52 &  00:54:44.704 &  -37:42:25.6  &  0.9570 \\   
iZ06    &  238 &  258 &  00:54:41.050 &  -37:41:44.6  &  1.0847 \\   
iZ07    &    33 &  212 &  00:54:44.510 &  -37:41:53.7  &  1.1889 \\   
iZ08a  &  153 &  194 &  00:54:42.488 &  -37:41:57.3  &  1.2187 \\   
iZ08b  &  155 &  193 &  00:54:42.455 &  -37:41:57.5  &  1.2193 \\   

jZ01   &    57 &  194 &  00:54:50.029 &  -37:42:06.7  &  0.8271  \\   
jZ02   &  298 &  160 &  00:54:45.959 &  -37:42:13.4  &  1.3362  \\   
jZ03   &    83 &    69 &  00:54:49.584 &  -37:42:31.6  &  4.3093: \\   
jZ04   &    57 &  146 &  00:54:50.029 &  -37:42:16.1  &  0.5483  \\   

\hline     
\end{tabular}
\end{small}   
\end{table}

\subsubsection*{Background galaxies}
During the extensive use of the {\textsc P3D} visualization tool, we incidentally noticed emission line signatures in the display of colour coded stacked spectra, that did not seem to match any familiar pattern. 
As it turned out, we discovered several background galaxies in each of our fields, with redshifts of $z = 0.13\ldots1.33$, and even two candidates for Ly$\alpha$ emitting galaxies at redshifts $z > 4$. These objects are listed in Table~\ref{bg_galaxies}. IDs with an index a,b,... indicate systems with kinematically distinct components, that could either correspond to a rotating disk, or otherwise to a group of galaxies. Although the study of background galaxies was not considered an immediate objective of this work, we note that such galaxies, in particular AGN, may serve as astrometric references for proper motion studies with future instruments delivering very high astrometric resolution and accuracy, such as MICADO for the E-ELT \citep{davies2016}.

\section{Summary and conclusions}

This paper is presenting the first results obtained from MUSE observations of the nearby galaxy NGC\,300 in the wide field mode, utilizing the novel technique of crowded field 3D spectroscopy. The dataset used at this stage comprises five complete pointings at the nucleus of the galaxy and adjacent regions to the west with a total exposure time of 1.5~h per field, two of which are compromised by poor seeing. Two more pointings were as yet incomplete with exposure times of 0.5 and 1~h, respectively. The major purpose of this first paper is to demonstrate the feasibility of crowded field 3D spectroscopy in nearby galaxies, and to inform the astronomical community about the legacy value of MUSE data\-cubes, in particular concerning the multitude of scientific results over a variety of different objects one can expect from such data. 


\begin{table*}[!th]
 \caption{Summary of measurements for individual objects (a colon ":" indicates preliminary analysis)}             
\label{Summary}      
\centering
\begin{tabular}{ l  c c c c c c c r }  
\hline\hline
field                                               &   (a)     &   (b)   &    (c)   &   (d)    &   (e)     &   (i)    &   (j)     &   total     \\
\hline
Seeing                                          &  0.7''    &  1.2''   &  1.0''  &  0.8''   &  0.75''  &  0.6''   &  0.85''  &          \\
\hline
Planetary nebulae (bona fide)      &    5      &     7    &    6    &    4      &      9    &    3    &    2     &     36~   \\
Planetary nebula candidates        &    4      &     0    &    0    &    1      &      4    &    0    &    0     &       9~   \\
{\hiiregs}                                       &   10     &   11    &     5   &   13     &      4    &    13   &    5     &     61~   \\
 compact  {\hiireg} candidates     &    8      &     4    &     5   &   19     &      5   &      2   &    8      &     51~    \\
Supernova remnant candidates  &  14      &     5    &     3   &     5     &      3    &     6   &    2      &     38~    \\
Emission line stars                      &  18      &     4    &     4   &  15     &    30    &   40   &    7      &    118~    \\
Background galaxies                   &   4       &     3    &     1   &    8     &      2    &     8   &    4      &     30~    \\
Stars                                            &  445:  &  77:  & 152:  &  265: & 299:  &  517  &   91:   &  1846:  \\
\hline     
\end{tabular}
\end{table*}

Table~\ref{Summary} summarizes the different classes of objects for which individual spectra and images were obtained.
The total number of objects, such as stars, PNe, \hiiregs, SNR, etc., including background galaxies, amounts to 2187. At this early stage of exploration of the new method of crowded field 3D spectroscopy, we have decided to visually inspect each and every object, assigning quality flags, and reassuring ourselves to not be trapped by immature or not fully reliable automatic algorithms. Concerning the classification of stars, this task has so far fully been accomplished for field (i).  Numbers flagged with a colon for the remaining fields in Table~\ref{Summary}  indicate that the classification is preliminary.
However, significant parts of the analysis have already been automized: the extraction of stellar spectra with {\textsc PampelMUSE}, and spectral type classification as well as radial velocity measurements with ULySS and GLIB fits. The significant level of human interaction at this stage is time-consuming and cumbersome. We are actively working on the improvement of the currently available tools and expect that future developments will enable fully automatic procedures, which is a pre\-requisite for the analysis of surveys larger than the present study. 

\vspace{1mm}

As the major outcome of this pilot study, that can be considered a proof-of-principle for MUSE crowded-field 3D spectroscopy in nearby galaxies, we summarize the following results:

\begin{enumerate}
\item It is possible to decompose nearby galaxies into individual giant stars and an unresolved component of faint background stars.
From the analysis of field (i), which is the best one in our dataset in terms of seeing, we present spectra for a total of 517 stars whose  S/N was sufficient for spectral type classification and the determination of radial velocities ( S/N~$\ge$~3). As inferred from simulations, the limiting S/N for a reliable spectral type classification was found to be in the range of 3\ldots5, depending on spectral type.
From the remaining, as yet uninspected fields, 1329 stars have yielded spectra, part of which were used to measure  preliminary radial velocities, subject to quality control and confirmation in the near future. Of the immediate findings from this study, we point out the discovery and classification of $\sim$100 blue supergiants, tens of yellow giants/supergiants, and more than 300 K and M giants/supergiants in field (i). The radial velocities measured for individual stars follow the rotation curve of the galaxy in accord with published velocity fields for H$\alpha$ and H\,I. Likewise, the velocity field determined from ULySS fits to the unresolved background stellar population is consistent with this data.
\vspace{3mm}

\item From the single field (i) with an exposure time of 1.5~h, we identified a total of 23 carbon stars.  This number compares with a total of 115 carbon stars found in the disk of M31 \citep{hamren2015} and the surrounding halo and satellite dwarf galaxies  \citep{hamren2016}, that were obtained from a total of 24762 stellar spectra secured over 10 years worth of Keck DEIMOS observations as part of the SPLASH survey. 
\vspace{3mm}

\item In addition to the stars selected and extracted on the basis of the ANGST catalogue, we found 82  blue emission lines stars, most of which show broad H$\alpha$ emission lines, that are characteristic for evolved massive stars with strong stellar winds, e.g. LBV stars. Such stars are very rare and difficult to find \citep{massey2015}. For comparison, \citet{massey2016} reported the discovery of 4/5 LBVs and 0/1 WR star in M31/M33, respectively. Point sources that show a continuum spectrum and unresolved Balmer lines in emission may represent circumstellar shells around stars that are small enough to remain spatially unresolved for ground-based observations. 
\vspace{3mm}

\item  Another unexpected discovery amongst the detected emission line stars is the finding that some of the emission line point sources are apparently associated with AGB stars, some of which are interpreted as the discovery of candidate symbiotic stars. We identified a total of 4 symbiotic star candidates that would match such a spectral pattern. It is interesting to note that according to the NASA HEASARC archive merely 188 such stars are known in the Milky Way \citep{belczynski2000}. According to \citet{mikolajewska2004}, 6 such objects are known in the SMC, and 8 objects in the LMC.
\vspace{3mm}

\item Both the analysis of stellar spectra through automized fits to the MIUSCAT and GLIB libraries, and the interactive measurement of emission line objects using {\textsc P3D}, have yielded LOSV estimates for hundreds of individual objects. While the stars match very well the velocity field of H$\,$II, which should be characteristic for the kinematics of the disk, amongst the sample of PNe, that generally follow the same trend, there are several  distinct outliers with extreme radial velocities that may be attributed to halo objects. The velocity field, that was obtained from local background estimate spectra as output from the {\textsc PampelMUSE} code, although as yet uncorrected for gaseous emission systematics, is also indistinguishable from the H$\,$II velocity field. 
\vspace{3mm}

\item From the complete analysis in field (i), so far only one foreground star was found on the basis of luminosity class and radial velocity, in accord with the expectations.
\vspace{3mm}

\item Amongst emission line point sources we have registered 36 bona fide PNe (out of which 13 objects were known from previous studies), and 9 more uncertain candidates at faint magnitudes. Extragalactic PNe can easily be confused with unresolved compact \hiiregs that show emission line spectra similar to the ones of PNe. From previous PN surveys in NGC\,300, we found several cases where objects were misclassified as PN and better match the criteria of cHII. In total, we registered 61 cHII. 
\vspace{3mm}

\item We discovered a total of 53 new extended \hiiregs and confirmed 8 that were already catalogued previously. On the basis of the IPM technique \citep{pellegrini2012}, we assigned categories of optically thick or thin \hiiregs, and shells. 
\vspace{3mm}

\item Supernova remnants were identified on the basis of the [S$\,$II]/H$\alpha$ line ratio. We registered a total of 38 SNR candidates, with sizes ranging between a few and more than 100~pc. Some of the SNR candidates exhibit very low surface brightness, and therefore a possible confusion with the DIG. \vspace{1mm}

\item We measured DIG emission lines in the nuclear region of NGC\,300 down to surface brightness levels as low as a few $10^{-18}$erg~cm$^{-2}$~s$^{-1}$~arcsec$^{-2}$ and identified filamentary structures that show diagnostic emission line ratios that are characteristic for shock excitation, whereas only a few areas with very low surface brightness were found whose spectra are characteristic for photoionization. 
\vspace{3mm}

\item Two young clusters within a projected distance of less than 75~pc from the nucleus of NGC\,300 were discovered.
\vspace{3mm}

\item Of serendipitous discoveries made in the process of data analysis, we mention that 28 background galaxies  with redshifts z=0.127$\ldots$1.33 were identified, that otherwise would likely have remained undetected from imaging surveys.
\end{enumerate}

It is obvious that these early results have merely scratched the surface of a potentially very powerful new tool to probe resolved stellar populations in nearby galaxies, a research topic that we are planning to expand in the future. 

The data products from this work will be made publicly available through the Strasbourg astronomical Data Center (CDS) and the MUSE website\footnote{http://muse-vlt.eu/science/}.

\section{Outlook}
As a technical shortcoming at the present stage, we have found that the {\textsc PampelMUSE} technique to account for contributions from the unresolved background as a constant average over  a finite area (the "tiles"), which works well for globular clusters, is a major limitation in regions with strong nebular emission lines. Of the stars with acceptable S/N, for which however no trustworthy classification could be obtained, the majority was found to be suffering from strong nebular contamination. We concluded that mainly massive stars or star clusters within bright \hiiregs are affected by this approximation. In previous experiments with the PMAS instrument published by \citet{roth2004}, we have separately modeled the 2-dimensional background surface brightness distribution in the continuum and in emission. For an instrument like MUSE that has a factor of $\sim$350 times more spaxels than PMAS, such an approach is computationally very expensive. The further development of {\textsc PampelMUSE} will specifically address this issue. 

Another limitation was imposed by the incomplete coverage of the HRD by the MIUSCAT and GLIB libraries, especially for hot stars, emission line stars, and also for carbon stars. We are planning to resolve this problem by complementing the libraries with simulations and new observational data. The MaStar library currently being planned as part of the SDSS-IV collaboration \citep{yan2017} may become a promising solution to this end.

Future work shall address the completed dataset of stellar spectra of all fields with secured quality checks,  including two new pointings at larger galactocentric distance that are not shown in this paper. We are also planning new observations of the fields that have suffered from poor seeing with support of the adaptive optics facility, which should substantially increase the number of useful stellar spectra. Moreover, we are planning to improve our methods towards automatic data analysis to more efficiently process the expected number of thousands of spectra. 

From the entire data set, we intend to study the stellar population from the nuclear region to the north-western spiral arm, including the leading and trailing interarm regions covered by our pointings, with the goal to infer the star formation history on the basis of individual stars and their relation to \hiiregs, SNR, and PNe, complemented by spectral synthesis modelling of the background of unresolved stars. Furthermore, from  a detailed kinematic study we expect insight into phenomena like migration, mergers, and runaway stars. Another challenging goal will be the determination of chemical abundances for individual stars, with interesting prospects concerning young and old stellar populations, e.g. blue and red supergiants vs. RGB and AGB stars.

Again on the basis of the complete dataset, we shall study the origin of the PNLF and influences on its shape owing to the underlying stellar population, in particular M and C stars as their progenitors. This goal has become more accute with the revised picture of post-AGB evolution from the latest models of \citet{miller2016}, that are three to ten times faster and $\sim$0.1-0.3 dex brighter than the previous models by \citet{vassiliadis1994} and \citet{bloecker1995}, that were being used in many studies. 

With the hypothesis that DIG emission may be dominated by shock excitation, we are planning a follow-up paper to systematically investigate all of our pointings in more detail in order to obtain a  more global view of the characteristics and origin of DIG in different parts of the galaxy.

For the modest investment in observing time (9~h) the yield as summarized in Table~\ref{Summary} is impressive. It has been a fundamental goal of the development of the MUSE instrument to break the conventional sequence of photometry --- follow-up spectroscopy, and replace it with a single-step observation, that provides spectra and images from a single homogenous dataset. Beyond this already fundamental achievement, it is worthwhile to stress that not just quantitatively, but also qualitatively it has been only the format and performance of MUSE that has enabled the efficient use of crowded field spectroscopy, both in terms of multiplex advantage, as well as in utilizing the PSF fitting technique to deblend overlapping stellar images. Combined with high instrumental throughput, the light collecting power of an 8m telescope,  and the superb image quality of VLT-UT4, we conclude that MUSE is an ideal instrument to study (partially) resolved stellar populations, gas, and dust in nearby galaxies. We expect that this capability will be even more enhanced with the new adaptive optics facility for VLT-UT4 that has become available recently. Our findings also strongly support new survey telescope concepts suitable for massively-multiplexed spectroscopy as presented by \citet{pasquini2016}.

\begin{acknowledgements} 
The authors would like to thank the anonymous referee for helping to improve the quality of this paper.
 SK, PMW, and CS received funding through BMBF Verbundforschung
(grants 05A14BAC, 05A14MGA, ERASMUS-F, grant 05A08BA1, and ELT-MOS, grant 05A15BA1). 
SD acknowledges BMBF support through grant 05A14MGA, and from DFG through DR281/35-1.
 AMI acknowledges support from the Spanish MINECO through project AYA2015-68217-P.
 MMR acknowledges fruitful discussions with PER and TJR. 
 The authors are thankful to A. Vazdekis for providing the (unpublished) MIUSCAT library.
 Based on observations made with the NASA/ESA Hubble Space Telescope, 
 and obtained from the Hubble Legacy Archive, which is a collaboration 
 between the Space TelescopevScience Institute (STScI/NASA), 
 the Space Telescope European Coordinating Facility (ST-ECF/ESA), 
 and the Canadian Astronomy Data Centre (CADC/NRC/CSA). 
 This research has made use of the Spanish Virtual Observatory (http://svo.cab.inta-csic.es) 
 supported from the Spanish MINECO/FEDER through grant AyA2014-55216
\end{acknowledgements}

%
%



%



\end{document}